\documentclass[]{aa} 

\usepackage{natbib}
\bibpunct{(}{)}{;}{a}{}{,} 
\usepackage{epsfig}

\newcommand{\beq}{\begin{equation}} 
\newcommand{\eeq}{\end{equation}} 
\newcommand{\beqa}{\begin{eqnarray}} 
\newcommand{\eeqa}{\end{eqnarray}} 
\newcommand{\bea}{\begin{array}} 
\newcommand{\ea}{\end{array}} 

\newcommand{\dd}{{\rm d}}

\newcommand{\lag}{\langle} 
\newcommand{\rag}{\rangle} 

\newcommand{\ii}{{\rm i}}

\newcommand{\rhob}{\overline{\rho}}

\newcommand{\ve}{{\bf e}}
\newcommand{\vk}{{\bf k}}

\newcommand{\vx}{{\bf x}}
\newcommand{\vOmega}{{\bf \Omega}}

\newcommand{\vr}{{\bf r}}

\newcommand{\tdelta}{{\tilde{\delta}}}

\newcommand{\vtheta}{\vec{\theta}}

\begin{document}

\title{Source-lens clustering and intrinsic-alignment bias of weak-lensing estimators}    
\author{Patrick Valageas}   
\institute{Institut de Physique Th\'eorique, CEA Saclay, F-91191 Gif-sur-Yvette, C\'edex, France\\
CNRS, URA 2306, F-91191 Gif-sur-Yvette, C\'edex, France}
\date{Received / Accepted }

\abstract
{}
{We estimate the amplitude of the source-lens clustering bias and of the
intrinsic-alignment bias of weak lensing estimators of the two-point and three-point
convergence and cosmic-shear correlation functions.}
{We use a linear galaxy bias model for the galaxy-density correlations, as well 
as a linear intrinsic-alignment model.
For the three-point and four-point density correlations, we use analytical or
semi-analytical models, based on a hierarchical ansatz or a combination of
one-loop perturbation theory with a halo model.}
{For two-point statistics, we find that the source-lens clustering bias is typically
several orders of magnitude below the weak lensing signal, except when we
correlate a very low-redshift galaxy ($z_2 \la 0.05$) with a higher redshift galaxy
($z_1 \ga 0.5$), where it can reach $10\%$ of the signal for the shear.
For three-point statistics, the source-lens clustering bias is typically of order
$10\%$ of the signal, as soon as the three galaxy source redshifts are not
identical. The intrinsic-alignment bias is typically about $10\%$ of the signal
for both two-point and three-point statistics. Thus, both source-lens clustering
bias and intrinsic-alignment bias must be taken into account for three-point
estimators aiming at a better than $10\%$ accuracy.}
{}

\keywords{weak gravitational lensing; cosmology: large-scale structure of Universe}

\maketitle

\section{Introduction} 
\label{Introduction}

Weak gravitational lensing of background galaxies by foreground large-scale
structures is an important probe of both the geometry of the Universe and the
growth of these large-scale structures. This makes it a powerful tool when studying
the distribution of dark matter and the nature of dark energy \citep{Albrecht2006}.
This effect arises from the deflection of light rays from distant galaxies by the
fluctuations of the gravitational potential along the line of sight
\citep{Bartelmann2001,Munshi2008}. This yields both a deformation of the
shape of the images of distant galaxies (associated with the ``cosmic shear''
$\gamma$, at lowest order) and a magnification of their flux (associated with
the ``convergence'' $\kappa$). 
Because we do not know a priori the shape or luminosity of individual 
background galaxies, cosmological studies use statistical averages over many
galaxies to detect the coherent shear due to large-scale structures (typically 
on angular scales of a few arcmin), assuming that background galaxies are 
statistically isotropic.
Thus, in practice we use the cosmic shear (through the coherent orientation of
galaxies on arcmin scales that it induces) rather than the convergence as
a probe of gravitational lensing (because it is difficult to predict the luminosity
distribution of background galaxies with a good accuracy
and we lack a standard candle).
Moreover, we usually do not measure a shear or convergence map from a galaxy
survey, that is, the fields $\gamma(\vtheta)$ or $\kappa(\vtheta)$ on some
region of the sky, but the two-point correlation
$\lag \gamma_i \gamma_j\rag(\theta)$, by averaging over all galaxy pairs $(i,j)$
separated by an angular distance $\theta$.

More precisely, weak gravitational lensing is measured from the ellipticities of galaxies,
$\epsilon_{\rm obs}$, which are related to the cosmological shear distortions $\gamma$
by $\epsilon_{\rm obs}=\epsilon_{\rm s}+\gamma$, where $\epsilon_{\rm s}$ is the
intrinsic galaxy ellipticity.
Then, assuming that intrinsic galaxy ellipticities are independent and decorrelated
from the shear, one measures the gravitational lensing signal by averaging over pairs
of galaxies. This gives estimators (that we denote with a hat) of the form
\beq
\hat\xi^{\gamma\gamma^*}(\theta) = \frac{\sum_{i,j} w_i w_j \; \epsilon(\vtheta_i)
\epsilon^*(\vtheta_j)}{\sum_{i,j} w_i w_j} ,
\label{hxi-1}
\eeq
where we sum over all galaxy pairs in the survey with an angular distance
$|\vtheta_i-\vtheta_j|$ that falls within some bin around $\theta$.
(In this example, we correlate $\epsilon$ with its complex conjugate
$\epsilon^*$ to obtain nonzero results, as the shear and the
ellipticity are spin-2 quantities.)
The weights $w_i$ may be chosen to diminish the importance of noisy objects,
to improve the signal-to-noise ratio.
This provides an estimator of the real-space two-point shear correlation function
$\xi^{\gamma\gamma^*}(\theta)=\lag\gamma_i\gamma_j^*\rag$. 
By summing over all pairs that are separated by a distance shorter than some
angular radius $\theta$, one also obtains the integral of  $\xi^{\gamma\gamma^*}$
within this scale or the variance of the smoothed shear or of the aperture
mass (e.g., \citet{Bartelmann2001,Munshi2008,Kilbinger2013}).

In this fashion, the two-point correlation 
\citep{Bacon2000,VanWaerbeke2000,Wittman2000,Hamana2003,Jarvis2006,Semboloni2006,Fu2008,Schrabback2010}
and three-point correlation \citep{Bernardeau2002a,Semboloni2011} of the cosmic
shear have been detected and measured on scales of a few arcmin.

In Eq.(\ref{hxi-1}), $\epsilon$ may be taken
as the tangential or cross-component or as the component along a given axis.
One can also consider different redshift bins for $z_i$ and $z_j$, as in tomographic
studies that make use of the redshift dependence of the lensing signal
\citep{Hu1999,Heymans2013}.
Next, one may take the Fourier transform of $\hat\xi^{\gamma\gamma^*}(\theta)$
to obtain an estimator of the weak lensing power spectrum.
This is more convenient than first taking the Fourier transform of the ellipticity
field and second taking its variance, because the masks and intricate boundaries
of galaxy surveys make it difficult to compute the Fourier transform of the field.

In practice, different sources of noise can bias the estimator (\ref{hxi-1}).
In addition to the instrumental noise itself (that may be included in
the statistical properties of $\epsilon_{\rm s}$), intrinsic galaxy alignments
must be taken care of. They come either through the correlation between nearby
galaxies, $\lag \epsilon_{\rm s}^i \epsilon_{\rm s}^j \rag$
\citep{Heavens2000,Croft2000,Catelan2001,Brown2002}, or the correlation
between the ellipticity of a foreground galaxy $i$ and the local density field, which
contributes to the shear of a background galaxy $j$ and gives rise to a correlation
$\lag \epsilon_{\rm s}^i \gamma^j \rag$
\citep{Hirata2004,Hirata2007,Mandelbaum2011,Heymans2013}.

Another source of bias arises from the fact that galaxies are not located at random
in space. Indeed, they are correlated with the density field, and this gives rise to
a ``source-lens clustering'' bias \citep{Bernardeau1998,Hamana2002}. 
For instance, in terms of the convergence $\kappa$, if galaxies
were only located behind overdense regions their luminosity would appear
systematically enhanced. This effect is expected to be rather small for 
measures of the two-point shear correlation, as compared with the full gravitational
lensing signal, because it is restricted to density fluctuations close to the observed
galaxies whereas the full signal arises from density fluctuations along the whole line
of sight. It is further suppressed by the fact that the lensing efficiency
[the kernel $g$ in Eq.(\ref{g-def}) below] vanishes at the source plane.
Moreover, on large scales this bias scales as $\xi^2$ whereas the weak lensing
signal scales as $\xi$ (where $\xi$ is the matter density correlation), so that this
bias should be subdominant.
Nevertheless, in view of the increasing accuracy of future surveys, it is interesting
to have an estimate of the magnitude of this systematic effect, to check that
it can indeed be neglected (as in all current studies).

On the other hand, for measures of the shear three-point correlation, both the
signal and the source-lens clustering bias scale as $\xi^2$ and one can
expect a significant contamination, especially for triplets of galaxies that are
at different redshifts, so that the lensing kernel $g$ is nonzero. 
Moreover, at leading order the bias writes as a sum of product of two terms.
The first term again involves the correlation between a foreground galaxy and nearby
density fluctuations along another line of sight, but the second term now involves
the correlations between density fluctuations along two full lines of sight as in
the cosmological signal and is not suppressed by the ratio of the density
correlation length to the Hubble length.

The source-lens clustering effect has already been investigated in
\citet{Bernardeau1998} and \citet{Hamana2002}, using perturbation theory and
numerical simulations, for the skewness of the smoothed convergence
as derived from a convergence map.
In this paper, to keep close to current observational procedures, we investigate
the source-lens clustering effect on estimators of the form of Eq.(\ref{hxi-1}),
without assuming a convergence map is first measured in the data analysis.
As we explain in Sect.~\ref{Comparison} below, the source-lens clustering bias
associated with such two-point and three-point estimators is rather different from
the one associated with the one-point estimator studied in
\citet{Bernardeau1998} and \citet{Hamana2002}. In particular, it has the opposite
sign.
Moreover, we consider both the convergence $\kappa$
(this allows us to introduce our approach in a simple fashion) and the
cosmic shear $\gamma$.
Then, we consider the bias due to galaxy intrinsic alignments, both for two-point 
and three-point shear correlation estimators. This allows us to extend previous works
that focused on the two-point shear correlation to the case of the three-point shear
correlation and to compare with the source-lens clustering bias.

We develop an analytical formalism to estimate these weak lensing biases 
and for numerical computations we use a linear bias model for the galaxy distribution
and a linear intrinsic alignment model.
We use semi-analytic models for the matter density two-, three- and four-point 
correlations.

This paper is organized as follows.
We first study the source-lens clustering bias for estimators of the two-point
correlation of the convergence in Sect.~\ref{galaxy-location-bias} and of the
cosmic shear in Sect.~\ref{galaxy-location-bias-shear}.
Then, we consider estimators of the three-point correlation of the convergence in
Sect.~\ref{three-point-kappa} and of the cosmic shear in
Sect.~\ref{three-point-gamma}.
We compare our approach with some previous works in Sect.~\ref{Comparison}.
Next, we investigate the intrinsic-alignment bias in Sect.~\ref{intrinsic}, for both
estimators of the two- and three-point cosmic shear correlations.
We conclude in Sect.~\ref{Conclusion}.

\section{Two-point convergence correlation function}
\label{galaxy-location-bias}

\subsection{Weak lensing convergence $\kappa$}
\label{bias-kappa}

For simplicity, we first consider the estimator of the two-point correlation function
of the convergence $\kappa$, which is a scalar rather than a spin-2 quantity. Then,
Eq.(\ref{hxi-1}) becomes
\beq
\hat\xi^{\kappa\kappa}(\theta) =  \frac{\int\dd\chi_1 \dd\vOmega_1 \, \chi_1^2 \; n_1 
\int \dd\chi_2 \dd\vtheta_2 \, \chi_2^2 \; n_2 \; \kappa_1  \kappa_2}
{\int\dd\chi_1 \dd\vOmega_1 \, \chi_1^2 \; n_1 \int \dd\chi_2 \dd\vtheta_2 \, 
\chi_2^2 \; n_2} ,
\label{hkappa-1}
\eeq
where $\chi(z)$ is the comoving radial and angular distance (we assume a flat
universe) and $n_i=n(\vx_i)$ is the observed galaxy number density (a sum of Dirac
peaks at the observed galaxy positions). We did not write weighting factors $w_i$,
which are not important for our purposes (and may be absorbed within the
number densities $n_i$). In Eq.(\ref{hkappa-1}), we count all pairs by first
counting all galaxies $i$ in the survey, of total angular area $(\Delta\Omega)$,
and next integrating over all their neighbors $j$ at the angular distance $\theta_2$.
We did not explicitly write the boundaries of the redshift and angular bins in the
integration signs.

The estimator (\ref{hkappa-1}) is somewhat academic, because in practice
we do not measure the convergences $\kappa_i$ but only the
ellipticities $\epsilon_i$ (which boil down to $\gamma_i$ if we discard
intrinsic alignments and instrumental noise). 
However, it provides a simpler presentation of our approach.
Moreover, it will be interesting to compare the results obtained for the convergence
and the shear, to see whether the latter could be estimated from the former ones.

For a sufficiently wide survey, we can neglect the fluctuations of the denominator
in Eq.(\ref{hkappa-1}). Indeed, defining
\beq
D = \int\dd\chi_1 \dd\vOmega_1 \, \chi_1^2 \; n_1 \int \dd\chi_2 \dd\vtheta_2 \, 
\chi_2^2 \; n_2 ,
\label{D-def}
\eeq
we obtain (assuming for simplicity thin redshift and angular bins $\Delta\chi$
and $\Delta\theta$)
\beqa
\lag D \rag \!\! & = & \!\!\! \int \!\! \dd\chi_1 \dd\vOmega_1 \, \chi_1^2 \; \bar{n}_1 
\int \!\! \dd\chi_2 \dd\vtheta_2 \, \chi_2^2 \; \bar{n}_2 
\; \lag (1 \!+\! b_1 \delta_1) (1 \!+\! b_2 \delta_2) \rag \nonumber \\
& = & \!\! (\Delta \chi_1) (\Delta\Omega) \chi_1^2 \bar{n}_1 (\Delta \chi_2)
(2\pi \theta \Delta\theta) \chi_2^2 \bar{n}_2 (1 \!+\! b_1 b_2 \xi_{1,2}) .
\eeqa
Here, $\delta_i=\delta(\vx_i)$ is the matter density contrast,
$\delta=(\rho-\rhob)/\rhob$, $\xi_{1,2}=\lag \delta_1 \delta_2 \rag$ its two-point
correlation function, and $b_i$ is the mean bias of galaxies at redshift $i$, assuming
a linear bias model, $n_i = \bar{n} (1+b_i \delta_i)$.
The second order moment of $D$ reads as
\beqa
\lag D^2 \rag \!\! & = & \!\!\! \int \!\! \dd\chi_1 \dd\vOmega_1 \, \chi_1^2 \; \bar{n}_1 
\int \!\! \dd\chi_2 \dd\vtheta_2 \, \chi_2^2 \; \bar{n}_2 \int 
\!\! \dd\chi_1' \dd\vOmega_1' \, \chi_1'^2 \; \bar{n}_1' \nonumber \\
&& \hspace{-1cm} \times \!\! \int \!\! \dd\chi_2' \dd\vtheta_2' \, \chi_2'^2 \; \bar{n}_2'
\; \lag (1 \!+\! b_1 \delta_1) (1 \!+\! b_2 \delta_2)  (1 \!+\! b_1' \delta_1') 
(1 \!+\! b_2' \delta_2') \rag \nonumber \\
& = & \lag D \rag^2 + \mbox{cross terms} ,
\eeqa
where the cross terms correspond to contributions that involve correlations between
$\delta_i$ and $\delta_j'$. They are negligible if $|\vx'-\vx| \gg x_0$, where $x_0$
is the correlation length, and this restricts the integral over $\vOmega_1'$ to
$\Delta\Omega_1' \sim x_0^2/\chi_1^2$.
Therefore, in terms of the scaling with respect to the survey width, we have
\beq
\lag D \rag \propto (\Delta\Omega) , \;\; \sigma_D^2 \propto (\Delta \Omega) , \;\;
\frac{\sigma_D}{\lag D\rag} \propto \frac{1}{\sqrt{(\Delta\Omega)}} ,
\label{D-sig}
\eeq
where $\sigma_D^2 = \lag D^2\rag - \lag D\rag^2$ is the variance of the denominator
of Eq.(\ref{hkappa-1}). Thus, the relative amplitude of the fluctuations of this denominator 
vanish as $1/\sqrt{(\Delta\Omega)}$ for large surveys.

Then, neglecting the fluctuations of the denominator\footnote{This is possible
for a sufficiently wide survey because the denominator sums all
pairs of separation $\theta$ over the survey. In contrast, if we consider
the one-point estimator for a convergence map $\kappa(\vtheta)$, as in the analysis of
\citet{Bernardeau1998}, for each direction $\vtheta$ on the sky the numerator and 
denominator only sum the small number of galaxies included within the smoothing
radius $\theta_s$. Then, the numerator and denominator show significant
correlated fluctuations that must be taken into account. See the discussion in
Sect.~\ref{Comparison} below.}
(there is no shot noise because of the nonzero angular separation), the
expectation value of Eq.(\ref{hkappa-1}) reads as
\beq
\lag\hat\xi^{\kappa\kappa}(\theta)\rag = \frac{\lag (1+b_1\delta_1) 
(1+b_2\delta_2)  \kappa_1 \kappa_2\rag}{1+b_1 b_ 2 \, \xi_{1,2}} .
\label{hkappa-2}
\eeq
Here, we have chosen infinitesimally thin redshift and angular bins, to avoid being too
specific. Averaging over finite redshift bins for $z_1$ and $z_2$,
and a finite angular bin for $\theta$, gives the appropriate results for a given
survey strategy.

Next, the weak lensing convergence $\kappa_i$ of a distant galaxy $i$ due to
density fluctuations along the line of sight is given by (in the Born approximation,
\citet{Bartelmann2001,Munshi2008})
\beq
\kappa_i = \int_0^{\chi_i} \dd\chi_{i'} \; g_{i',i} \; \delta(\chi_{i'}) ,
\label{kappa-def}
\eeq
 where the lensing kernel $g_{i',i}$ is given by
\beq
g(\chi_{i'},\chi_i) = \frac{3\Omega_{\rm m0} H_0^2}{2c^2} \;
\frac{\chi_{i'}  (\chi_i - \chi_{i'})}{\chi_i} \; (1+z_{i'}) ,
\label{g-def}
\eeq
and $i'$ denotes the point along the line of sight to the galaxy $i$.
(Hereafter, primed indices or coordinates refer to points along the line of sight,
which contribute to the weak lensing signal, whereas unprimed indices or coordinates
refer to the background source galaxies.)
Then, the average (\ref{hkappa-2}) can be split into four components,
\beq
\lag\hat\xi^{\kappa\kappa}\rag = \xi^{\kappa\kappa} + \xi^{\delta\kappa\delta\kappa}
+ \zeta^{\delta\kappa\kappa} + \eta^{\delta\delta\kappa\kappa} .
\label{split}
\eeq
The first component, which does not include cross-correlations between the
galaxies and the density fluctuations along the lines of sight, is the weak lensing
signal,
\beq
\xi^{\kappa\kappa}(\theta) = \lag \kappa_1 \kappa_2 \rag
= \int_0^{\chi_1} \dd\chi_{1'} g_{1',1} \int_0^{\chi_2} \dd\chi_{2'} g_{2',2} \; \xi_{1',2'} .
\label{xi-1}
\eeq
The second component involves products of the two-point correlations between
a galaxy and a line of sight,
\beqa
\xi^{\delta\kappa\delta\kappa} & = & \frac{b_1 b_2}{1+b_1 b_ 2 \, \xi_{1,2}} \;
[  \lag \delta_1 \kappa_1 \rag \lag \delta_2 \kappa_2 \rag 
+ \lag \delta_1 \kappa_2 \rag \lag \delta_2 \kappa_1 \rag ]  
\label{xi-deltakappa-01} \\
& = & \frac{b_1 b_2}{1\!+\!b_1 b_ 2 \, \xi_{1,2}} \int \dd\chi_{1'}  
\dd\chi_{2'} \, g_{1',1} g_{2',2} \nonumber \\
&& \times [ \xi_{1,1'}  \xi_{2,2'} \! + \xi_{1,2'}  \xi_{2,1'} ] ,
\label{xi-deltakappa-1}
\eeqa
while the third and fourth components involve the three- and four-point
density correlations $\zeta$ and $\eta$,
\beqa
\zeta^{\delta\kappa\kappa} & = & \frac{b_1 \lag \delta_1 \kappa_1 \kappa_2 \rag
+ b_2 \lag \delta_2 \kappa_1 \kappa_2 \rag}{1+b_1 b_ 2 \, \xi_{1,2}} \\
& = & \frac{1}{1\!+\!b_1 b_ 2 \, \xi_{1,2}} \int \dd\chi_{1'}  
\dd\chi_{2'} \, g_{1',1} g_{2',2} \nonumber \\
&& \times [ b_1 \zeta_{1,1',2'} \! + b_2 \zeta_{2,2',1'} ] ,
\label{zeta-1}
\eeqa
\beqa
\eta^{\delta\delta\kappa\kappa} & = & \frac{b_1 b_2}{1+b_1 b_ 2 \, \xi_{1,2}} \;
\lag \delta_1 \delta_2 \kappa_1 \kappa_2 \rag_c  \\
& = & \frac{b_1 b_2}{1\!+\!b_1 b_ 2 \, \xi_{1,2}} \int \dd\chi_{1'}  
\dd\chi_{2'} \, g_{1',1} g_{2',2} \; \eta_{1,2,1',2'} .
\label{eta-1}
\eeqa
Thus, the last three terms in Eq.(\ref{split}) bias the estimator (\ref{hkappa-1})
of cosmological gravitational lensing. Of course, they vanish when the galaxy bias
$b_i$ goes to zero, that is, when the galaxy positions are uncorrelated with
the density fluctuations along the lines of sight.
This source-lens clustering bias does not depend on the size of the survey
(we assumed a survey window that is large as compared with the angular scale
$\theta$ at which we probe the gravitational lensing correlation), because it is due
to the intrinsic correlations of the galaxy and matter distributions and not to shot
noise effects.

\subsection{Analytical approximations}
\label{approximations}

To estimate the source-lens clustering bias for estimators
of the two-point correlations, as in Eq.(\ref{split}), we need the matter density
three- and four-point correlation functions $\zeta$ and $\eta$.
Because we are only interested in orders of magnitude estimates and do not
require a $10\%$ or better accuracy, we use a simple 
hierarchical ansatz for these high-order density correlations
\citep{Groth1977,Peebles1980}.
Thus, we write the three-point density correlation as a sum of products of
two-point correlations,
\beq
\zeta_{1,2,3} = \frac{S_3}{3} [ \xi_{1,2} \xi_{1,3} + \xi_{2,1} \xi_{2,3} 
+ \xi_{3,1} \xi_{3,2} ] ,
\label{zeta-def}
\eeq
and in a similar fashion for the four-point density correlation,
\beq
\eta_{1,2,3,4} = \frac{S_4}{16} [ \xi_{1,2} \xi_{1,3} \xi_{1,4} + 3 {\rm cyc.} 
+ \xi_{1,2} \xi_{2,3} \xi_{3,4} + 11 {\rm cyc.} ] ,
\label{eta-def} 
\eeq
where ``3 cyc.'' and ``11 cyc.'' stand for three and eleven terms that are
obtained from the previous one by permutations over the labels ``1,2,3,4''
of the four points.
For the normalization factors $S_n$ in Eqs.(\ref{zeta-def})-(\ref{eta-def}), we
interpolate from the large-scale quasilinear limit
\citep{Bernardeau2002} (where we take an angular average to neglect the
angular dependence of $S_3$ that would arise in the exact leading-order
perturbative result)
\beq
S_3^{\rm QL} = \frac{34}{7} - (n+3) , 
\label{S3-QL}
\eeq
\beq
S_4^{\rm QL} = \frac{60712}{1323} - \frac{62}{3}(n+3) + \frac{7}{3} (n+3)^2 ,
\label{S4-QL}
\eeq
to the highly nonlinear HEPT approximation \citep{Scoccimarro1999a}
\beq
S_3^{\rm HEPT} = 3 \frac{4-2^n}{1+2^{n+1}} ,
\label{S3-HEPT}
\eeq
\beq
S_4^{\rm HEPT} = 8 \frac{54-27 \times 2^n + 2 \times 33^n+ 6^n}
{1+6 \times 2^n+3 \times 33^n + 6 \times 66^n} ,
 \label{S4-HEPT}
\eeq
as
\beq
S_n = S_n^{\rm QL} + \frac{\xi^2}{1+\xi^2} \, ( S_n^{\rm HEPT} - S_n^{\rm QL} ) .
\label{Sn-def}
\eeq
Here, $\xi$ is the two-point correlation at the scale of interest and $n$ the local
slope of the linear matter power spectrum. Thus, in the quasilinear regime, where
$\xi\ll 1$, we have $S_n \rightarrow S_n^{\rm QL}$, while in the highly
nonlinear regime, where $\xi\gg 1$, we have $S_n \rightarrow S_n^{\rm HEPT}$.
Since the density correlations only contribute on much smaller scales
($\sim 8 h^{-1}$Mpc) than the cosmological scales ($c/H_0 \sim 3000 h^{-1}$Mpc),
in the integrals such as (\ref{xi-1}) or (\ref{eta-1}) it is sufficient to use for the
two-point correlations $\xi_{i,j}$ and the coefficients $S_n$ the mean redshift of the
relevant points. [For $S_n$, we also use the geometrical mean of the relevant
scales to compute $n$ and $\xi$ in Eqs.(\ref{S3-QL})-(\ref{Sn-def}).]

\begin{figure}
\begin{center}
\epsfxsize=8.5 cm \epsfysize=6. cm {\epsfbox{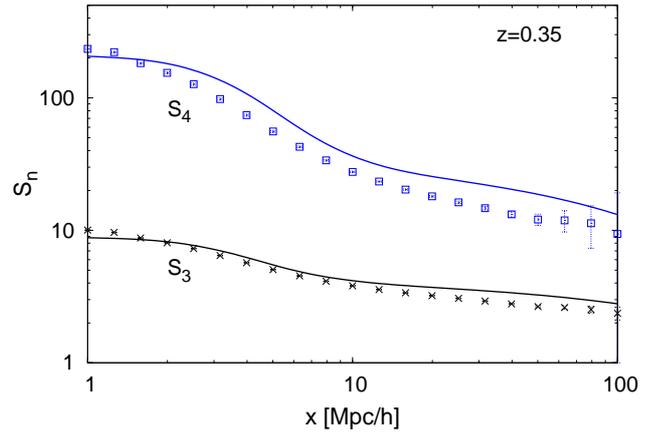}}
\end{center}
\caption{Skewness $S_3$ and kurtosis $S_4$ at redshift $z=0.35$.
The solid lines are the analytical approximation (\ref{Sn-def}) and the symbols
are results from numerical simulations \citep{Nishimichi2011} for top-hat cumulants.}
\label{fig_Sn}
\end{figure}

This ansatz is the simplest model that is in qualitative agreement with large-scale
theoretical predictions [because $\zeta \sim \xi^2$ and $\eta \sim \xi^3$ at
leading order in perturbation theory, see \cite{Goroff1986,Bernardeau2002}]
and with numerical simulations on nonlinear scales \citep{Colombi1996}.
It was already used to estimate the covariance matrices of
galaxy surveys \citep{Bernstein1994,Szapudi1996} or X-ray cluster surveys
\citep{Valageas2011c,Valageas2012}.
Its generalization to all-order density correlations was also used to compute the
high-order cumulants and the probability distributions of the smoothed convergence
and cosmic shear, providing a good agreement with results from ray-tracing
in N-body simulations 
\citep{Valageas2004a,Barber2004,Munshi2004,Munshi2005c}. 

For completeness, we check the approximation (\ref{Sn-def}) in Fig.~\ref{fig_Sn},
where we plot the coefficients $S_n$ as defined by Eq.(\ref{Sn-def})
and the skewness and kurtosis of the density contrast within spherical cells of
radius $x$ measured in numerical simulations \citep{Nishimichi2011}.
The latter are defined from the cumulants of the density contrast as
$S_n^{\rm T.H.}(R) = \lag \delta_R^n \rag_c/\lag \delta_R^2\rag^{n-1}$,
where the superscript ``T.H.'' refers to the top-hat filter.
These two definitions only coincide
if we neglect the scale dependence of the two-point correlations and of the
coefficients $S_n$ when we compute the cumulants $\lag \delta_R^n \rag_c$
from the three- and four-point correlations (\ref{zeta-def}) and (\ref{eta-def}).
However, this is sufficient for our purpose because in our numerical computations
of the cosmic shear bias below, we also factor out the coefficients $S_n$, using their
value at the typical angular scale of interest, so that geometrical integrations 
over angles, including the typical spin-2 factor $e^{2\ii\alpha}$, can be
done analytically. Thus, Fig.~\ref{fig_Sn} shows that our approximation provides
the correct order of magnitude for three- and four-point correlations.
On these scales, the match is better than $20\%$ for the skewness and
$35\%$ for the kurtosis.
It might be possible to improve the agreement with the simulations by
using another interpolation form, such as $\alpha \xi^{\beta}/(1+\alpha \xi^{\beta})$
where $\alpha$ and $\beta$ are best-fit parameters, but in this paper
we keep the simple interpolation (\ref{Sn-def}), which is sufficient for our purposes.

Since galaxies have a bias of order unity, and we are only interested in
general-purpose estimates, we take $b_i=1$ in our numerical computations
[and our results may be multiplied by the appropriate factors $b_i$ if needed,
as in Eqs.(\ref{xi-deltakappa-1})-(\ref{eta-1})].
For the nonlinear density correlation function $\xi(x)$, we use the semi-analytic
model developed in \citet{Valageas2013}, which combines one-loop perturbation
theory with a halo model to predict the matter density power spectrum and
correlation function with a percent accuracy on quasilinear scales and
a ten-percent accuracy on highly nonlinear scales.
For cosmological parameters, we use the best fit $\Lambda$CDM cosmology
from Planck observations \citep{Planck-Collaboration2013}.

In numerical computations, we keep the real-space expressions
(\ref{xi-1})-(\ref{eta-1}), rather than going to Fourier space. 
This avoids integrations over oscillatory kernels, such as Bessel functions, and
the use of Limber's approximation.
Indeed, in configuration-space expressions such as Eq.(\ref{xi-1}), which involves
the correlation $\xi_{1',2'}$ between density fluctuations $\delta_{1'}$ and
$\delta_{2'}$ along two lines of sight, Limber's approximation corresponds to
setting $\chi_{1'} = \chi_{2'}$ in cosmological kernels such as the lensing efficiency
$g(\chi',\chi)$ \citep{Limber1953,Munshi2008}.
This is because the density correlation is negligible beyond $x_0 \sim 8 h^{-1}$Mpc
whereas cosmological kernels vary on much larger scales 
$\sim c/H_0 \sim 3000 h^{-1}$Mpc.
However, if we use this approximation in 
Eqs.(\ref{xi-deltakappa-1})-(\ref{eta-1}), we obtain
$\xi^{\delta\kappa\delta\kappa}=\zeta^{\delta\kappa\kappa}
=\eta^{\delta\delta\kappa\kappa}=0$, because $g(\chi',\chi)=0$ at $\chi'=\chi$.
This also means that the source-lens clustering contributions in Eq.(\ref{split})
will be suppressed by a factor $\sim x_0/(c/H_0)$ (which vanishes in Limber's
limit) and that the computation of this source-lens clustering bias requires going
beyond Limber's approximation.
(This is no longer the case for the three-point estimators studied in
Sects.~\ref{three-point-kappa} and \ref{three-point-gamma}, where we use Limber's
approximation because the source-lens clustering contributions do not vanish in
this limit).

\subsection{Numerical results}
\label{results-convergence}

\begin{figure}[h]
\begin{center}
\epsfxsize=8.5 cm \epsfysize=5.8 cm {\epsfbox{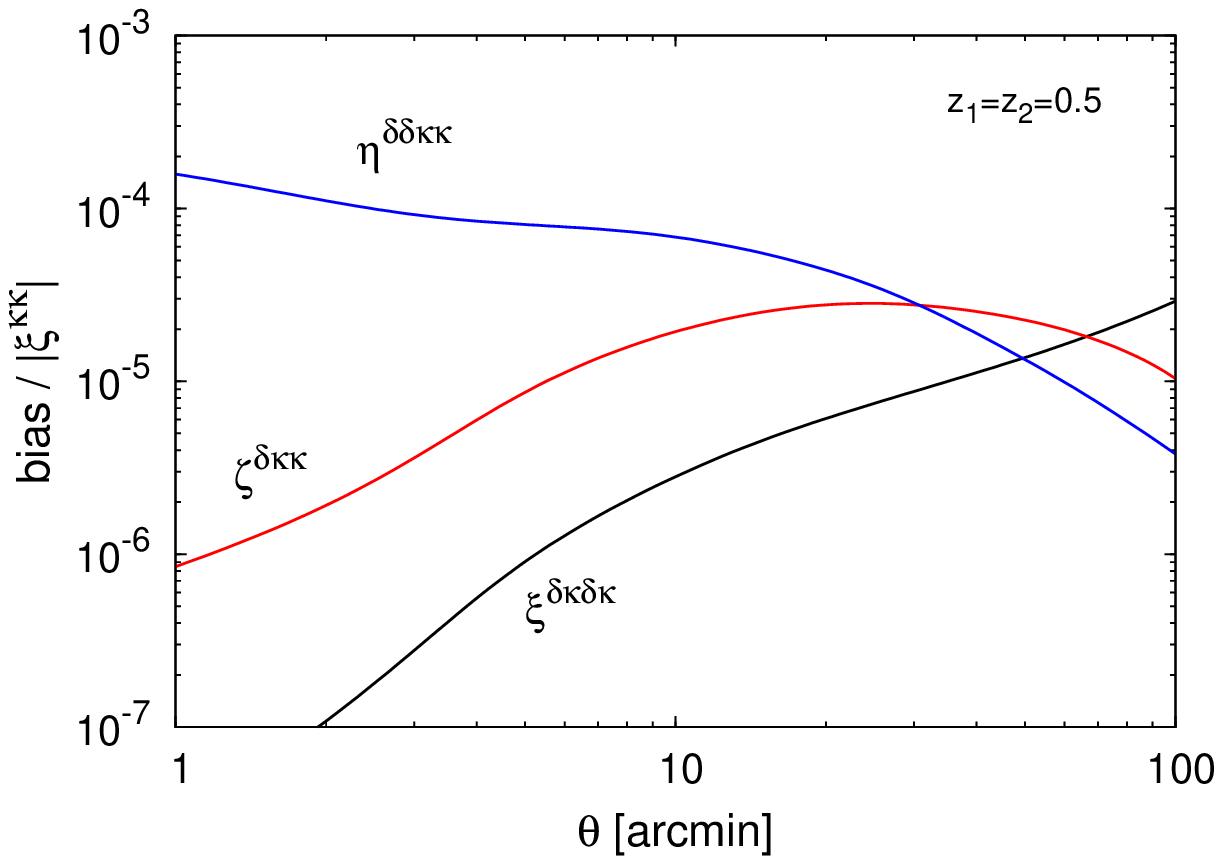}}
\epsfxsize=8.5 cm \epsfysize=5.8 cm {\epsfbox{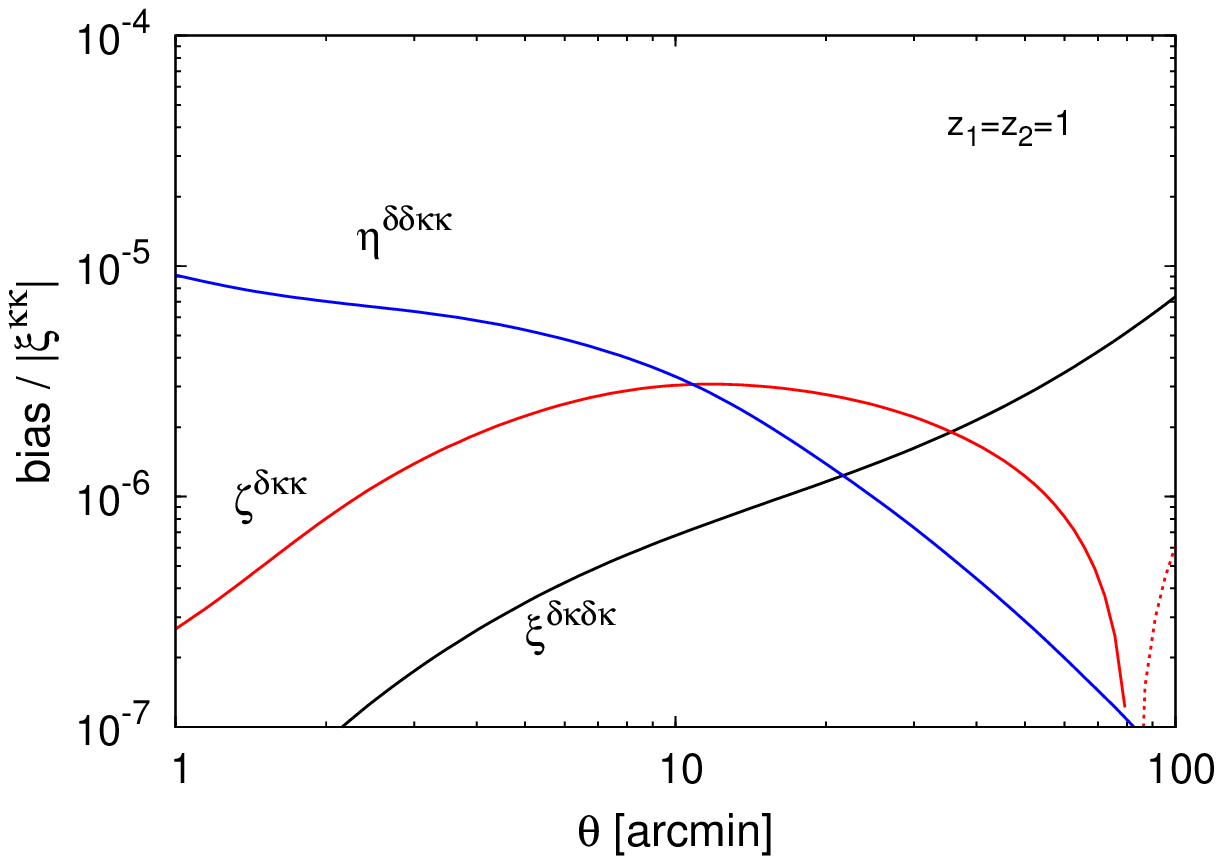}}
\epsfxsize=8.5 cm \epsfysize=5.8 cm {\epsfbox{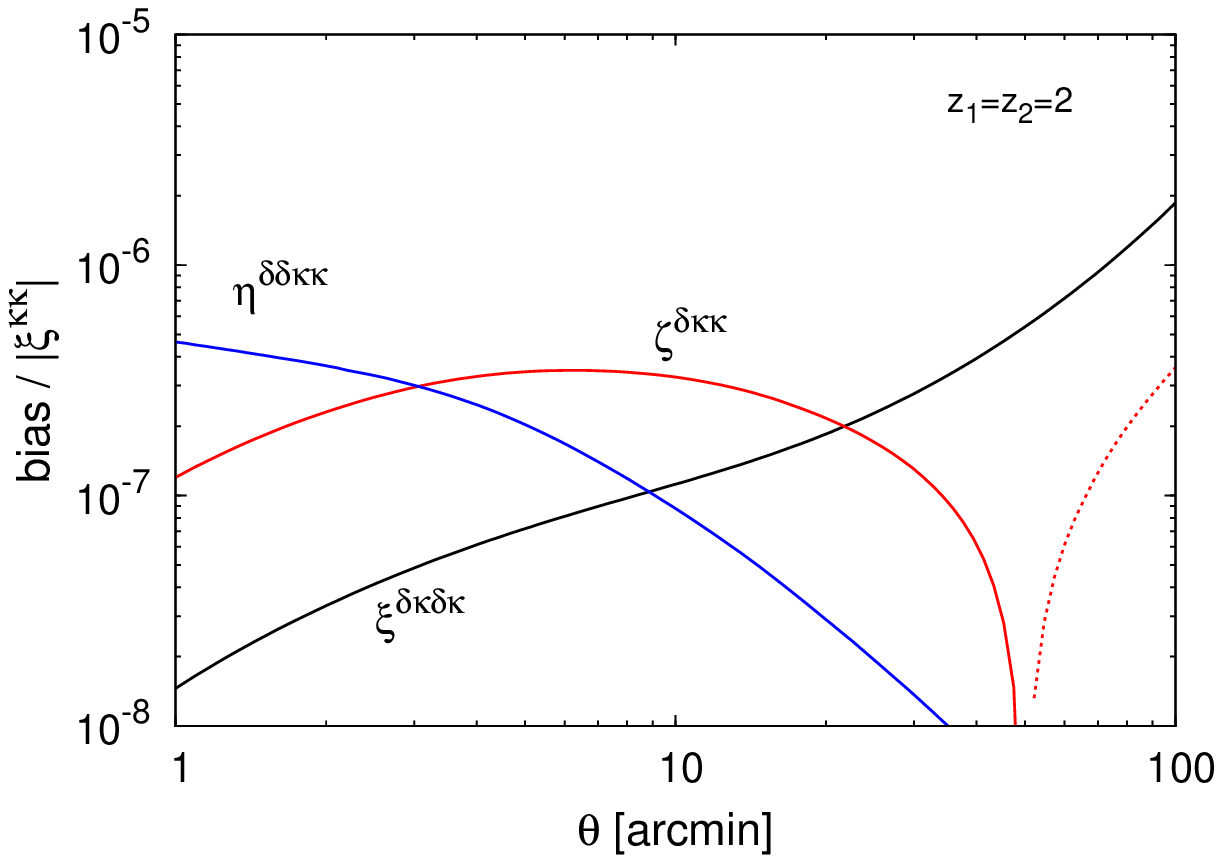}}
\end{center}
\caption{Relative source-lens clustering 
bias of the weak lensing convergence two-point correlation
$\xi^{\kappa\kappa}$, as a function of the angular scale $\theta$, for the
three pairs of coincident source redshifts $z_1=z_2=0.5$, $1$, and $2$, from
top to bottom. In each panel, we show the ratios
$\xi^{\delta\kappa\delta\kappa}/|\xi^{\kappa\kappa}|$ (lower black line),
$\zeta^{\delta\kappa\kappa}/|\xi^{\kappa\kappa}|$ (middle red line), and
$\eta^{\delta\delta\kappa\kappa}/|\xi^{\kappa\kappa}|$ (upper blue line).
The spike for $\zeta^{\delta\kappa\kappa}$ is due
to a change of sign and this contribution to the bias is negative at large angles
(in all figures in this paper, a positive/negative bias is shown by a solid/dotted line).}
\label{fig_xi_kappa_z}
\end{figure}

\begin{figure}
\begin{center}
\epsfxsize=8.5 cm \epsfysize=6. cm {\epsfbox{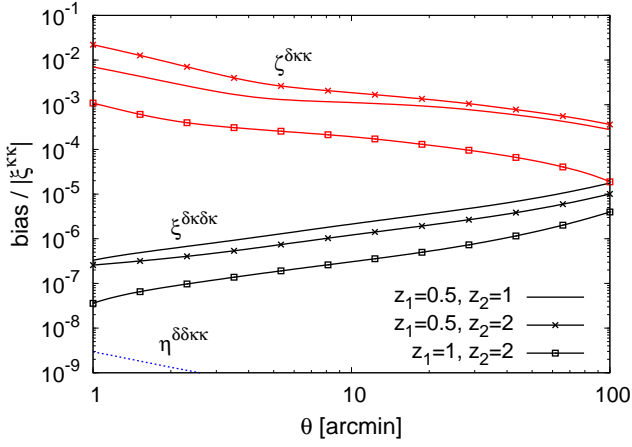}}
\end{center}
\caption{Same as in Fig.~\ref{fig_xi_kappa_z}, but for pairs of different source
redshifts, $(z_1,z_2)=(0.5,1)$, $(0.5,2)$, and $(1,2)$. The contribution
$\eta^{\delta\delta\kappa\kappa}$ is negative as shown by the dotted line.}
\label{fig_xi_kappa_z_z}
\end{figure}

\begin{figure}[h]
\begin{center}
\epsfxsize=8.5 cm \epsfysize=5.8 cm {\epsfbox{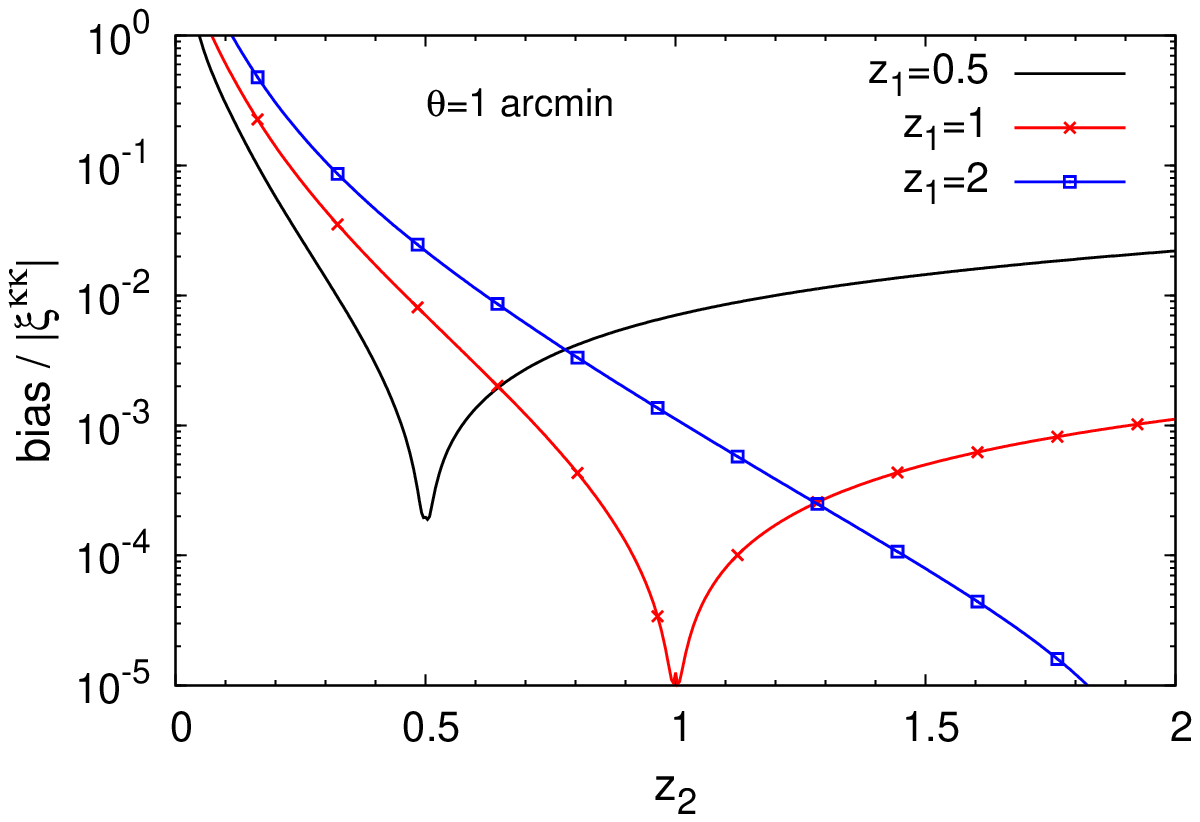}}
\epsfxsize=8.5 cm \epsfysize=5.8 cm {\epsfbox{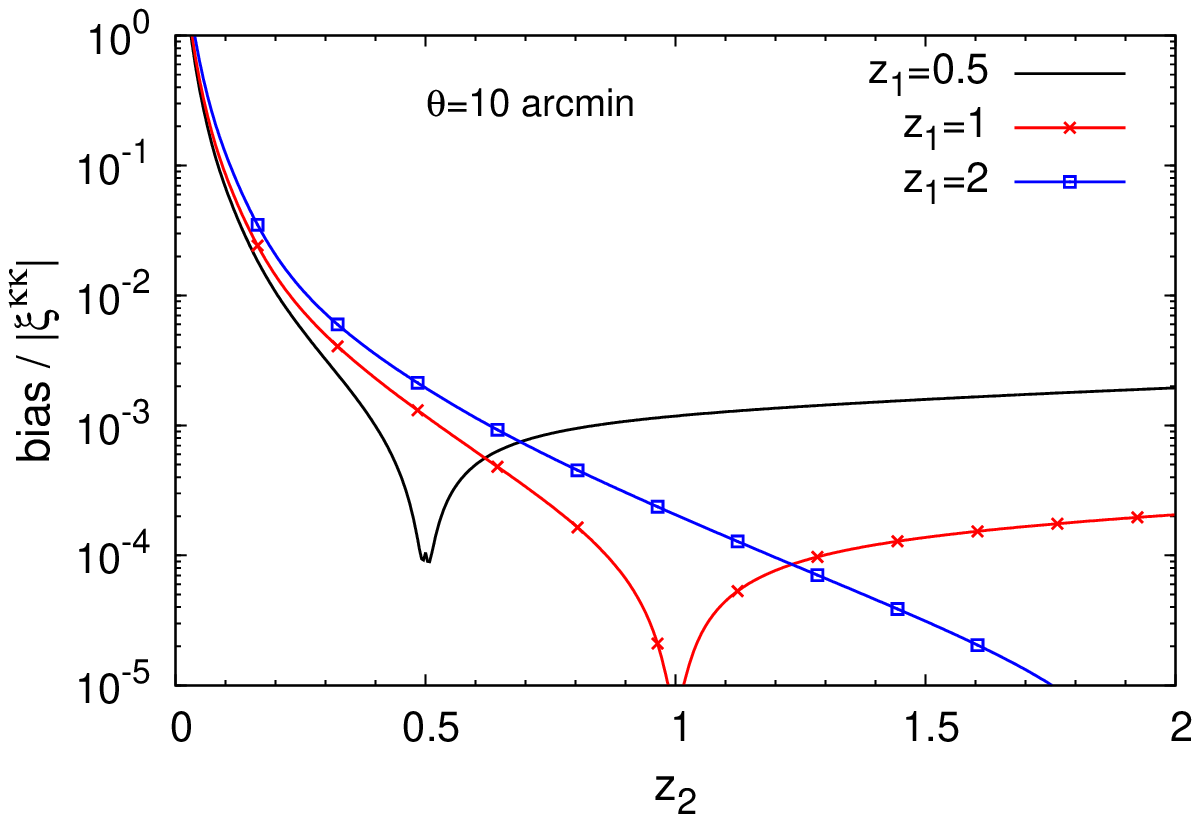}}
\epsfxsize=8.5 cm \epsfysize=5.8 cm {\epsfbox{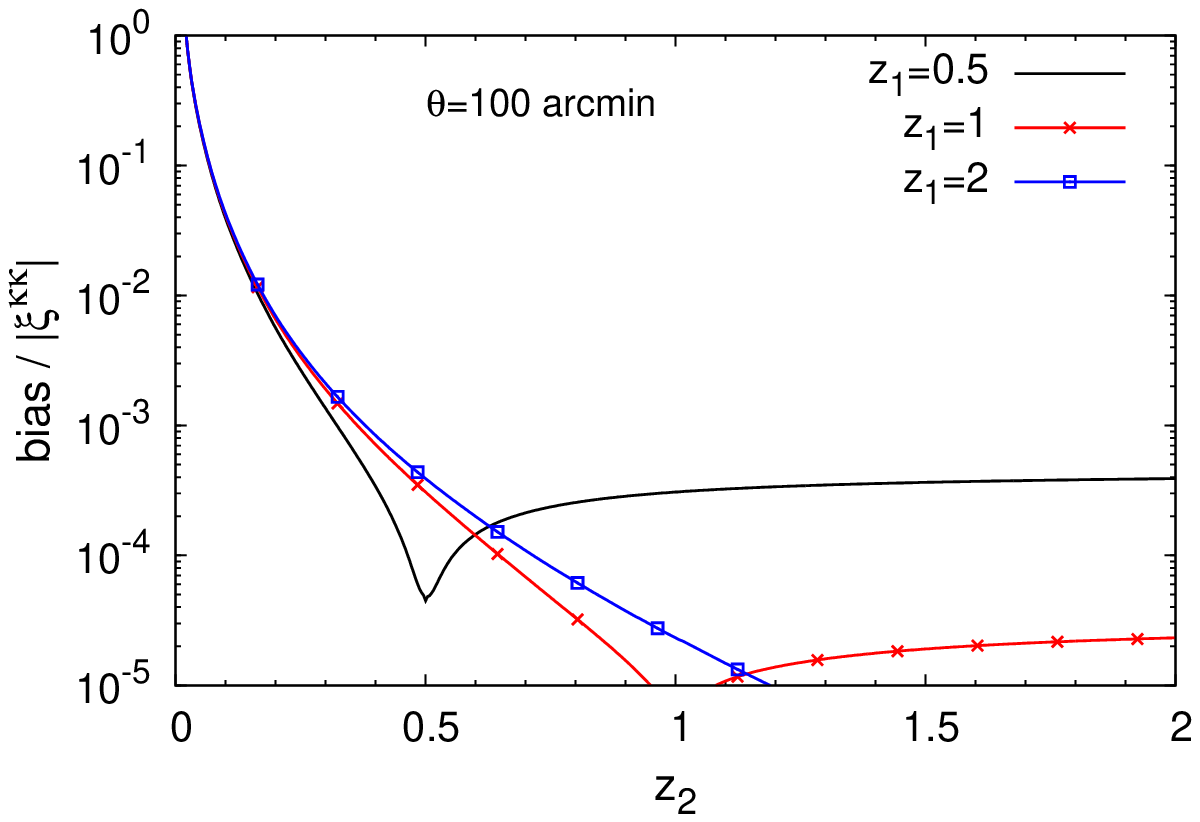}}
\end{center}
\caption{Relative source-lens clustering bias of the weak lensing convergence
two-point correlation,
$(\xi^{\delta\kappa\delta\kappa}+\zeta^{\delta\kappa\kappa}+\eta^{\delta\delta\kappa\kappa})/|\xi^{\kappa\kappa}|$,
as a function of the second galaxy redshift $z_2$, for a fixed first galaxy redshift 
$z_1=0.5, 1$, or $2$.
We consider the angular scales $\theta=1$, $10$, and $100$ arcmin, from the upper
to lower panel. All total biases are positive.}
\label{fig_xi_kappa_theta}
\end{figure}

We show our results for cases where the two galaxy redshifts are the same 
in Fig.~\ref{fig_xi_kappa_z}.
[Throughout this paper, a positive (resp. negative) bias is shown by a solid
(resp. dotted) line in the figures.]

On small angular scales, the total bias is dominated by the four-point correlation
contribution (\ref{eta-1}), because it scales as $\xi^3$ and grows faster than the
other terms in the nonlinear regime.
This arises from correlations between the two nearby source galaxies and close
density fluctuations on the two lines of sight.
On very large scales, the bias is dominated by the first term in
Eq.(\ref{xi-deltakappa-1}), which does not depend on the angular scale $\theta$
because it arises from the correlation between each galaxy and density fluctuations
along its line of sight, whereas the other terms and the weak lensing signal
$\xi^{\kappa\kappa}$ itself decrease for larger angles $\theta$ as they involve
correlations between the two lines of sight.

In all cases, the last three terms in Eq.(\ref{split}) only give rise to a relative bias
of the weak lensing estimator that is smaller than about $10^{-4}$.
This can be safely neglected for all practical purposes.
This is due to the fact that:

(a) this bias only arises
from density fluctuations close to the source galaxies, whereas the
weak lensing signal is generated by density fluctuations along the whole line
of sight, 

(b) the lensing efficiency $g(\chi_{i'},\chi_i)$ of Eq.(\ref{g-def})
vanishes at the source plane, $\chi_{i'}=\chi_i$, which further suppresses the bias
by a factor of order $x_0/(c/H_0)$, where $x_0 \sim 8 h^{-1}$Mpc is the typical
correlation length, and 

(c) the bias scales as $\xi^2$ whereas the signal scales
as $\xi$.

We show our results for cases where the two galaxy redshifts are different
in Fig.~\ref{fig_xi_kappa_z_z}.
Let us choose for instance $z_1<z_2$.
Then, the contribution $\xi^{\delta\kappa\delta\kappa}$ is now always dominated
by the first term in Eq.(\ref{xi-deltakappa-1}), because the correlation
$\xi_{2,1'}$ in the second term in Eq.(\ref{xi-deltakappa-1}) is very small
since $z_{1'} \leq z_1 < z_2$. Nevertheless, this gives an overall contribution 
$\xi^{\delta\kappa\delta\kappa}$ to the relative bias
that is again smaller than about $10^{-5}$, as for the similar coincident redshifts
of Fig.~\ref{fig_xi_kappa_z}.
In contrast, the four-point correlation
$\eta_{1,2,1',2'}$ and its bias contribution (\ref{eta-1}) are now very small, several
orders of magnitude below the corresponding contribution obtained for similar
coincident redshifts, as expected because it involves correlations between
density fields at different redshifts.
The only significant contribution that is left is the first term in Eq.(\ref{zeta-1}),
which involves the three-point correlation $\zeta_{1,1',2'}$, because it is still possible
for the three points $\{1,1',2'\}$ to be located at about the same redshift.
Moreover, this contribution is now much greater than for coincident redshifts
and can reach $1\%$.
This is because this contribution is dominated
by configurations where the points $1'$ and $2'$ are at about the redshift $z_1$,
and while the lensing kernel $g_{1',1}$ is still suppressed by a factor
of order $x_0/(c/H_0)$, the kernel $g_{2',2}$ is now of the same order as its typical
value along the line of sight to galaxy $2$, because $z_{2'}$ is now significantly
different from $z_2$ since $z_{2'} \simeq z_1 < z_2$.

To see more clearly the redshift dependence of the source-lens clustering bias,
we show our
results as a function of the second galaxy redshift $z_2$, for a fixed first galaxy
redshift $z_1$, in Fig.~\ref{fig_xi_kappa_theta}.
In agreement with the discussion above, the relative bias
$(\xi^{\delta\kappa\delta\kappa}+\zeta^{\delta\kappa\kappa}+\eta^{\delta\delta\kappa\kappa})/\xi^{\kappa\kappa}$
is minimum for $z_2=z_1$ because of the vanishing of both lensing kernels
$g_{1',1}$ and $g_{2',2}$ for $z_{1'} = z_{2'} = z_1=z_2$.
For $z_2 \gg z_1$, the lensing signal $\xi^{\kappa\kappa}$ saturates because it
is dominated by density fluctuations in the common redshift range,
$z_{1'} \simeq z_{2'} \leq \min(z_1,z_2)=z_1$, whereas the bias is dominated
by the three-point correlation $\zeta_{1,1',2'}$ (with a similar kernel $g_{2',2}$),
so that the relative bias also saturates and remains small.
For $z_2 \ll z_1$, the lensing signal decreases with the length of the common
redshift range, $z_{1'} \simeq z_{2'} \leq \min(z_1,z_2)=z_2$, 
while the bias is dominated by the three-point correlation $\zeta_{2,2',1'}$.
This leads to a steep growth of the relative bias for very small $z_2$
(because the short line of sight diminishes the signal, which only arises from scales
where the three-point correlation is significant and contributes to the bias).
The comparison between the panels shows that the relative amplitude of the
bias decreases on larger angular scales, because the three-point correlation
is smaller.

Thus, most of the source-lens clustering bias arises from the three-point correlation
between a low-redshift source galaxy, nearby density fluctuations on its line of sight,
and density fluctuations at about the same redshift on the line of sight to a second
higher redshift source galaxy.
Figure~\ref{fig_xi_kappa_theta} shows that this source-lens clustering bias is
almost always negligible, except when we cross-correlate the gravitational
lensing distortions of a low-redshift galaxy, $z_2 \la 0.2$, with a
higher redshift galaxy, $z_1 \ga 0.5$
(the effect being larger for higher $z_1$ and smaller $\theta$).
There, the bias can actually dominate the weak lensing signal.

\section{Two-point cosmic shear correlation function}
\label{galaxy-location-bias-shear}

\subsection{Weak lensing shear $\gamma$}
\label{bias-gamma}

The measure of the convergence $\kappa$ from galaxy surveys is not easy,
because galaxies do not have a unique luminosity that could serve as a
standard candle. In practice, one rather measures the two-point correlation
function of the cosmic shear from the galaxy ellipticities, as in Eq.(\ref{hxi-1}).
Using in the following the flat sky approximation (which is sufficient
for small angular scales), the shear $\gamma$ is given by
\beq
\gamma = \int_0^{\chi_i} \dd \chi_{i'} \, g_{i',i} \, \int\dd\vk_{i'} \, 
e^{\ii\vk_{i'}\cdot\vx_{i'}} \, e^{2\ii \alpha_{\vk_{i'}}} \, \tdelta(\vk_{i'}) ,
\label{gamma-def}
\eeq
where $\tdelta(\vk) = (2\pi)^{-3} \int \dd\vx \, e^{-\ii\vk\cdot\vx} \delta(\vx)$ is the
Fourier transform of the density contrast, $\vx_{i'}=(\chi_{i'},\chi_{i'}\vtheta_{i'})$
is the position of point $i'$ along the line of sight, and $\alpha_{\vk}$ is the polar angle
of the wave vector $\vk$ in the plane that is orthogonal to the line of sight.
The shear $\gamma$ is a complex quantity, $\gamma=\gamma_x+\ii\gamma_y$,
where $\gamma_x$ and $\gamma_y$ are the real components along the two
axis $\ve_x$ and $\ve_y$ in the transverse plane. Because of the factor
$e^{2\ii \alpha_{\vk}}$, it is also a spin-2 quantity
\citep{Bartelmann2001,Munshi2008}. 
This additional factor makes the computations somewhat heavier than for the
convergence $\kappa$.
From the shear $\gamma$ one may compute several correlation functions,
such as $\lag\gamma\gamma^*\rag$, $\lag\gamma_x\gamma_x\rag$,
$\lag\gamma_y\gamma_y\rag$, $\lag\gamma_{\rm t}\gamma_{\rm t}\rag$,
$\lag\gamma_{\times}\gamma_{\times}\rag$, where $\gamma_{\rm t}$ and
$\gamma_{\times}$ are the tangential and cross-components, with respect to
the direction of the pair in the transverse plane.
They can all be expressed in terms of the convergence two-point correlation
(\ref{xi-1}) and have similar magnitudes 
\citep{Kaiser1992,Bartelmann2001,Munshi2008}. For our purposes we focus on the
full shear correlation, $\lag\gamma\gamma^*\rag$, as in Eq.(\ref{hxi-1}).
Then, in a fashion similar to Eq.(\ref{hkappa-2}), the average of this estimator
writes as
\beq
\lag\hat\xi^{\gamma\gamma^*}(\theta)\rag = \frac{\lag (1+b_1 \delta_1) 
(1+b_2 \delta_2)  \gamma_1 \gamma_2^*\rag}{1+b_1 b_ 2 \, \xi_{1,2}} ,
\label{hgamma-2}
\eeq
where again the indices $1$ and $2$ refer to the two lines of sight separated by the
angular distance $\theta$.
This average can be split into four components,
\beq
\lag\hat\xi^{\gamma\gamma^*}\rag = \xi^{\gamma\gamma^*} 
+ \xi^{\delta\gamma\delta\gamma^*} + \zeta^{\delta\gamma\gamma^*} 
+ \eta^{\delta\delta\gamma\gamma^*} .
\label{split-gamma}
\eeq
The first component is again the weak lensing signal and it is equal to
the convergence correlation (\ref{xi-1}),
\beq
\xi^{\gamma\gamma^*}(\theta) = \xi^{\kappa\kappa}(\theta) 
= \int_0^{\chi_1} \dd\chi_{1'} g_{1',1} \int_0^{\chi_2} \dd\chi_{2'} g_{2',2} \; \xi_{1',2'} .
\label{xi-gamma-1}
\eeq
The second component involves products of the two-point correlations between
the galaxies and the density fluctuations along the line of sight,
\beq
\xi^{\delta\gamma\delta\gamma^*} = \frac{b_1 b_2 }{1+b_1 b_ 2 \, \xi_{1,2}} \;
\lag \delta_1 \gamma^*_2\rag \lag \delta_2 \gamma_1\rag .
\label{xi-deltagamma-1}
\eeq
As compared with Eq.(\ref{xi-deltakappa-01}), there is no term
$\lag \delta_1 \gamma_1\rag \lag \delta_2 \gamma^*_2\rag$ because it vanishes
thanks to the spin-2 factor $e^{\ii2\alpha}$. This is not the case for the
cross term $\lag \delta_1 \gamma^*_2\rag \lag \delta_2 \gamma_1\rag$, where each
product breaks the rotational invariance as it connects two different lines of sight
(which defines a prefered direction) and the final contribution is nonzero
(this direction is the same for the two terms so that averaging over the
direction $\vtheta$ does not yield a null result).

Whereas scalar quantities like the convergence only involve the
matter density two-point correlation function, $\xi(x)$, which is the
the Fourier transform of the power spectrum,
\beq
\xi(x) = \int \dd\vk \, e^{\ii\vk\cdot\vx} \, P(k) ,
\label{P-xi-def}
\eeq
for quantities that involve the spin-2 cosmic shear, we also need the integral with
a spin-2 factor $e^{2\ii \alpha_{\vk}}$,
\beq
\int \dd\vk \, e^{\ii\vk\cdot\vx} \,  e^{2\ii \alpha_{\vk}} \, P(k) = 
e^{2\ii \alpha_{\vx}} \; \xi^{(2)}(x_{\parallel},x_{\perp})  ,
\label{2_xi-def}
\eeq
where $x_{\parallel}$ and $\vx_{\perp}$ are the longitudinal and transverse
components of the separation vector $\vx$ with respect to the line of sight.
As shown in App.~\ref{density-shear-2pt}, this correlation function reads
as
\beq
\xi^{(2)}(\vx) = \xi(x) - \int_0^{x_{\perp}} \frac{\dd r_{\!\perp} \, 
2r_{\!\perp}}{x_{\perp}^2} \; \xi(x_{\parallel},r_{\!\perp})  ,
\label{2_xi-1}
\eeq
and $\xi^{(2)}(\vx)=0$ if $x_{\perp}=0$.
Then, Eq.(\ref{xi-deltagamma-1}) also reads as (see App.~\ref{density-shear-2pt})
\beq
\xi^{\delta\gamma\delta\gamma^*} = \frac{b_1 b_2 }{1+b_1 b_ 2 \, \xi_{1,2}} \;
\int \!\! \dd\chi_{1'}  \dd\chi_{2'} \, g_{1',1} g_{2',2} \; \xi^{(2)}_{1,2'} \, 
\xi^{(2)}_{2,1'}  \; .
\label{xi-deltagamma-2}
\eeq
As compared with Eq.(\ref{xi-deltakappa-1}), the first product $\xi_{1,1'}  \xi_{2,2'}$
vanishes because of the spin-2 factor $e^{\ii2\alpha}$, as explained above, whereas
in the second product $\xi_{1,2'}  \xi_{2,1'}$ the scalar correlation $\xi$ is replaced
by the ``spin-2 correlation'' $\xi^{(2)}$.
Because of the subtraction in Eq.(\ref{2_xi-1}), associated with the constraint
$\xi^{(2)}(\vx)=0$ if $x_{\perp}=0$, $|\xi^{(2)}|$ is usually smaller than
$|\xi|$. Therefore, the spin-2 factor $e^{\ii2\alpha}$ decreases the
amplitude of the source-lens clustering bias of the cosmic shear, as compared with the
convergence.
 
The third and fourth components involve the three- and four-point
density correlations and read as
\beq
\zeta^{\delta\gamma\gamma^*} = \frac{b_1 \lag \delta_1 \gamma_1 \gamma_2^* \rag 
+ b_2 \lag \delta_2 \gamma_1 \gamma_2^* \rag}
{1+b_1 b_ 2 \, \xi_{1,2}} ,
\label{zeta-gamma-1}
\eeq
\beq
\eta^{\delta\delta\gamma\gamma^*} = \frac{b_1 b_2}{1+b_1 b_ 2 \, \xi_{1,2}} 
\lag \delta_1 \delta_2 \gamma_1 \gamma_2^* \rag_c  .
\label{eta-gamma-1}
\eeq
For instance, using Eq.(\ref{gamma-def}), the first average that enters the
numerator in Eq.(\ref{zeta-gamma-1}) reads as
\beqa
\lag \delta_1 \gamma_1 \gamma_2^* \rag  & = & 
\int \!\! \dd\chi_{1'} \dd\chi_{2'} \, g_{1',1} \, g_{2',2} \int\dd\vk_1 \dd\vk_{1'}\dd\vk_{2'}
\nonumber \\
&& \times \, e^{\ii (\vk_1\cdot\vx_1+\vk_{1'}\cdot\vx_{1'}+\vk_{2'}\cdot\vx_{2'})} 
\, e^{2\ii (\alpha_{\vk_{1'}}-\alpha_{\vk_{2'}})} \nonumber \\
&& \times \, \delta_D(\vk_1+\vk_{1'}+\vk_{2'}) \, B(k_1,k_{1'},k_{2'})  ,
\label{zeta-gamma-2}
\eeqa
where $B(k_1,k_2,k_3;z)$ is the matter density bispectrum.
Because this contribution is dominated by configurations where the points
$\{1,1',2'\}$ are nearby and at almost the same redshift (otherwise the three-point
correlation is negligible), the bispectrum can be taken at the mean redshift
of these three points.

\begin{figure}
\begin{center}
\epsfxsize=8.5 cm \epsfysize=6. cm {\epsfbox{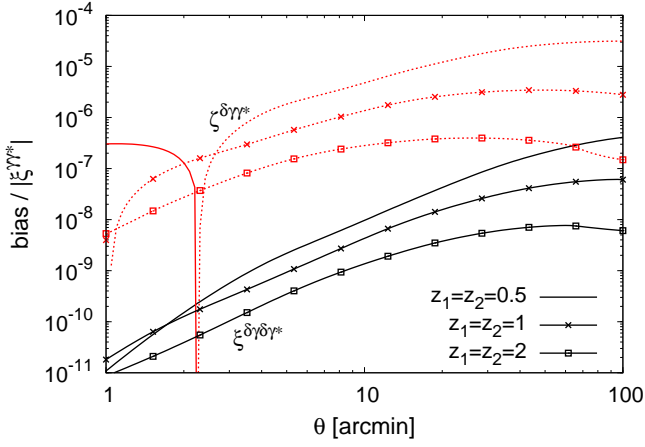}}
\end{center}
\caption{Relative source-lens clustering bias of the weak lensing shear two-point 
correlation $\xi^{\gamma\gamma^*}$, as a function of the angular scale $\theta$,
for the three pairs of coincident source redshifts $z_1=z_2=0.5$, $1$, and $2$.
The lower curves show the two-point contribution (\ref{xi-deltagamma-2}) and
the upper curves the three-point contribution (\ref{zeta-gamma-1}).
The spike for $\zeta^{\delta\gamma\gamma^*}$ is due to a change of sign and at
large angles this contribution to the bias is negative (dotted lines).}
\label{fig_xi_gamma_z}
\end{figure}

\begin{figure}
\begin{center}
\epsfxsize=8.5 cm \epsfysize=6. cm {\epsfbox{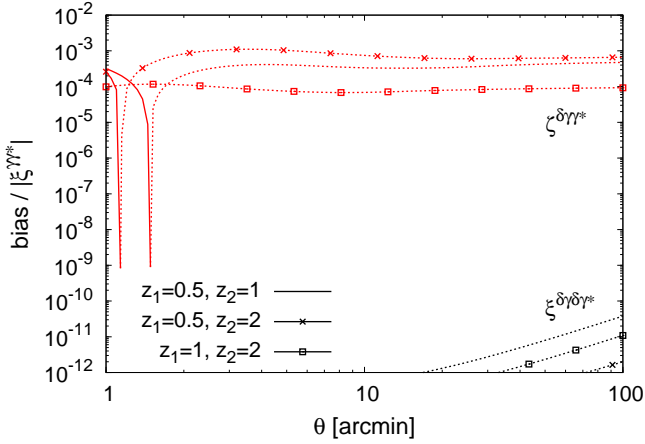}}
\end{center}
\caption{Same as in Fig.~\ref{fig_xi_gamma_z}, but for pairs of different source
redshifts, $(z_1,z_2)=(0.5,1)$, $(0.5,2)$, and $(1,2)$. Solid lines correspond to
positive bias and dotted lines to negative bias.}
\label{fig_xi_gamma_z_z}
\end{figure}

\begin{figure}
\begin{center}
\epsfxsize=8.5 cm \epsfysize=6. cm {\epsfbox{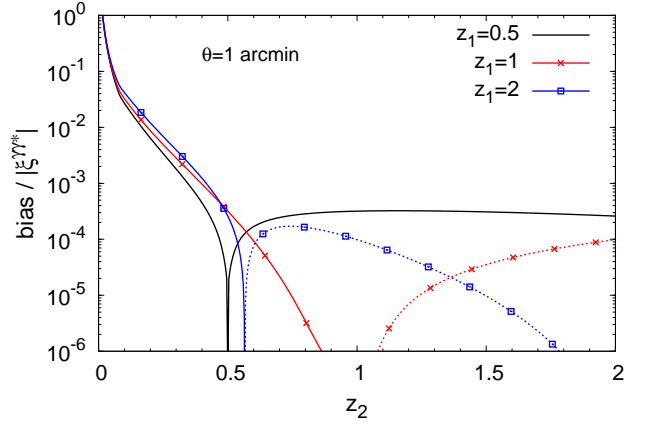}}
\epsfxsize=8.5 cm \epsfysize=6. cm {\epsfbox{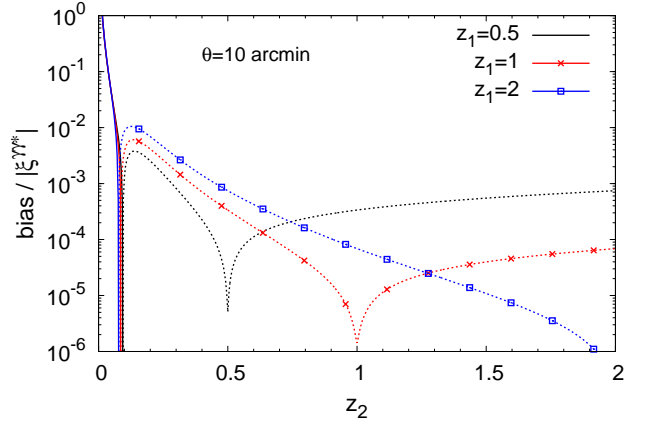}}
\epsfxsize=8.5 cm \epsfysize=6. cm {\epsfbox{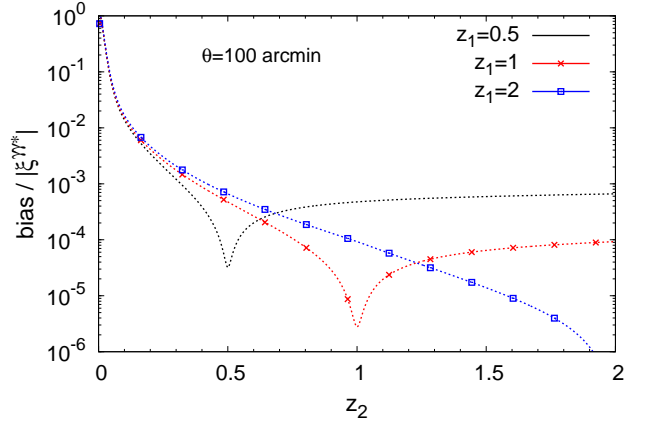}}
\end{center}
\caption{Relative source-lens clustering bias of the weak lensing shear two-point 
correlation,
$(\xi^{\delta\gamma\delta\gamma^*}+\zeta^{\delta\gamma\gamma^*})
/|\xi^{\gamma\gamma^*}|$ (we neglect the four-point contribution), as a function
of the second
galaxy redshift $z_2$, for a fixed first galaxy redshift  $z_1=0.5,1$, or $2$.
We consider the angular scales $\theta=1$, $10$, and $100$ arcmin, from the upper
to lower panel.}
\label{fig_xi_gamma_theta}
\end{figure}

\subsection{Analytical approximations}
\label{approximations-shear}

For the shear, computations are not as straightforward because of the
spin-2 factor $e^{2\ii \alpha}$. As we have seen in Sect.~\ref{results-convergence},
the source-lens clustering bias is only important when we correlate a high-redshift
galaxy, $z_2 \ga 0.5$, with a low-redshift galaxy, $z_1 \la 0.2$.
Then, the bias is dominated by
the first term in Eq.(\ref{zeta-gamma-1}), which involves the three-point
correlation between the low-redshift galaxy with density fluctuations at almost
the same redshift on the two lines of sight.
Therefore, we neglect the four-point contribution (\ref{eta-gamma-1}) and we
only consider the contributions (\ref{xi-deltagamma-2}) and (\ref{zeta-gamma-1}).
In Fourier space, neglecting the scale dependence of the coefficient $S_3$,
the ansatz (\ref{zeta-def}) yields the factorized bispectrum
\beqa
B(k_1,k_2,k_3) & = & \frac{S_3}{3}  [ P(k_2) P(k_3) + P(k_1) P(k_3) 
\nonumber \\
&& + P(k_1) P(k_2) ] .
\label{bispectrum-def}
\eeqa
As described in App.~\ref{density-shear-shear-3pt}, substituting the ansatz
(\ref{bispectrum-def}) into Eq.(\ref{zeta-gamma-2}) gives
\beq
\lag \delta_1 \gamma_1 \gamma_2^* \rag  =  
\int \!\! \dd\chi_{1'} \dd\chi_{2'} \, g_{1',1} \, g_{2',2} \frac{S_3}{3} \left[
\zeta_{1,1',2'}^{(1,1')} + \zeta_{1,1',2'}^{(1,2')} \right] ,
\label{zeta-gamma-3}
\eeq
where $\zeta_{1,1',2'}^{(1,1')}$ and $\zeta_{1,1',2'}^{(1,2')}$ are given by
Eqs.(\ref{zeta1-3}) and (\ref{zeta2-3}) (and the contribution $\zeta_{1,1',2'}^{(1',2')}$
vanishes because of the spin-2 factor $e^{2\ii \alpha}$).
A symmetric expression gives
$\lag \delta_2 \gamma_1 \gamma_2^* \rag$ and this yields
the three-point contribution (\ref{zeta-gamma-1}).

\subsection{Numerical results}
\label{results-shear}

We show our results for galaxy pairs at the same redshift in
Fig.~\ref{fig_xi_gamma_z}. As compared with the case of the convergence
shown in Fig.~\ref{fig_xi_kappa_z}, the three-point contribution is somewhat
smaller while the two-point contribution is several orders of magnitude smaller.
This is because of the spin-2 factor that replaces the density correlation $\xi$
by the smaller correlation $\xi^{(2)}$ in Eq.(\ref{xi-deltagamma-2}) and
removes the contribution
$\lag \delta_1 \gamma_1\rag \lag \delta_2 \gamma^*_2\rag$.
This implies that the contribution $\xi^{\delta\gamma\delta\gamma^*}$
now decreases at large angles $\theta$ so that the relative bias does not show
the faster growth found in Fig.~\ref{fig_xi_kappa_z}.
In any case, Fig.~\ref{fig_xi_gamma_z} shows that as for the convergence
the source-lens clustering bias is negligible for same-redshift sources.

We show our results for cases where the two galaxy redshifts are different in
Fig.~\ref{fig_xi_gamma_z_z}.
We find again that the three-point contribution is somewhat smaller than for the
case of the convergence shown in Fig.~\ref{fig_xi_kappa_z_z}, especially on small 
scales, where it only reaches $0.1\%$ instead of $1\%$.
The two-point contribution is several orders of magnitude smaller than for the
convergence.
This is because it only involves the cross correlation
$\lag \delta_1 \gamma_2^*\rag \lag \delta_2 \gamma_1\rag$, which correlates
the high-redshift galaxy $2$ with low-redshift density fluctuations $1'$
with $z_{1'} \leq z_1 < z_2$, as the term 
$\lag \delta_1 \gamma_1\rag \lag \delta_2 \gamma^*_2\rag$ is zero by symmetry.

We show the dependence of the source-lens clustering bias on the second
galaxy redshift $z_2$, for a fixed first galaxy redshift $z_1$, in
Fig.~\ref{fig_xi_gamma_theta}.
We obtain behaviors that are similar to those found in Fig.~\ref{fig_xi_kappa_theta}
for the convergence, with a minimum at the coincident redshift $z_2=z_1$,
a saturation at high redshift $z_2 \gg z_1$, and a steep increase for
$z_2 \rightarrow 0$ (but the bias is no longer always positive).
The amplitude of the bias is somewhat smaller than in Fig.~\ref{fig_xi_kappa_theta},
especially for small angular scales.
This leads to an even smaller range of redshifts at $z_2 \la 0.05$ where the
bias reaches $10\%$ of the signal or more.

Therefore, as for the convergence, we find that the source-lens clustering
bias of estimators of the cosmic shear two-point correlation function
is almost always negligible. It is only relevant when we cross-correlate the
shear of a low-redshift galaxy, $z_2 \la 0.05$, with the shear of a higher redshift
galaxy, $z_1 \ga 0.5$ 
(the effect being larger for higher $z_1$ and smaller $\theta$).
For $z_1 \la 0.01$, the bias can actually dominate the weak lensing signal.
In practice, it would be sufficient to remove such pairs from the data analysis,
because they are a very small fraction of the pairs measured in a survey
and their cosmological information is highly redundant with other pairs
(where both galaxies are at the same low redshift or at possibly different
redshifts above $0.05$).

\section{Three-point convergence correlation function}
\label{three-point-kappa}

\subsection{Source-lens clustering bias}
\label{kappa-3pt}

We now consider the impact of the source-lens clustering bias on estimators
of the three-point weak lensing correlation functions.
As for the two-point correlation, we first investigate the simpler case of the
weak lensing convergence $\kappa$.
Then, the generalization of Eqs.(\ref{hkappa-1})-(\ref{hkappa-2}) to three-point
statistics, obtained by measuring triplets of galaxies, gives
\beq
\lag\hat\zeta^{\kappa\kappa\kappa}\rag = \frac{\lag (1+b_1\delta_1) 
(1+b_2\delta_2) (1+b_3\delta_3)  \kappa_1 \kappa_2 \kappa_3 \rag}
{\lag (1+b_1\delta_1) (1+b_2\delta_2) (1+b_3\delta_3) \rag} ,
\label{hzeta-kappa-1}
\eeq
for the estimator of the three-point convergence correlation.
As seen in Sect.~\ref{galaxy-location-bias}, because the lensing kernel
$g(\chi',\chi)$ vanishes for $\chi'=\chi$, the source-lens clustering bias is
only significant when a foreground galaxy $i$ correlates with the density
fluctuations $j'$ along the line of sight to a background galaxy $j$,
with $z_{j'} \simeq z_i < z_j$. 
Therefore, to simplify the analysis, we neglect correlations that correspond
to a vanishing lensing efficiency kernel $g$ [that at next order 
give a damping factor $x_0/(c/H_0)$ instead of zero] or that involve different
redshifts.
This allows us to use Limber's approximation (which gives zero for the
discarded terms).
Then, assuming without loss of generality $z_1 \leq z_2 \leq z_3$, 
$\delta_1$ can only be correlated with $\{\delta_2,\delta_3,\kappa_2,\kappa_3\}$,
$\delta_2$ with $\{\delta_1,\delta_3,\kappa_3\}$, and $\delta_3$ with
$\{\delta_1,\delta_2\}$.
Then, the average (\ref{hzeta-kappa-1}) reads as
\beq
z_1 \leq z_2 \leq z_3 :  \;\;\; \lag\hat\zeta^{\kappa\kappa\kappa}\rag \simeq
\zeta^{\kappa\kappa\kappa} + \zeta^{\delta} ,
\label{hzeta-kappa-1-1}
\eeq
where the source-lens clustering contribution (that we denote with the
superscript $\delta$) writes as
\beqa
\zeta^{\delta} & = & \left[ b_1 b_2 \lag\delta_1\delta_2\kappa_3\rag \lag\kappa_1\kappa_2\rag + (1 \!+\! b_1 b_3 \xi_{1,3} ) 
b_2 \lag\delta_2\kappa_3\rag \lag\kappa_1\kappa_2\rag \right. \nonumber \\
&& \hspace{-0.3cm} \left. + ( 1 \!+\! b_2 b_3 \xi_{2,3} ) b_1 ( \lag\delta_1\kappa_2\rag 
\lag\kappa_1\kappa_3\rag \!+\! \lag\delta_1\kappa_3\rag \lag\kappa_1\kappa_2\rag ) 
\right]  \nonumber \\
&& \hspace{-0.3cm} \times \left[ 1 \!+\! b_1 b_2 \xi_{1,2} \!+\! b_2 b_3 \xi_{2,3} 
\!+\! b_1 b_3 \xi_{1,3} \!+\! b_1 b_2 b_3 \zeta_{1,2,3} \right]^{-1} .
\label{hzeta-kappa-2}
\eeqa
Here we note again $\xi_{i,j}=\lag\delta_i \delta_j\rag$ the density-density
correlation (which arises from the galaxy-galaxy correlations).

In contrast with the case of the two-point estimator (\ref{split}), the
source-lens clustering bias is no longer dominated by contributions that
involve the density three-point correlation, but by contributions that
involve products of the density two-point correlation.
In particular, for the generic case of three different source redshifts, the galaxy-galaxy
correlations are negligible and Eq.(\ref{hzeta-kappa-2}) simplifies as
\beqa
z_1 \!<\! z_2 \!<\! z_3 : \;\;\;  \zeta^{\delta} & \! \simeq \! &
b_1 ( \lag\delta_1\kappa_2\rag 
\lag\kappa_1\kappa_3\rag \!+\! \lag\delta_1\kappa_3\rag \lag\kappa_1\kappa_2\rag ) 
\nonumber \\
&& + b_2 \lag\delta_2\kappa_3\rag \lag\kappa_1\kappa_2\rag .
\label{hzeta-kappa-3}
\eeqa
Therefore, the source-lens clustering bias is much easier to evaluate for the
three-point convergence correlation functions than for the two-point statistics
studied in Sects.~\ref{galaxy-location-bias} and \ref{galaxy-location-bias-shear}.
Nevertheless, in the following we use Eq.(\ref{hzeta-kappa-2}) to include
the case where galaxy redshifts coincide.

On large scales, whereas in Eq.(\ref{split})
the two-point weak lensing signal
$\xi^{\kappa\kappa}$ scales as the linear density correlation $\xi_L$
and the bias $\zeta^{\delta\kappa\kappa}$ obeys the higher-order scaling
$\xi_L^2$, in Eq.(\ref{hzeta-kappa-3}) the three-point weak lensing signal
$\zeta^{\kappa\kappa\kappa}$ and its bias $\lag\delta\kappa\rag\lag\kappa\kappa\rag$
show the same scaling $\xi_L^2$.
Therefore, the impact of the source-lens clustering bias is expected to be
greater for measures of three-point lensing correlations than for two-point
lensing correlations (see also \citet{Bernardeau1998}).

The first contribution in Eq.(\ref{hzeta-kappa-1-1}) is the weak lensing signal,
\beqa
\zeta^{\kappa\kappa\kappa} = \lag \kappa_1 \kappa_2 \kappa_3 \rag
& = & \int_0^{\chi_1} \dd\chi_{1'} g_{1',1} \int_0^{\chi_2} \dd\chi_{2'} g_{2',2} 
\nonumber \\
&& \times \int_0^{\chi_3} \dd\chi_{3'} g_{3',3} \; \zeta_{1',2',3'} .
\label{zeta-kappa-1}
\eeqa
Using Limber's approximation \citep{Limber1953,Kaiser1992,Munshi2008},
that is, neglecting the variation of the lensing kernels $g$ on scales where the
density correlations are not negligible, this writes as
\beq
\zeta^{\kappa\kappa\kappa} \simeq \int_0^{\chi_1} \dd\chi_{1'} g_{1',1} g_{1',2}
g_{1',3} \; \zeta^{\rm 2D}_{1',2',3'} ,
\label{zeta-kappa-2}
\eeq
where we introduced the 2D three-point density correlation obtained by
integrating along two lines of sight,
\beqa
\zeta^{\rm 2D}(\vx_{1'\!\perp},\vx_{2'\!\perp},\vx_{3'\!\perp};z) & \!\! = \!\! & 
\int_{-\infty}^{\infty} \! \dd x_{2'\parallel} \dd x_{3'\parallel} \; 
\zeta(\vx_{1'},\vx_{2'},\vx_{3'};z) . \nonumber \\
&&
\label{zeta-2D-def}
\eeqa

Using again Limber's approximation, the two-point functions that enter
Eq.(\ref{hzeta-kappa-2}) read as (for $z_1 \leq z_2$),
\beq
\lag\delta_1\kappa_2\rag \simeq g_{1,2} \; \xi^{\rm 2D}_{1,2'}
\label{xi-delta1-kappa2}
\eeq
and
\beq
\lag\kappa_1\kappa_2\rag \simeq \int_0^{\chi_1} \dd\chi_{1'} \, g_{1',1} g_{1',2} 
\; \xi^{\rm 2D}_{1',2'}   \;  ,
\label{xi-kappa1-kappa2}
\eeq
where we introduced the 2D two-point density correlation obtained by
integrating along one line of sight,
\beqa
\xi^{\rm 2D}(x_{\perp}) & = & \int_{-\infty}^{\infty} \dd x_{\parallel} \, 
\xi(x_{\parallel},x_{\perp}) \nonumber \\
& = & (2\pi)^2 \int_0^{\infty} \dd k_{\perp} k_{\perp} P(k_{\perp})
J_0(k_{\perp} x_{\perp}) .
\label{xi-2D-def}
\eeqa

The term $\lag\delta_1\delta_2\kappa_3\rag$ in Eq.(\ref{hzeta-kappa-2})
is only significant for the rare cases where $|z_2 - z_1| \la x_0/(c/H_0)$.
Therefore, it is sufficient to use the ansatz (\ref{zeta-def})
(which we compare with a more precise model in App.~\ref{comparison-3pt}).
This gives
\beqa
\lag\delta_1\delta_2\kappa_3\rag \!\! & \simeq & \!\! \frac{S_3}{3} \left[ \xi_{1,2} 
\left( \lag\delta_1\kappa_3\rag \!+\! \lag\delta_2\kappa_3\rag \right) \! + \!\!
\int \!\! \dd\chi_{3'} \, g_{3',3} \, \xi_{3',1} \xi_{3',2} \right] \! . \nonumber \\
&&
\label{zeta-d1d2k3}
\eeqa

\subsection{Numerical computations}
\label{numerical-computations-kappa-3pt}

\begin{figure}
\begin{center}
\epsfxsize=8.5 cm \epsfysize=6. cm {\epsfbox{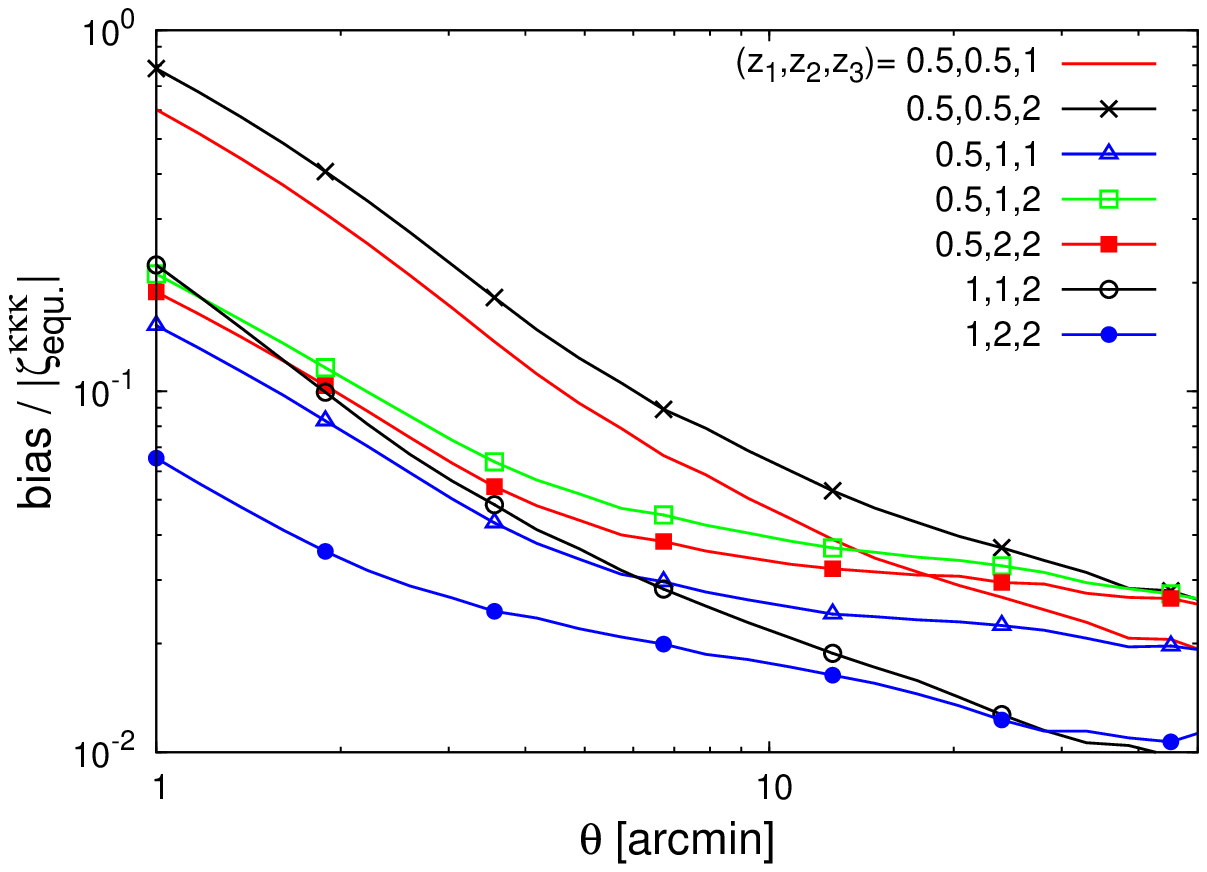}}
\end{center}
\caption{Relative source-lens clustering bias of the equilateral weak lensing 
convergence three-point correlation $\zeta^{\kappa\kappa\kappa}_{\rm equ.}$,
as a function of the angular scale $\theta$, for a few redshift triplets
$z_1 \leq z_2 \leq z_3$. All total biases are positive.}
\label{fig_zeta_kappa_z_z_z}
\end{figure}

\begin{figure*}
\begin{center}
\epsfxsize=8.5 cm \epsfysize=6. cm {\epsfbox{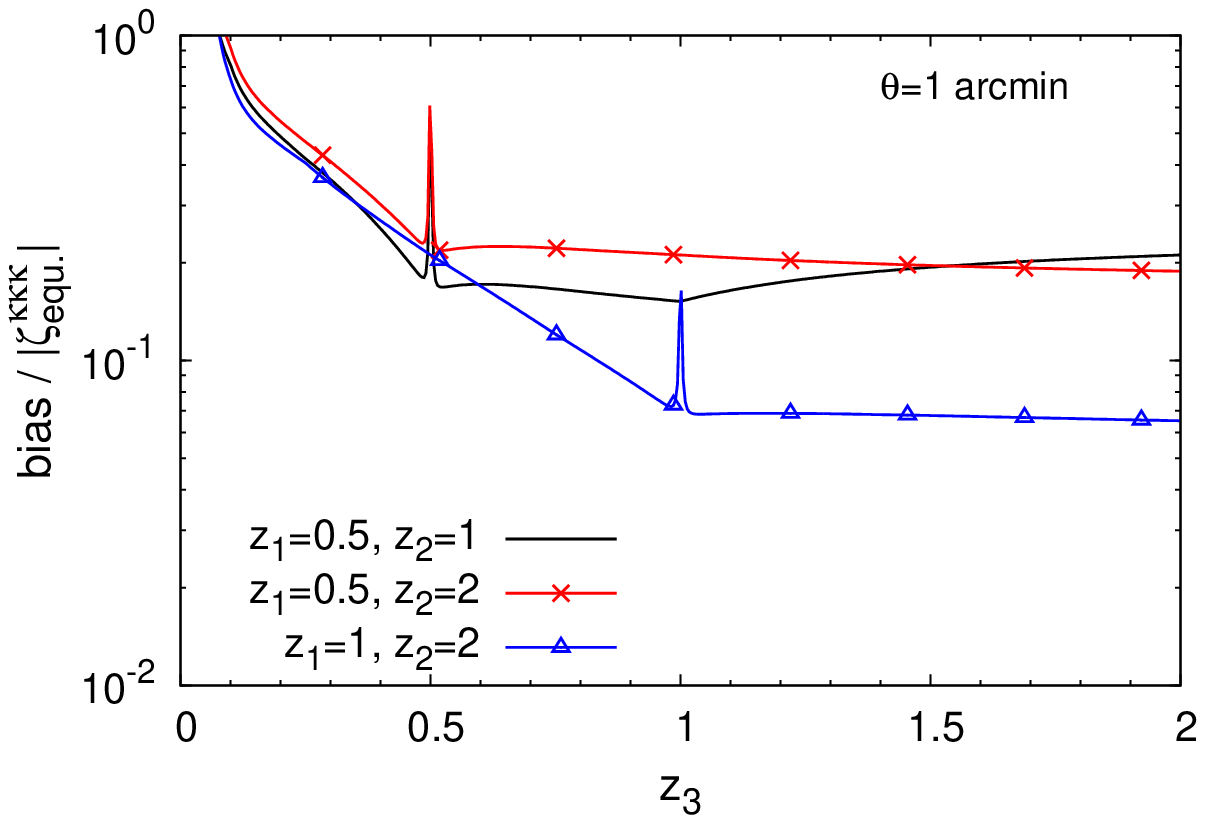}}
\epsfxsize=8.5 cm \epsfysize=6. cm {\epsfbox{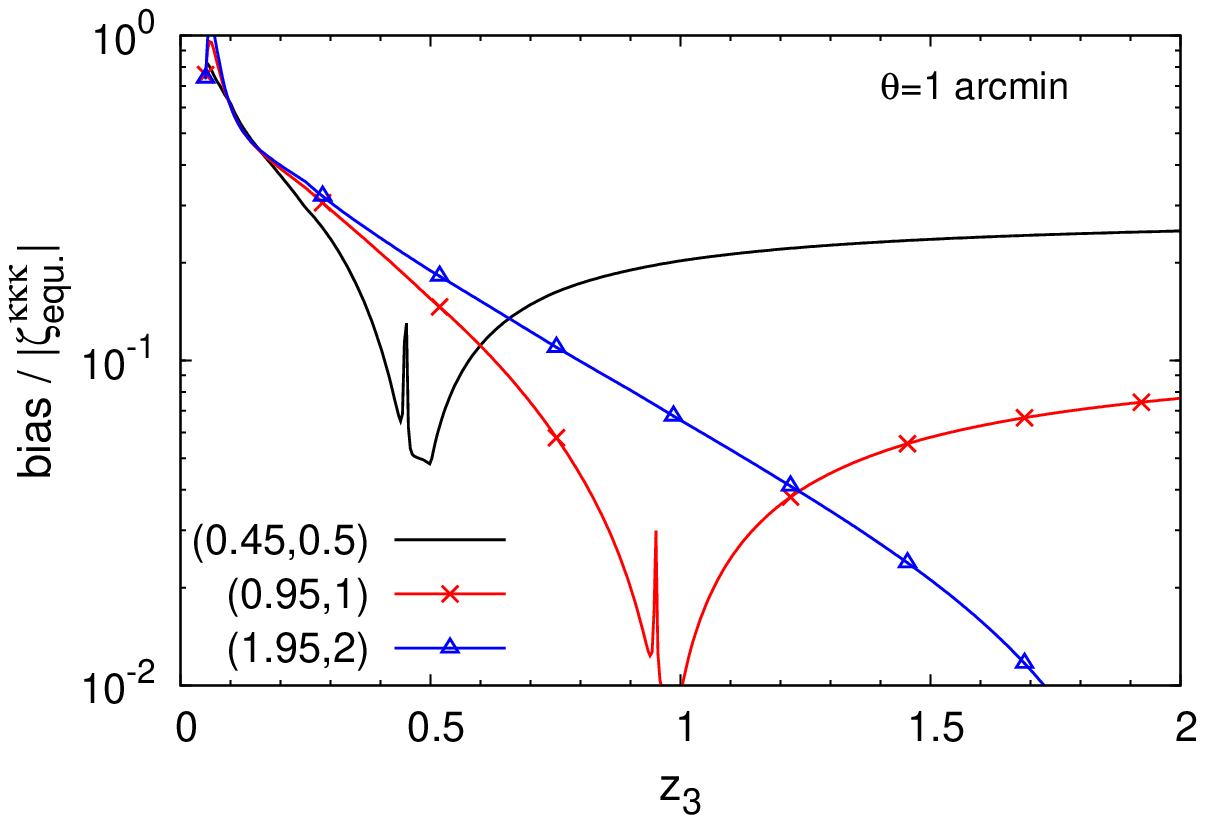}}\\
\epsfxsize=8.5 cm \epsfysize=6. cm {\epsfbox{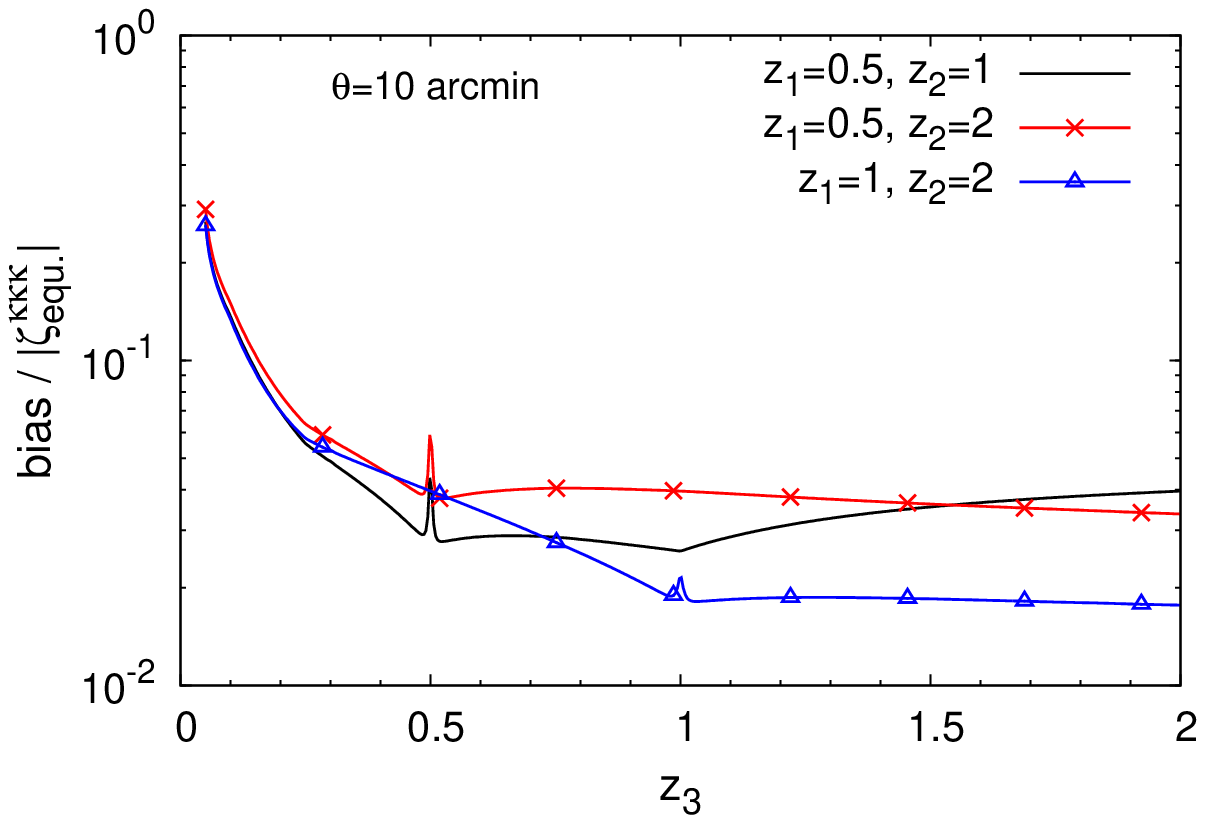}}
\epsfxsize=8.5 cm \epsfysize=6. cm {\epsfbox{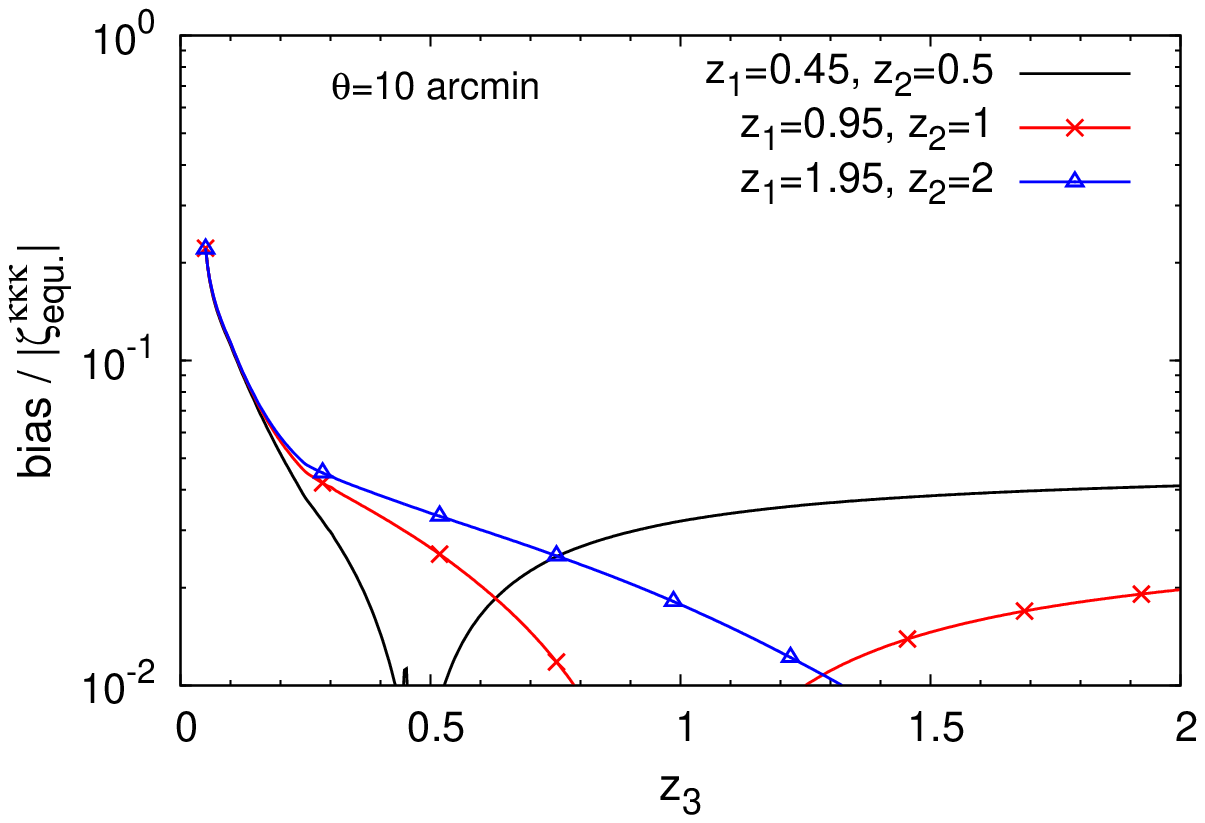}}
\end{center}
\caption{Relative source-lens clustering bias of the equilateral weak lensing 
convergence three-point correlation $\zeta^{\kappa\kappa\kappa}_{\rm equ.}$,
as a function of the third galaxy redshift $z_3$ for a fixed pair of redshifts
$\{z_1,z_2\}$. We consider the angular
scales $\theta=1$ (upper panels) and $10$ arcmin (lower panels).
All total biases are positive.}
\label{fig_zeta_kappa_theta}
\end{figure*}

As shown in Eqs.(\ref{hzeta-kappa-2})-(\ref{hzeta-kappa-3}), the source-lens
clustering bias of three-point lensing statistics is dominated by terms that only
involve the two-point matter and galaxy correlation functions.
Therefore, the density three-point correlation is mainly needed to compute
the lensing signal to estimate the relative amplitude of the source-lens clustering
bias.
Nevertheless, because this bias is no longer negligible, in contrast with the
case of two-point statistics studied in Sects.~\ref{galaxy-location-bias}
and \ref{galaxy-location-bias-shear}, it is useful to go beyond orders of
magnitude estimates for the three-point lensing correlation. Therefore,
instead of the ansatz (\ref{zeta-def}) we now use the more accurate
modeling described in \citet{Valageas2011e}.
It combines one-loop standard perturbation theory and a halo model to predict
the density 3D bispectrum and three-point correlation.
This provides in turn the weak lensing bispectrum and three-point correlation
and gives a good agreement with ray-tracing numerical simulations
\citep{Valageas2012a,Valageas2012b}.
[In this model, the bispectrum is split as usual as a sum of three-halo, two-halo,
and one-halo contributions. The three-halo term is identified with the perturbative
contribution and is given by the standard one-loop perturbation theory.
The two-halo term involves the correlation between halos (taken proportional to the
linear correlation), as well as the halo profiles (we use the NFW profile
from \citet{Navarro1997}) and mass function (as in \cite{Valageas2011d}),
which also fully determine the one-halo term.]

We also compared these results with the ansatz (\ref{zeta-def}),
where 
\beq
\zeta^{\rm 2D}_{1',2',3'} \simeq \frac{S_3}{3} \left[ \xi^{\rm 2D}_{1',2'} 
\xi^{\rm 2D}_{1',3'} + \xi^{\rm 2D}_{2',1'} \xi^{\rm 2D}_{2',3'} 
+ \xi^{\rm 2D}_{3',1'} \xi^{\rm 2D}_{3',2'}  \right] ,
\label{zeta-kappa-3}
\eeq
and found that both approximations agree to better than a factor $1.5$
for $\theta<10'$ and a factor $3$ for $\theta<40'$, as illustrated in 
Fig.~\ref{fig_zeta_ansatz_z_z_z} in App.~\ref{comparison-3pt}.
The agreement degrades on large scales because they do not change sign
at exactly the same scale, which gives rise to large relative deviations around
the scale where either one prediction goes through zero.
However, this is not a serious problem for such purposes, because the three-point
convergence correlation becomes very small on large scales and
most of the information from weak lensing surveys comes from smaller
scales, $\theta \la 10'$, where the signal can be discriminated from the
different sources of noise \citep{Semboloni2011}.
Thus, the simple approximation (\ref{zeta-kappa-3}), which enables fast 
and simple numerical computations, would actually be sufficient to estimate the
magnitude of the convergence three-point correlation, whence of the
relative bias.

\subsection{Numerical results}
\label{numerical-results-kappa-3pt}

We show our results for equilateral configurations in
Fig.~\ref{fig_zeta_kappa_z_z_z}, as a function of the angular width $\theta$ of
the triangle sides.
On these scales, the source-lens clustering bias is typically of order $10\%$ for the
convergence three-point correlation. 
Within the Limber approximation, the source-lens clustering bias is nonzero
as soon as the three galaxy redshifts are not identical [as explained in
Sect.~\ref{kappa-3pt}, to obtain a better estimate at identical redshifts one
needs to go beyond the Limber approximation, which will however give
a small bias because of the suppression factor $x_0/(c/H_0)$].

On small scales, the relative bias is somewhat higher for the cases where
$z_1=z_2$. This is due to the term 
$\lag\delta_1\delta_2\kappa_3\rag \lag\kappa_1\kappa_2\rag$ in
Eq.(\ref{hzeta-kappa-2}), which is only significant when $z_1 \simeq z_2$.
Moreover, this contribution typically scales as $\xi^3$ because it involves
the product of a three-point and a two-point correlations, see also
Eq.(\ref{zeta-d1d2k3}). This leads to a significant growth on small scales provided
the condition $z_1 \simeq z_2$ is satisfied.

We show the redshift dependence of the source-lens clustering bias
in Fig.~\ref{fig_zeta_kappa_theta} (here we no longer have the ordering
$z_1 \leq z_2 \leq z_3$ since we let $z_3$ vary from $0$ to $2$ at fixed
$\{z_1,z_2\}$).
The upward spikes correspond to redshifts $z_3$ such that the two
lowest redshifts of the triplet $\{z_1,z_2,z_3\}$ are equal.
As explained above for Fig.~\ref{fig_zeta_kappa_z_z_z}, this is due to the
clustering of the two foreground galaxies with nearby density fluctuations on
the third line of sight, through the factor
$\lag\delta_1\delta_2\kappa_3\rag \lag\kappa_1\kappa_2\rag$ in 
Eq.(\ref{hzeta-kappa-2}).
This effect occurs in a narrow redshift band of width $x_0/(c/H_0)$ set by
the galaxy correlation length.
Therefore, for generic galaxy redshifts drawn from actual surveys, this amplification
should be rare and the typical bias is of order $10\%$.
The right panels in Fig.~\ref{fig_zeta_kappa_theta} correspond to closer
pairs $\{z_1,z_2\}$. We obtain similar results as in the left panels but
with a broad valley as $z_3$ becomes close to the pair $\{z_1,z_2\}$.
This is due to the effect of the lensing kernels $g_{i',i}$ that vanish on the
source plane and lead to a zero bias when $z_1=z_2=z_3$ within our
approximations. This yields suppression factors $x_0/(c/H_0)$ and a decrease
of the bias for $z_1 \simeq z_2 \simeq z_3$, as found in
Sects.~\ref{galaxy-location-bias} and \ref{galaxy-location-bias-shear} in the case
$z_1 \simeq z_2$ for two-point estimators.
When $z_3=\min(z_1,z_2)$ we recover the localized upward spike.
When $z_3$ is far from the pair $\{z_1,z_2\}$ we recover a bias of about
$10\%$ of the signal as in the left panels.

As for the case of the convergence two-point correlation shown
in Fig.~\ref{fig_xi_kappa_theta}, the bias becomes more important as
one of the galaxies lies at a small redshift. Indeed, the shorter line of sight
decreases the signal, as in Eq.(\ref{zeta-kappa-2}), and lessens the
impact of the suppression factor $x_0/(c/H_0)$ associated with the
source-lens clustering.
Then, for $\theta \la 1'$ the bias becomes of the same order as the
signal for $z_1 \la 0.1$.

\section{Three-point cosmic shear correlation function}
\label{three-point-gamma}

\subsection{Source-lens clustering bias}
\label{gamma-3pt}

We now consider the three-point correlation function of the cosmic shear $\gamma$.
Because of the spin-2 factor $e^{2\ii \alpha}$ computations are somewhat
heavier. To simplify the analysis we focus on the geometrical average
$\zeta^{\gamma\gamma\gamma}_{\rm circ.}(\theta)$,
\beqa
\zeta^{\gamma\gamma\gamma}_{\rm circ.}(\theta) \!\! & = & \!\!\! \int_0^{2\pi} \! 
\frac{\dd\alpha_{\vx_1}\dd\alpha_{\vx_2}\dd\alpha_{\vx_3}}{(2\pi)^3}
\lag \gamma_1 \gamma_2 \gamma_3 \, 
e^{-2\ii(\alpha_{\vx_1} \!+\! \alpha_{\vx_2} \!+\! \alpha_{\vx_3})} \rag 
\label{zeta-gamma-circ-def} \\
& = & \biggl \lag \left( \int_0^{2\pi} \frac{\dd\alpha_{\vx}}{2\pi} \; \gamma \; 
e^{-2\ii\alpha_{\vx}}\right)^{\!3} \biggl \rag .
\label{zeta-gamma-circ-M3}
\eeqa
Here $\alpha_{\vx_i}$ is again the polar angle of the line of sight to the galaxy
$i$. The exponential factors in the term $\lag .. \rag$ ensure that the full product
is a spin-0 quantity and does not vanish by symmetry. In
Eq.(\ref{zeta-gamma-circ-def}), the three lines of sight are at the same angular
separation $\theta$ from a fixed center $O$ and we integrate over their angles
$\alpha_{\vx_i}$ with respect to this central point. Therefore,
$\zeta^{\gamma\gamma\gamma}_{\rm circ.}$ is the geometrical mean of the
shear three-point correlation $\zeta^{\gamma\gamma\gamma}$ over all
triangles with a circumcircle of radius $\theta$.
[It is identical to the correlations
$\zeta^{\gamma\gamma\gamma^*}_{\rm circ.}$ (where the factor
$e^{-2\ii\alpha_{\vx_3}}$ is changed to $e^{2\ii\alpha_{\vx_3}}$),
$\zeta^{\gamma\gamma^*\gamma^*}_{\rm circ.}$, and
$\zeta^{\gamma^*\gamma^*\gamma^*}_{\rm circ.}$.]
The fully symmetric three-point correlation (\ref{zeta-gamma-circ-def})
provides simpler expressions than the correlation
$\zeta^{\gamma\gamma\gamma}(\theta_{12},\theta_{23},\theta_{31})$, 
associated with a single triangular shape, thanks to the independent
integrations over the polar angles $\alpha_{\vx_i}$.
As shown by the second equality (\ref{zeta-gamma-circ-M3}),
this is also the third-order cumulant of the complex aperture mass $M$
(with a Dirac weight),
the usual E-mode aperture mass $M_{\rm ap}$ being defined as
$M_{\rm ap}=\mbox{Re}(M)$, which can be expressed in terms of the
tangential shear $\gamma_{\rm t}$
\citep{Schneider1998,Jarvis2004,Schneider2005}.
Here we have
\beqa
M = M_{\rm ap}+\ii M_{\times} & = & - \int \dd^2 \vartheta \, Q_{\theta}(|\vartheta|)
\, \gamma \, e^{-2\ii\alpha} \label{Map-Q} \\
& = & \int \dd^2 \vartheta \, U_{\theta}(|\vartheta|) \,
\kappa(\vartheta) , \label{Map-U}
\eeqa
with
\beq
Q_{\theta}(\vartheta) = - \frac{\delta_D(\vartheta \!-\! \theta)}{2\pi\theta} ,
U_{\theta}(|\vartheta|) = \frac{\delta_D(\vartheta \!-\! \theta)}{2\pi\theta} - 
\frac{2\Theta(\vartheta \!<\! \theta)}{2\pi\theta^2} ,
\label{Q-U-def}
\eeq
where $\Theta(\vartheta \!<\! \theta)$ is the unit top-hat.
As is well known, gravitational lensing only gives rise to E modes, so that
$M_{\times}=0$ as seen from Eq.(\ref{Map-U}), within the Born approximation.
Then $\lag M_{\rm ap}^3 \rag = \lag M^3 \rag$.
However, this is no longer the case when we include additional observational
effects, such as source clustering \citep{Schneider2002}
or galaxy intrinsic alignments (this depends on the properties of the latter,
e.g., whether they follow a linear or quadratic dependence on the density
field, \citet{Crittenden2001}).
Here we do not investigate the E/B modes separation and focus on the
overall amplitude of the source-lens clustering bias as compared
with the gravitational lensing signal and the intrinsic-alignment bias.

Using the vanishing of the lensing kernel $g$ on the source plane as in
Sect.~\ref{kappa-3pt}, Eqs.(\ref{hzeta-kappa-1-1}) and (\ref{hzeta-kappa-2}) 
become
\beq
z_1 \leq z_2 \leq z_3 : \;\;\;  \lag\hat\zeta^{\gamma\gamma\gamma}_{\rm circ.}\rag
\simeq \zeta^{\gamma\gamma\gamma}_{\rm circ.} + 
\zeta^{\delta}_{\rm circ.} ,
\label{hzeta-gamma-1-1}
\eeq
and
\beqa
\zeta^{\delta}_{\rm circ.} \!\!\!\! & = & \!\!  \left[ b_1 b_2 
\lag\delta_1\delta_2\gamma_3\rag_{\alpha} \lag\gamma_1\gamma_2\rag_{\alpha}
\!+\! (1 \!+\! b_1 b_3 \xi_{1,3} ) b_2 \lag\delta_2\gamma_3\rag_{\alpha} 
\lag\gamma_1\gamma_2\rag_{\alpha} \right. \nonumber \\
&& \hspace{-0.4cm} \left. + ( 1 \!+\! b_2 b_3 \xi_{2,3} ) b_1 
( \lag\delta_1\gamma_2\rag_{\alpha} \lag\gamma_1\gamma_3\rag_{\alpha}
 \!+\! \lag\delta_1\gamma_3\rag_{\alpha} \lag\gamma_1\gamma_2\rag_{\alpha} ) 
\right]  \nonumber \\
&& \hspace{-0.4cm} \times \left[ 1 \!+\! b_1 b_2 \xi_{1,2} \!+\! b_2 b_3 \xi_{2,3} 
\!+\! b_1 b_3 \xi_{1,3} \!+\! b_1 b_2 b_3 \zeta_{1,2,3} \right]^{-1} \! . 
\label{hzeta-gamma-2}
\eeqa
The subscripts ``$\alpha$'' denote the factors $e^{-2\ii\alpha_{\vx_i}}$ and the
integrations over the angles $\alpha_{\vx_i}$, as in Eq.(\ref{zeta-gamma-circ-def}).
To simplify the computations, we only perform the geometrical average
(\ref{zeta-gamma-circ-def}) for the terms that involve the shear $\gamma$ and
we factor out the galaxy-galaxy correlations $\xi_{i,j}$ by simply using their values
at the angular scale $\theta$. This should be sufficient for our purposes because, as
seen in the previous sections, these terms are only important when $z_i=z_j$.
Again, for the generic case where the three galaxy redshifts are different,
Eq.(\ref{hzeta-gamma-2}) simplifies as
\beqa
z_1 \!<\! z_2 \!<\! z_3 : \;\;  \zeta^{\delta}_{\rm circ.} \!\! & \simeq & \!\!
b_1 ( \lag\delta_1\gamma_2\rag_{\alpha} \lag\gamma_1\gamma_3\rag_{\alpha} 
\!+\! \lag\delta_1\gamma_3\rag_{\alpha} 
\lag\gamma_1\gamma_2\rag_{\alpha} ) \nonumber \\
&& + b_2 \lag\delta_2\gamma_3\rag_{\alpha} \lag\gamma_1\gamma_2\rag_{\alpha} .
\label{hzeta-gamma-3}
\eeqa
Thus, the source-lens clustering bias of the shear three-point correlation is dominated
by contributions that only involve the density two-point correlation.

The first contribution in Eq.(\ref{hzeta-gamma-1-1}) is the weak lensing signal. 
Using Limber's approximation, it reads as (for $z_1 \leq z_2 \leq z_3$)
\beq
\zeta^{\gamma\gamma\gamma}_{\rm circ.} = 
\lag\gamma_1\gamma_2\gamma_3\rag_{\alpha} = \int_0^{\chi_1} \dd \chi_{1'} \;
g_{1',1} g_{1',2} g_{1',3} \; \zeta_{\rm circ.}^{\rm 2D} ,
\label{zeta-ggg-0}
\eeq
with
\beqa
\zeta_{\rm circ.}^{\rm 2D} & = & \int_{-\infty}^{\infty} \dd x_{2'\parallel} \dd x_{3'\parallel} 
\int_0^{2\pi} \! \frac{\dd\alpha_{\vx_1}\dd\alpha_{\vx_2}\dd\alpha_{\vx_3}}{(2\pi)^3}
\nonumber \\
&& \hspace{-0.cm} \times \int \dd\vk_{1'}\dd\vk_{2'}\dd\vk_{3'} \;
e^{\ii [ \vk_{1'}\cdot\vx_{1'}+\vk_{2'}\cdot\vx_{2'}+\vk_{3'}\cdot\vx_{3'}]}
\nonumber \\
&& \hspace{-0.cm} \times \; \delta_D(\vk_{1'}+\vk_{2'}+\vk_{3'}) \; 
B(k_{1'},k_{2'},k_{3'}) \;  \nonumber \\
&& \hspace{-0.cm} \times \; e^{2\ii (\alpha_{\vk_{1'}}+\alpha_{\vk_{2'}}+\alpha_{\vk_{3'}}
- \alpha_{\vx_1} - \alpha_{\vx_2} - \alpha_{\vx_3})} .
\label{zetap-circ-B}
\eeqa
Expressing the bispectrum in terms of the density three-point correlation and
using Eq.(\ref{Jn-1Jn+1}) this also writes as
\beqa
\zeta_{\rm circ.}^{\rm 2D} & = & \int \frac{\dd \vr_{1\perp} \dd \vr_{2\perp} 
\dd \vr_{3\perp}}{(2\pi)^3} \; \zeta^{\rm 2D}(\vr_{1\perp},\vr_{2\perp},\vr_{3\perp})
\nonumber \\
&& \times \prod_{i=1}^3 \left[ \frac{\delta_D(r_{i\perp}-d)}{d}
- \frac{2 \, \Theta(r_{i\perp}<d)}{d^2} \right] ,
\label{zetap-circ-zeta2D}
\eeqa 
where $d=\chi_{1'} \theta$ is the radius of the circumcircle at radial distance
$\chi_{1'}$, $\Theta(r_{i\perp}<d)$ is the unit top-hat, and $\zeta^{\rm 2D}$
is the 2D three-point density correlation introduced in Eq.(\ref{zeta-2D-def}).
This could be obtained at once from Eqs.(\ref{Map-U}) and (\ref{Q-U-def}).

Thus, as compared with the circular average $\zeta^{\kappa\kappa\kappa}_{\rm circ.}$
of the three-point convergence correlation, the shear introduces a non-local
dependence. As for the two-point ``spin-2'' correlation (\ref{2_xi-1}), this comes
through a counterterm that involves the integral of the two- or three-point correlation
over smaller angular scales.
This decreases the amplitude of the weak lensing signal, as compared with the
convergence case. For instance, if we generalize the geometrical average
(\ref{zeta-gamma-circ-def}) to $\zeta^{\gamma\gamma\gamma}_{\rm circ.}(\theta_1,\theta_2,\theta_3)$, so that the three lines of sight $\{\vx_1,\vx_2,\vx_3\}$
are at different radii $\{\theta_1,\theta_2,\theta_3\}$ from a given center,
we can see from the generalization of Eq.(\ref{zetap-circ-zeta2D}) that
$\zeta^{\gamma\gamma\gamma}_{\rm circ.}$ goes to zero when one radius
$\theta_i$ vanishes while the other two radii remain finite
(because $\zeta^{\rm 2D}(\vr_{1\perp},\vr_{2\perp},\vr_{3\perp})$ remains finite
in this limit). This is related to the well-known property that the aperture mass
(\ref{Map-Q}) can be written in terms of the convergence with a compensated
window function $U_{\theta}$, as in Eq.(\ref{Map-U}).

Therefore, in contrast with measures of the two-point correlation, where
$\xi^{\gamma\gamma^*} = \xi^{\kappa\kappa}$ in Eq.(\ref{xi-gamma-1}),
for the three-point statistics the spin-2 factor $e^{2\ii\alpha}$ of the shear
does not fully cancel (because the three-point Dirac factor
$\delta_D(\vk_1+\vk_2+\vk_3)$ no longer ensures
$e^{2\ii (\alpha_{\vk_{1}}+\alpha_{\vk_{2}}+\alpha_{\vk_{3}})}=1$).
This yields a smaller signal and a non-local dependence in terms of real-space
correlations. Nevertheless, for the circular statistics
$\zeta^{\gamma\gamma\gamma}_{\rm circ.}$ this non-locality does not extend to the
whole transverse plane. It only involves the three point function $\zeta^{\rm 2D}$
within radius $\theta$, with simple weights, and
$\zeta^{\gamma\gamma\gamma}_{\rm circ.}$ could still provide a good probe of
$\zeta^{\rm 2D}$.

The term $\lag\delta_1\gamma_2\rag_{\alpha}$ of Eq.(\ref{hzeta-gamma-2}) writes
as (for $z_1 \leq z_2$)
\beqa
\lag\delta_1\gamma_2\rag_{\alpha} \!\! & = & \! g_{1,2} \int_{-\infty}^{\infty} \!\!
\dd x_{2'\parallel} \int_0^{2\pi} \frac{\dd\alpha_{\vx_2}}{2\pi} 
\int \! \frac{\dd\vr}{(2\pi)^3} \int \!\! \dd\vk_1 \dd\vk_{2'} \nonumber \\
&& \times \; e^{\ii[\vk_1\cdot(\vx_1+\vr)+\vk_{2'}\cdot(\vx_{2'}+\vr)]} P(k_1) 
e^{2\ii(\alpha_{\vk_{2'}}-\alpha_{\vx_2})} ,
\eeqa
where we again assumed that variations of $g_{2',2}$ can be neglected on scales
where the correlation $\xi_{1,2'}$ is significant.
Integrating over the longitudinal components
$\{x_{2'\parallel},k_{2'\parallel},r_{\parallel},k_{1\parallel}\}$, and the angles
$\{\alpha_{\vx_2},\alpha_{\vk_{2'}},\alpha_{\vr},\alpha_{\vk_1}\}$, we obtain
\beqa
\lag\delta_1\gamma_2\rag_{\alpha} \!\! & = & - g_{1,2} (2\pi)^2 \int_0^{\infty} \!\!
 \dd r_{\!\perp} \dd k_{1\perp} \dd k_{2'\!\perp} \; r_{\!\perp} k_{1\perp} k_{2'\!\perp} P(k_{1\perp}) 
\nonumber \\
&& \times \; J_2(k_{2'\!\perp} d) J_0(k_{2'\!\perp} r_{\!\perp}) J_0(k_{1\perp} r_{\!\perp})
J_0(k_{1\perp} d) ,
\eeqa
where $d=\chi_1 \theta$.
Then, writing the transverse power spectrum as in Eq.(\ref{Pkperp-2})
and using the properties (\ref{J0JnJn}) and (\ref{Jn-1Jn+1}), we can perform
the integrations over wavenumbers. This yields
\beqa
\lag\delta_1\gamma_2\rag_{\alpha} & = & g_{1,2} \! \int_0^{2d} 
\frac{\dd r \; \xi^{\rm 2D}(r)}{\pi d \sin(\varphi)} - g_{1,2} \int_0^d \frac{\dd r}{d} 
\nonumber \\
&& \times \int_{d-r}^{d+r} \frac{\dd r' \; 2 \, \xi^{\rm 2D}(r')}{\pi d \sin(\varphi')}  ,
\label{xi-d1-g2}
\eeqa
where the angles $\varphi$ and $\varphi'$ are given by
\beq
\varphi = \mbox{Arccos} \left( \frac{r}{2d} \right) , \;\;\;
\varphi' = \mbox{Arccos} \left( \frac{r^2+r'^2-d^2}{2 r r'} \right).
\label{phi-phip}
\eeq

In a similar fashion, the term $\lag\gamma_1\gamma_2\rag_{\alpha}$ reads as
($z_1 \leq z_2$)
\beqa
\lag\gamma_1\gamma_2\rag_{\alpha} & \! = & - \int_0^{\chi_1} \dd\chi_{1'} \, 
g_{1',1} g_{1',2}
\; \biggl\lbrace \int_0^{2d} \frac{\dd r \, r \, \xi^{\rm 2D}(r)}{\pi d^2} \nonumber \\
&& \hspace{-0.5cm} \times \left[ 2 \varphi - \frac{d}{r \sin(\varphi)} \right] 
- \int_0^d \frac{\dd r \, 2r}{d^2} \left( \xi^{\rm 2D}(r) \left[ \frac{d-r}{d}\right]^2 \right. 
\nonumber \\
&& \hspace{-0.5cm} \left. + \int_{d-r}^{d+r} \frac{\dd r' \, r' \, \xi^{\rm 2D}(r')}{\pi d^2}
\left[ 2 \varphi' - \frac{d^2}{r \, r' \sin(\varphi')} \right] \right) \biggl \rbrace ,
\label{xi-g1-g2}
\eeqa
where we used Eq.(\ref{J2J0J0}) and the angles $\varphi$ and $\varphi'$ are given
by Eq.(\ref{phi-phip}).

The term $\lag\delta_1\delta_2\gamma_3\rag_{\alpha}$ in Eq.(\ref{hzeta-gamma-2})
is only significant for the rare cases where $|z_2 - z_1| \la x_0/(c/H_0)$.
Therefore, we simply use the ansatz (\ref{bispectrum-def}), which gives
\beqa
\lag\delta_1\delta_2\gamma_3\rag_{\alpha} & \! \simeq & \frac{S_3}{3} \biggl \lbrace
\xi_{1,2} \left( \lag\delta_1\gamma_3\rag_{\alpha} \!+\! 
\lag\delta_2\gamma_3\rag_{\alpha} \right) +
\int  \dd\chi_{3'} \, g_{3',3} \nonumber \\
&& \hspace{0cm} \times \int_0^{2\pi} \frac{\dd\alpha_1\dd\alpha_2}{(2\pi)^2}
\int_0^{\infty} \dd x_{3' \!\perp} \, x_{3'\!\perp} \;  \xi_{3',1} \xi_{3',2} 
\nonumber \\
&& \hspace{0cm} \times \left[ \frac{\delta_D(x_{3' \!\perp} \!-\! d)}{d} - 
\frac{2\Theta(x_{3' \!\perp} \!<\!d)}{d^2} \right] \biggl \rbrace .
\label{zeta-d1d2g3}
\eeqa
Here $d=(\chi_1+\chi_2)\theta/2$ and we made the simplifying approximation
of factorizing the angular integrations over $\alpha_{\vx_i}$ of the factors
$\lag\delta_1\delta_2\gamma_3\rag_{\alpha}$ and
$\lag\gamma_1\gamma_2\rag_{\alpha}$.

\begin{figure}
\begin{center}
\epsfxsize=8.5 cm \epsfysize=6. cm {\epsfbox{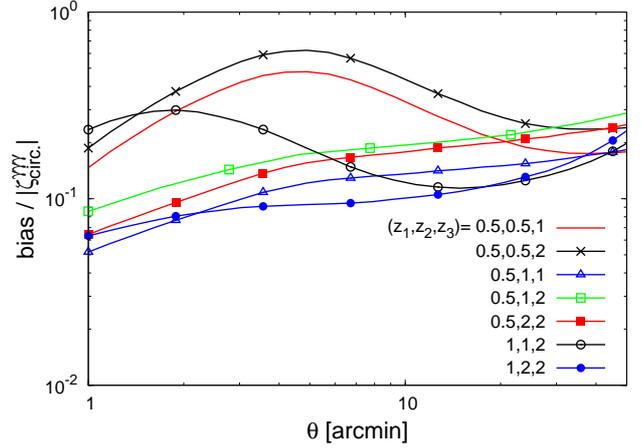}}
\end{center}
\caption{Relative source-lens clustering bias of the circular weak lensing shear
three-point correlation $\zeta^{\gamma\gamma\gamma}_{\rm circ.}$,
as a function of the angular scale $\theta$, for a few redshift triplets
$z_1 \leq z_2 \leq z_3$.
All total biases are positive.}
\label{fig_zeta_gamma_z_z_z}
\end{figure}

\subsection{Numerical computations}
\label{Numerical-computations-gamma-3pt}

As for the case of the weak lensing convergence, the source-lens clustering
bias in Eqs.(\ref{hzeta-gamma-2}) and (\ref{hzeta-gamma-3}) mainly depends
on the density two-point correlation function (within our approximations).
The density three-point
correlation is only needed to compute the weak lensing signal (\ref{zeta-ggg-0})
(whence the relative amplitude of the bias) and the term
$\lag\delta_1\delta_2\gamma_3\rag_{\alpha}$ in Eq.(\ref{hzeta-gamma-2}), which
is only significant for $z_1 \simeq z_2$.
To improve the accuracy of our computations, as in Sect.~\ref{three-point-kappa},
we use the more accurate modeling described in
\citet{Valageas2011e} and \citet{Valageas2012a,Valageas2012b},
instead of the hierarchical ansatz (\ref{zeta-def}), to compute the
weak lensing signal (\ref{zeta-ggg-0}).

Nevertheless, we also compared these results with the ansatz 
(\ref{bispectrum-def}). Then, Eq.(\ref{zetap-circ-B}) 
also writes as Eq.(\ref{zetap-circ-4}), see App.~\ref{shear-3pt-app}.
This provides an expression in terms of the real-space density correlation,
without oscillatory kernels and with lower-dimensional integrals.

Then, we found that both approximations agree to better than a factor $1.5$
for $\theta<30'$, as illustrated in 
Fig.~\ref{fig_zeta_ansatz_z_z_z} in App.~\ref{comparison-3pt}.
The three-point cosmic shear correlation function becomes very small
on large scales and most of the information from weak lensing surveys comes
from smaller scales, $\theta \la 10'$, where the signal can be discriminated from
the noise \citep{Semboloni2011}.
Therefore, the simple approximation (\ref{bispectrum-def}), whence
Eq.(\ref{zetap-circ-4}), would be sufficient to estimate the relative importance
of various sources of noise such as the source-lens clustering bias.

\subsection{Numerical results}
\label{Numerical-results-gamma-3pt}

\begin{figure*}
\begin{center}
\epsfxsize=8.5 cm \epsfysize=6. cm {\epsfbox{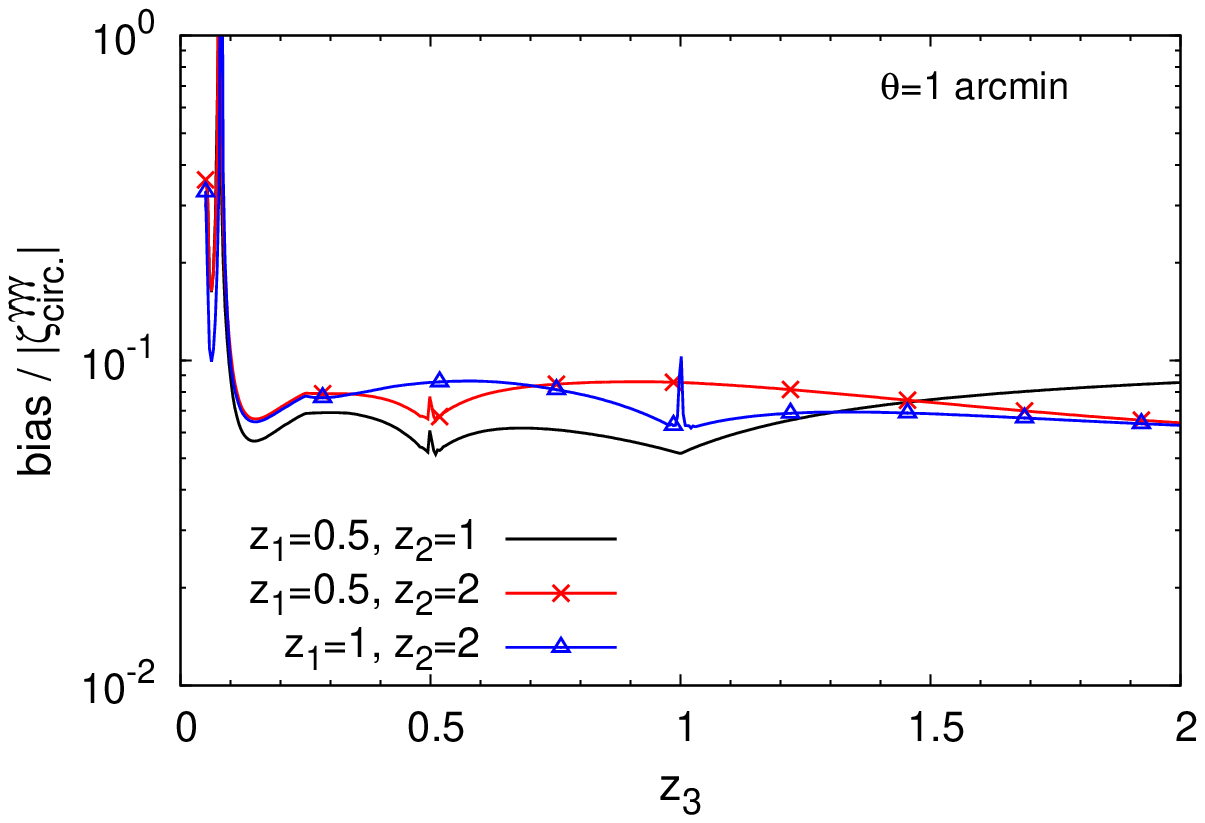}}
\epsfxsize=8.5 cm \epsfysize=6. cm {\epsfbox{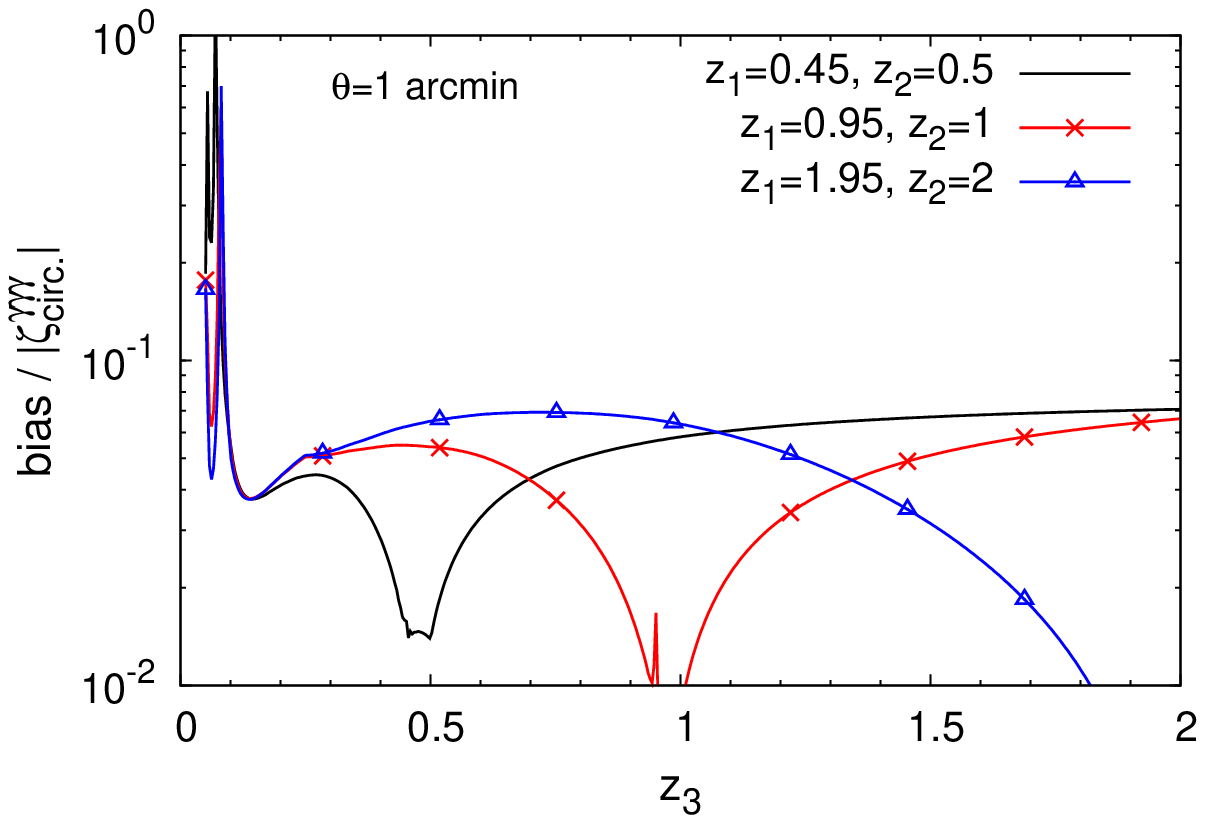}}\\
\epsfxsize=8.5 cm \epsfysize=6. cm {\epsfbox{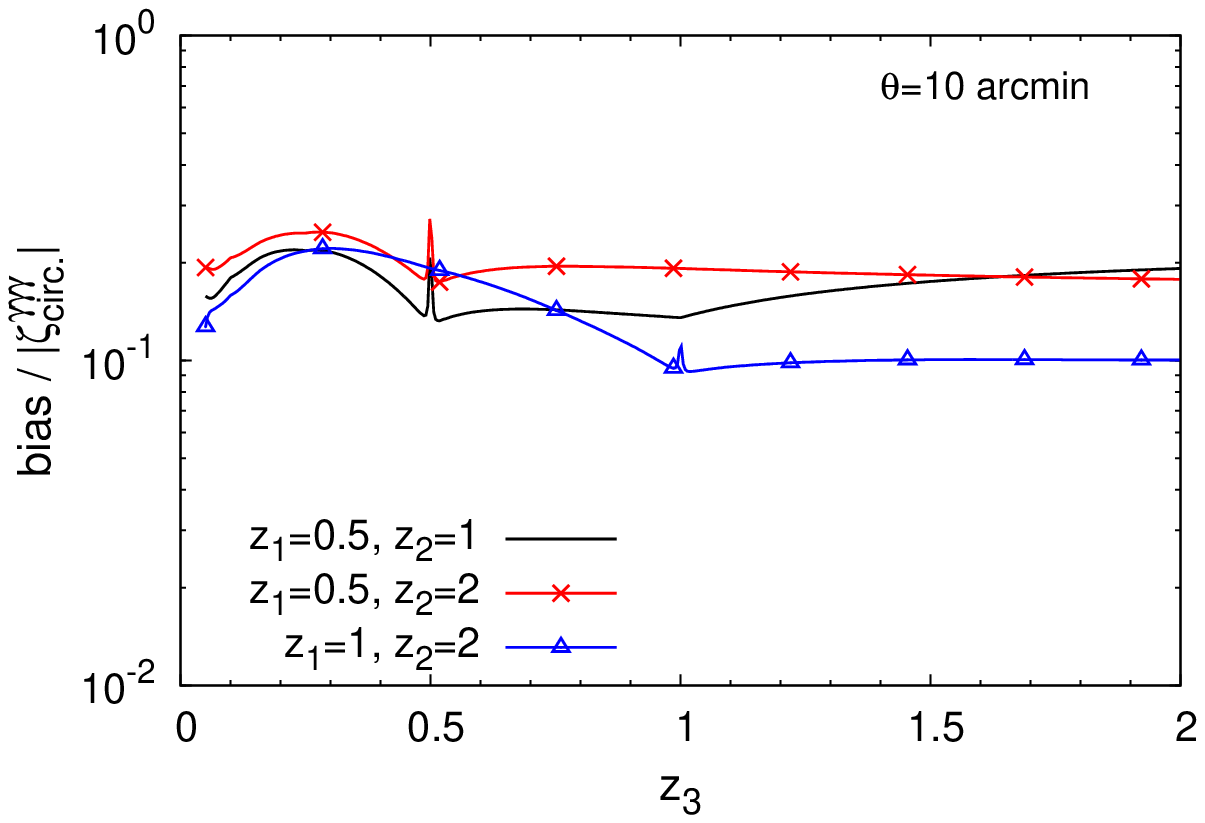}}
\epsfxsize=8.5 cm \epsfysize=6. cm {\epsfbox{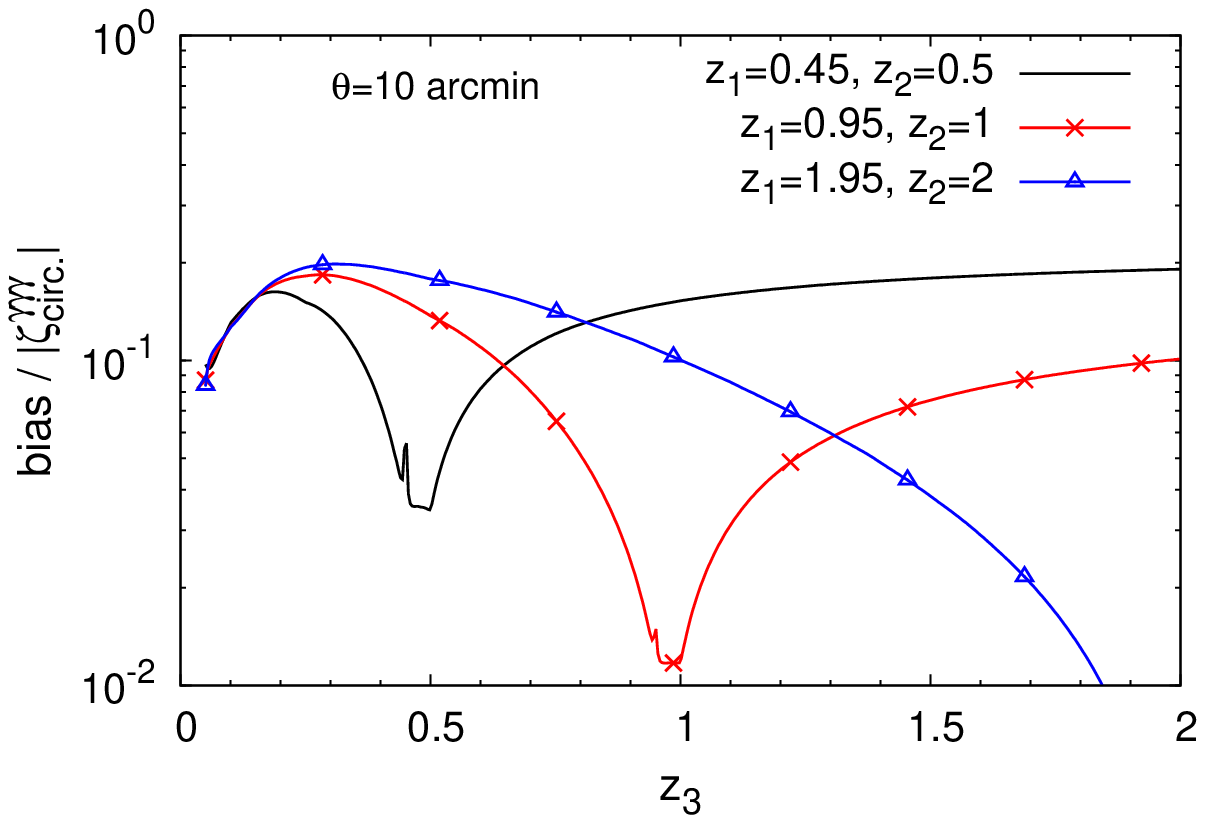}}
\end{center}
\caption{Relative source-lens clustering bias of the circular weak lensing shear
three-point correlation $\zeta^{\gamma\gamma\gamma}_{\rm circ.}$,
as a function of the third galaxy redshift $z_3$ for a fixed pair of redshifts
$\{z_1,z_2\}$. We consider the angular
scales $\theta=1$ (upper panels) and $10$ arcmin (lower panels).
All total biases are positive.}
\label{fig_zeta_gamma_theta}
\end{figure*}

We show our results as a function of the angular radius $\theta$ in
Fig.~\ref{fig_zeta_gamma_z_z_z}.
As for the convergence three-point correlation shown in
Fig.~\ref{fig_zeta_kappa_z_z_z}, the relative amplitude of the source-lens
clustering bias is of order $10\%$. However, it does not significantly decrease
on larger scales (in this range $1'<\theta<50'$) and is about ten times larger
than would be estimated from a study of the convergence alone on scales
$\theta \sim 50'$. This is due to the fact that the shear correlations are not
simply proportional to the convergence statistics because of the spin-2 factor
$e^{2\ii\alpha}$. This leads to counterterms as in Eq.(\ref{zetap-circ-zeta2D}),
associated with the non-local dependence of the shear, which make the shear
also depend on the slope of the lensing correlation functions.
For instance, as seen in Fig.~\ref{fig_zeta_ansatz_z_z_z} in
App.~\ref{comparison-3pt}, the ratio 
$|\zeta^{\gamma\gamma\gamma}_{\rm circ.}/\zeta^{\kappa\kappa\kappa}_{\rm equ.}|$
grows with the angular scale $\theta$ together with the slope of
$\zeta^{\kappa\kappa\kappa}_{\rm equ.}$.
The angular dependence of the density bispectrum also changes with scale, as
we go from the small scale one-halo regime to the large-scale perturbative
regime, and this also makes the ratio 
$|\zeta^{\gamma\gamma\gamma}_{\rm circ.}/\zeta^{\kappa\kappa\kappa}_{\rm equ.}|$
grows with $\theta$ (as shown by the comparison between the solid and dashed
lines in Fig.~\ref{fig_zeta_ansatz_z_z_z}).
Therefore, it is useful to go beyond the convergence case and to compute the
shear case itself (or the statistics of interest in each specific data analysis),
even though the computations are somewhat heavier.

As for the convergence, the relative bias on small scales is somewhat higher
if $z_1=z_2$, because of the term 
$\lag\delta_1\delta_2\gamma_3\rag_{\alpha} \lag\gamma_1\gamma_2\rag_{\alpha}$
in Eq.(\ref{hzeta-gamma-2}).

We show the redshift dependence of the source-lens clustering bias in
Fig.~\ref{fig_zeta_gamma_theta}.
Again, the upward spikes correspond to redshifts $z_3$ such that the two lowest
galaxy redshifts of the triplet $\{z_1,z_2,z_3\}$ are equal.
(The features at $z_3 \sim 0.1$ for $\theta=1'$ are due to changes of sign
of $\zeta^{\gamma\gamma\gamma}_{\rm circ.}$, which make the ratio
${\rm bias}/\zeta^{\gamma\gamma\gamma}_{\rm circ.}$ diverge.)
In agreement with Fig.~\ref{fig_zeta_gamma_z_z_z}, for both angular
scales $\theta=1'$ and $10'$ the relative amplitude of the source-lens
clustering bias is of order $10\%$.
Therefore, although it is subdominant and can be neglected in current 
weak lensing surveys, where the signal to noise ratio is about unity for
three-point statistics \citep{Semboloni2011}, it could be necessary to
take this bias into account in future surveys such as Euclid
\citep{Refregier2010}.

As for the convergence, the right panels, which correspond to closer
pairs $\{z_1,z_2\}$, show a decrease of the bias for $z_3$ close to
$\{z_1,z_2\}$, because of the vanishing of the lensing kernels on the source
plane, which yields a zero bias (in our approximations) for $z_1=z_2=z_3$.
When $z_3$ is farther from the pair $\{z_1,z_2\}$ we recover a $10\%$
bias as in the left panels.

\section{Comparison with some previous works on the source-lens clustering bias}
\label{Comparison}

The source-lens clustering bias has already been studied in 
\citet{Bernardeau1998} and \citet{Hamana2002} but from a different point
of view. They consider the measure of a $\kappa$-map from an estimator of
the form
\beq
\hat{\kappa}(\vtheta) = \frac{\sum_i \, w_i \,\kappa(\vtheta_i)}{\sum_{i'} \, w_{i'}}
=  \frac{\int\dd\chi_s \dd\vtheta_s \, \chi_s^2 \; n_s \, \kappa_s}
{\int\dd\chi_s' \dd\vtheta_s' \, \chi_s'^2 \; n_s'} ,
\label{kappa-map}
\eeq
where in the second equality we used notations similar to Eq.(\ref{hkappa-1}),
and $n_s$ is the observed galaxy density at position $(\chi_s,\vtheta_s)$.
[For the clarity of the discussion below, we denote the dummy variables
$i'$ and $(\chi_s',\vtheta_s')$ with a prime in the denominator, but both numerator
and denominator are integrated over the same source distribution.]
Thus, one measures the convergence $\kappa(\theta)$ in each direction on
the sky from the galaxies in a small area around this direction.
This is mostly suited for studies on large angular scales, so that there are
enough galaxies in each direction bin to obtain a meaningful average.
As shown by Eq.(\ref{kappa-map}), \citet{Bernardeau1998} and 
\citet{Hamana2002} actually consider a one-point statistics,
$\hat{\kappa}(\vtheta)$, which is smoothed on the scale $\theta_s$ of the angular
bin, and next estimate the variance, skewness, and kurtosis
of this one-point quantity.

In contrast, in this paper we directly consider two-point or three-point estimators
such as Eq.(\ref{hxi-1}), which read for the convergence as
\beq
\hat\xi^{\kappa\kappa}(\theta_{ij}) = \frac{\sum_{i,j} w_i w_j \; \kappa(\vtheta_i)
\kappa(\vtheta_j)}{\sum_{i',j'} w_{i'} w_{j'}} , 
\label{hxi-kap-1}
\eeq
\beq
\hat\zeta^{\kappa\kappa\kappa}(\theta_{ij},\theta_{jk},\theta_{ki}) =
\frac{\sum_{i,j,k} w_i w_j w_k \; \kappa(\vtheta_i) \kappa(\vtheta_j)
\kappa(\vtheta_k)}{\sum_{i',j',k'} w_{i'} w_{j'} w_{k'}} , 
\label{hzeta-kap-1}
\eeq
without building an intermediate $\kappa$-map, which is
closer to current observational practice (for the same reason, we also consider
estimators of the shear correlations in addition to the convergence).
[Again, we denote with a prime the dummy indices in the denominator,
but $\{i,j,k\}$ and $\{i',j',k'\}$ run over the same set.]
This also allows one to probe smaller angular scales, because the
statistical sum in Eq.(\ref{hxi-kap-1}), or Eq.(\ref{hzeta-kap-1}),
is taken over all pairs of separation $\theta_{ij}$, or triplets of separation
$\{\theta_{ij},\theta_{jk},\theta_{ki}\}$, over the full survey and one is not limited
by the binning width of the $\kappa$-map.

This also implies that contributions due to source-lens clustering appear both in the
numerator and denominator in Eq.(\ref{kappa-map}), through the fluctuations
of the galaxy densities in each direction bin (independently of the total survey
area), whereas in Eqs.(\ref{hxi-kap-1}) and (\ref{hzeta-kap-1}) we can neglect the
fluctuations of the denominator as they scale as $1/\sqrt{(\Delta\Omega)}$ over
the survey area $(\Delta\Omega)$ (and we assume a sufficiently large survey)
[the clustering of the sources contributes to the denominator, as shown by the terms
$\xi_{i,j}$ or $\zeta_{i,j,k}$ in the denominator of Eqs.(\ref{hkappa-2}) or
(\ref{hzeta-kappa-2}), but for a survey area that is much larger than the galaxy
correlation length, the sample variance of the denominator becomes negligible in
relative terms, as $1/\sqrt{(\Delta\Omega)}$, see Eq.(\ref{D-sig})].

From a physical point of view, this also means that the source-lens clustering
effects studied in \citet{Bernardeau1998} and \citet{Hamana2002} are
rather different from those studied in this article.
Two effects come into play for the one-point estimator (\ref{kappa-map}).
A first positive bias comes from the coupling in the numerator between
$n_s = \bar{n}_s (1+\delta_s)$ and $\kappa_s$, on the {\it same line of sight},
as large-scale overdensities around the source plane $\chi_s$ yield both
a positive convergence (due to matter overdensities in front of the source) and
a positive galaxy number density fluctuation $\delta_s$. A second negative
bias, due to the finite redshift width of the source distribution, comes from the
coupling between the convergence $\kappa_s$ in the numerator and the
galaxy number density fluctuations $\delta_s'$ in the denominator, with
$\chi_s'<\chi_s$. Indeed, a large-scale overdensity at $\chi_s'$
yields both a positive galaxy number density fluctuation $\delta_s'$ 
and a positive convergence $\kappa_s$. This damps the contributions with a
positive convergence to the estimator (\ref{kappa-map}), because the 
same effect associated with a large positive background $\kappa_i$ also produces
a great number of foreground galaxies $i'$ so that the contribution of $\kappa_i$
to Eq.(\ref{kappa-map}) is diluted.
[Mathematically, the negative sign of this bias comes
from the fact that expanding a fluctuation in the denominator gives a term
$1/(1+\delta_s') \simeq 1 - \delta_s'$.]
This is also briefly described in Sect.~2 in \citet{Hamana2002}.
The first effect, which involves correlations at the source plane, is suppressed
by a factor of order $x_0/(c/H_0)$ because of the vanishing of the lensing
efficiency kernel $g(\chi',\chi)$ at $\chi'=\chi$ (as also discussed in
Sect.~\ref{galaxy-location-bias} for our two-point estimators).
Therefore, the second effect dominates provided the redshift source distribution
is broad enough.
Thus, \citet{Bernardeau1998} and \citet{Hamana2002} find a
{\it negative bias} that becomes more important as the width of the redshift
source distribution increases (the convergence is defined with an opposite sign in
\citet{Bernardeau1998}).

In this paper, the source-lens clustering bias arises in a different manner.
Indeed, since we can neglect the fluctuations of the denominators in
Eqs.(\ref{hxi-kap-1}) and (\ref{hzeta-kap-1}), source-lens clustering effects
come from the coupling in the numerators between {\it different
lines of sight}. This difference from the case studied in \citet{Bernardeau1998}
and \citet{Hamana2002} is clearly due to the fact that (\ref{kappa-map}) is
a one-point estimator while (\ref{hxi-kap-1}) and (\ref{hzeta-kap-1}) are two-point
and three-point estimators.
Then, we typically obtain a {\it positive bias} on small angular scales,
see Figs.~\ref{fig_xi_kappa_theta} and \ref{fig_zeta_kappa_theta}, as a
large-scale overdensity around $(\chi_i,\vtheta_i)$ yields a
positive galaxy fluctuation $\delta_i$ and a positive convergence $\kappa_i$
along the same line of sight [this is also the first effect described for the one-point
estimator (\ref{kappa-map}), which is suppressed by a factor of order
$x_0/(c/H_0)$] but also positive convergences $\kappa_j$ and $\kappa_k$
along the other lines of sight of background galaxies along the directions
$\vtheta_j$ and $\vtheta_k$. The latter effect is the dominant one for the three-point
estimator (\ref{hzeta-kap-1}).

Thus, it is not possible to make a direct comparison between our results and those
obtained in \citet{Bernardeau1998} and \citet{Hamana2002}, because we consider
different estimators that lead to different source-lens clustering effects
(as clearly shown by the fact that they have different signs).
Indeed, as explained above, the one-point estimator (\ref{kappa-map}) is associated
with a dominant negative bias, so that its skewness
$S_3^{\kappa}=\lag \kappa^3\rag_c/\lag \kappa^2\rag_c^2$ 
is decreased by the source-lens clustering effects (see also the discussion
in Sect.~2 in \citet{Hamana2002}), whereas we find a positive bias for the
estimator (\ref{hzeta-kap-1}) of the three-point correlation function.
Nevertheless, we may note that for the skewness of the convergence, 
\citet{Bernardeau1998} finds a relative bias that goes from $-3\%$, for a source distribution with a redshift width of
$\Delta z_s = 0.15$, to $-27\%$ for $\Delta z_s= 0.87$ (and typical redshifts
$z_s \sim 1$). In our case, we can see from the right panels in
Fig.~\ref{fig_zeta_kappa_theta} that for three source redshifts at $z_s \simeq 1$
with $\Delta z_s \simeq 0.1$,
we obtain a relative bias of $+1\%$ for $\theta=1'$ and somewhat below $+1\%$ for
$\theta=10'$. From the left and right panels, we can see that we obtain a relative
bias that can reach more than $+50\%$ for $\theta=1'$, or $+10\%$ for $\theta=10'$,
when one source redshift is significantly lower than the other two with $z_1 \la 0.1$,
and $\Delta z_s \ga 0.5$.
Thus, we find similar behaviors, in terms of absolute amplitudes, as in 
\citet{Bernardeau1998}, but with a somewhat lower bias (if we consider our results
at $10'$). 
Apart from the different set of cosmological parameters, these differences and
the opposite signs come from the fact that we consider different estimators,
as explained above. Nevertheless, the similar orders of magnitude could be expected
from dimensional analysis, since we probe the same physical process.
Next, from the two-point estimator (\ref{hxi-kap-1}) we explicitly estimate the
source-lens clustering bias of two-point statistics, see Sect.~\ref{galaxy-location-bias},
which was not considered in \citet{Bernardeau1998} and \citet{Hamana2002}
because it is a higher-order effect ($\sim \xi^2$ while the signal scales as $\xi$).
This allows us to check at a quantitative level that this bias is indeed negligible
for most practical purposes.

\section{Intrinsic alignments}
\label{intrinsic}

As noticed in the introduction, another well-known source of noise for
weak-lensing measurements is the intrinsic alignment of galaxies. Indeed,
each galaxy ellipticity may be correlated with the ellipticities of other galaxies
and with the local density field that gives rise to the lensing distortion of background
galaxies.
When we average over the product of observed ellipticities of pairs of galaxies,  
the first effect leads to an ``intrinsic-intrinsic'' contribution (``II'') and the second
effect to a ``lensing-intrinsic'' effect (``GI'').
Following \citet{Heymans2013,Bridle2007}, we may use a version of the
linear tidal field alignment model described in \citet{Catelan2001,Hirata2004}, and 
write the observed galaxy ellipticity as
\beq
\epsilon_{\rm obs} = \epsilon^{\star} + \gamma^{\rm I} + \gamma ,
\label{eps-obs}
\eeq
where $\gamma$ is the gravitational lensing contribution as in previous
sections, $\epsilon^{\star}$ is a random uncorrelated component that does not give
rise to any bias, and $\gamma^{\rm I}$ is the component of the intrinsic galaxy
ellipticity that is correlated with the large-scale density field, as
\beq
\gamma^{\rm I}(\vx;z) = F_{\rm I}(z) \; \int \dd\vk \; e^{\ii\vk\cdot\vx} \;
e^{2\ii\alpha_{\vk}} \; \tdelta(\vk) .
\label{gamma_I-def}
\eeq
In particular, galaxy ellipticities are only cross-correlated through their
correlation to the same density field.
Equation (\ref{gamma_I-def}) may be seen as the simplest alignment model,
where we linearize the dependence of the intrinsic ellipticity $\gamma^{\rm I}$ on the
matter density field [as for the galaxy number density fluctuations $\delta_{\rm g}$,
where we often use a linear bias model with $\delta_{\rm g} = b(M,z) \delta$].
As compared with the spin-0 quantity $\delta_{\rm g}$, the spin-2 factor
$e^{2\ii\alpha_{\vk}}$ can be understood from the spin-2 character of the
ellipticity and it is identical to the factor that appears in the expression
(\ref{gamma-def}) of the cosmic shear.
Thus, the form of Eq.(\ref{gamma_I-def}) is rather generic, once we decide to
keep only the linear term, and follows from symmetry constraints
(we assume there is no intrinsic B mode). 
If we introduce a new scale, such as the galaxy radius $R$, we can consider
other contributions, where we replace $\tdelta(\vk)$ in Eq.(\ref{gamma_I-def})
by higher-derivative terms, such as $(kR)^2 \tdelta(\vk)$, or add some explicit
smoothing, such as $W(kR) \tdelta(\vk)$ with $W(kR) \sim e^{-(kR)^2}$ 
(they are also scalar quantities and the length $R$ is required for dimensional
reasons).
However, as compared with the contribution (\ref{gamma_I-def}), these new
contributions or modifications are suppressed by a factor $(R/d)^2$,
where $d=\chi\theta$ is the typical angular scale between the lines of sight.

However, in practice we never measure the intrinsic ellipticity $\gamma^{\rm I}(\vx)$
itself, seen as a continuous field in the limit $R\rightarrow 0$, but the
galaxy-density weighted intrinsic ellipticity $\hat{\gamma}^{\rm I}(\vx)$
\citep{Hirata2004},
\beq
\hat{\gamma}^{\rm I}(\vx) = (1+\delta_{\rm g}) \, \gamma^{\rm I} = (1+b\delta) \,
\gamma^{\rm I} .
\label{tgamma-I-def}
\eeq
Indeed, in a fashion similar to Eq.(\ref{hxi-1}), estimators of the
intrinsic-ellipticity -- galaxy correlation function write for instance as
\beq
\hat{\xi}^{\gamma^{\rm I}\delta_{\rm g}}(x_{\parallel},r_{\perp}) 
= \frac{\sum_{i,j} \gamma_i^{\rm I}}{\bar{n}^2 {\cal V} (\Delta V)} 
= \lag (1+b_1\delta_1) \gamma_1^{\rm I} (1+b_2\delta_2) \rag ,
\label{hxi-gI-d-def}
\eeq
where ${\cal V}$ is the total survey volume or simulation box (in the redshift
slice of interest) and $(\Delta V)$ the small volume of the separation bin
$\Delta x_{\parallel} 2\pi r_{\perp} \Delta r_{\perp}$.
In a similar fashion, the estimator of the intrinsic-ellipticity auto-correlation is of
the form
\beq
\hat{\xi}^{\gamma^{\rm I}\gamma^{\rm I*}}(x_{\parallel},r_{\perp}) 
= \frac{\sum_{i,j} \gamma_i^{\rm I}\gamma_j^{\rm *}}{\bar{n}^2 {\cal V} (\Delta V)} 
= \lag (1+\delta_1) \gamma_1^{\rm I} (1+\delta_2) \gamma_2^{\rm I*} \rag .
\label{hxi-gI-gI-def}
\eeq
In practice, one considers the tangential and cross components of the ellipticity
and may use Landy-Szalay estimators \citep{Landy1993} to reduce the noise
\citep{Mandelbaum2006,Mandelbaum2011}.
In any case, because we only measure the intrinsic ellipticities at galaxy locations,
observations and simulations constrain the galaxy-density weighted intrinsic ellipticity
$\hat{\gamma}^{\rm I}$ rather than $\gamma^{\rm I}$.
Then, following the usual practice (which is sometimes implicit in published
works), we assume that the galaxy-density weight $(1+\delta_{\rm g})$
only renormalizes the linear model (\ref{gamma_I-def}) and we write
\beq
\hat{\gamma}^{\rm I}(\vx;z) = \hat{F}_{\rm I}(z) \; \int \dd\vk \; e^{\ii\vk\cdot\vx} \;
e^{2\ii\alpha_{\vk}} \; \tdelta(\vk) .
\label{tgamma_I-def}
\eeq
Then, the ``II'' and ``GI'' power spectra are related to the matter power spectrum
as $P^{\rm II} = \hat{F}_{\rm I}(z)^2 P$ and $P^{\rm GI} = \hat{F}_{\rm I}(z) P$, 
as in \citet{Heymans2013}. In most previous works, one directly works with these
power spectra, which are typically fitted to simulations or observations, without
relying on the explicit model (\ref{tgamma_I-def}). This is more general, because
one could use different cross- and auto-correlation normalizations,
$P^{\rm II} = \hat{F}_{\rm II}(z) P$ and $P^{\rm GI} = \hat{F}_{\rm I}(z) P$, 
or even different power spectrum shapes, whereas the model (\ref{tgamma_I-def})
implies $\hat{F}_{\rm II} = \hat{F}_{\rm I}^2$.
Nevertheless, this is not a serious drawback as compared with previous works
that often assume $\hat{F}_{\rm II} = \hat{F}_{\rm I}^2$ (to avoid introducing too
many parameters).
In our case, we need the explicit model (\ref{tgamma_I-def}) to compute
three-point correlations (otherwise we would need an additional model for the
intrinsic-alignment bispectra).

This distinction between $\gamma^{\rm I}$ and $\hat{\gamma}^{\rm I}$ is
important in this paper because we explicitly consider the source-lens clustering
effects, that is, the factors $\delta_i$ in statistical averages such as
Eq.(\ref{hkappa-2}), and we cannot ``forget'' them when we include intrinsic-alignment
contributions. 
The linear tidal field model itself only gives Eq.(\ref{gamma_I-def}), because it
is based on the continuous dark matter tidal field \citep{Catelan2001}.
Then, the density weighted shear would read as 
$\hat{\gamma}^{\rm I} = (1+b\delta) \gamma^{\rm I}$ \citep{Hirata2004}.
However, if we use such a model we face small-scale problems when we
consider two- or three-point estimators such as Eq.(\ref{hgamma-2}).
Indeed, this gives rise to factors such as
$\lag (1+b_1\delta_1) \gamma_1^{\rm I} \gamma_2^*\rag$, which involve the
three-point correlation $\lag \delta_1 \gamma_1^{\rm I} \gamma_2^*\rag$.
However, this quantity is divergent for a power spectrum
that decreases more slowly than $k^{-3}$ at high $k$, that is, where $\xi(0) = \infty$.
Indeed, using for instance the ansatz (\ref{bispectrum-def}), we obtain a term
that behaves as
$\overline{\lag \gamma_1^{\rm I} \delta_1\rag \lag \gamma_1^{\rm I} \gamma_2^*\rag}$,
which corresponds to the coincident limit $\vx_{1'} \rightarrow \vx_1$ in
Eq.(\ref{zeta2-3}).
(The overbar denotes that the two averages cannot be separated and the
angular average requires some care and gives a nonzero result.
In contrast, coincident two-point correlations such as
$\lag \delta_1 \gamma_1^{\rm I}\rag$ vanish by symmetry, but this is no longer the
case for $\lag \delta_1 \gamma_1^{\rm I} \gamma_2^*\rag$ because of the explicit 
coupling to the extra factor $\gamma_2^*$, associated with the second line of sight, 
which breaks the previous symmetry.)

In the current $\Lambda$CDM cosmology, this quantity does not really diverge,
and $\xi(0)$ is finite, but this still means that the contribution
$\lag (1+b_1\delta_1) \gamma_1^{\rm I} \gamma_2^*\rag$ can be very large and 
depends on the shape of the power spectrum at very small scales, which does not
make much physical sense.
[We checked numerically that within this approach, terms such as
$\lag\delta_1\gamma_1^{\rm I}\gamma_2^*\rag$ would dominate over
$\lag\gamma_1^{\rm I}\gamma_2^*\rag$, even when we introduce a small-scale
cutoff at $R=1h^{-1}$Mpc, and would give rise to a too large intrinsic-alignment bias.]
In fact, this problem only warns us that the simple intrinsic-alignment model
(\ref{gamma_I-def}) is not sufficient to compute the three-point correlation
$\lag \delta_1 \gamma_1^{\rm I} \gamma_2^* \rag$.
Indeed, by discarding the galaxy length-scale $R$ in Eq.(\ref{gamma_I-def}),
we have assumed that we consider much larger scales, so that the limit
$R \rightarrow 0$ is meaningful. 
In the three-point correlation $\lag \delta_1 \gamma_1^{\rm I} \gamma_2^* \rag$
this is no longer the case, as one distance goes to zero (between $\delta_1$
and $\gamma_1^{\rm I}$, which correspond to the same galaxy). 
Then, we should include the galaxy scale $R$, for instance through a smoothing
kernel $W(kR)$, in the intrinsic alignment model (\ref{gamma_I-def}) (and also
in the galaxy bias model $\delta_g = b \delta$).

In practice, we consider the same-point combination
$(1+b\delta)\gamma^{\rm I}$ as a new quantity $\hat{\gamma}^{\rm I}$
with a ``renormalized'' amplitude $\hat{F}_{\rm I}$, as in
Eq.(\ref{tgamma_I-def}), rather than considering the ``bare'' intrinsic 
ellipticity $\gamma^{\rm I}$.
This makes sense from both theoretical and observational/practical points of view.
Indeed, three-point correlations such as $\lag (1+b\delta_1) \gamma_1 \gamma_2\rag$
are dominated by contributions of the form $\xi_{1,1} \xi_{1,2}$, which decay
linearly over $\xi_{1,2}$ with the distance between the two lines of sight while
$\xi_{1,1}$ behaves as a scale-independent prefactor, whereas contributions
of the form  $\xi_{1,2} \xi_{1,2}$ are negligible at large distance.
Thus, for two-point and higher-order correlations between different lines of sight, the
product $(1+b\delta_1)\gamma_1$ can be considered as a single quantity
and correlations with distant objects can be expected to scale as $\xi_{1,2}$, which
is satisfied by the renormalized model (\ref{tgamma_I-def}).
On the other hand, because measures only deal with the density-weighted
intrinsic ellipticity, as in Eqs.(\ref{hxi-gI-d-def})-(\ref{hxi-gI-gI-def}), they directly
constrain the ``renormalized'' factor $\hat{F}_{\rm I}$ of Eq.(\ref{tgamma_I-def}),
rather than the ``bare'' factor $F_{\rm I}$ of Eq.(\ref{gamma_I-def}).
Therefore, we can directly work with the model (\ref{tgamma_I-def}) and
forget (or rather bypass) the ``bare'' model (\ref{gamma_I-def}). 
Then, the structure of Eq.(\ref{tgamma_I-def}) can be motivated by the same
symmetry and long-distance arguments discussed below Eq.(\ref{gamma_I-def}). 

We choose for $\hat{F}_{\rm I}(z)$ the relation used in previous
works \citep{Hirata2004,Bridle2007,Joachimi2011,Heymans2013}
\beq
\hat{F}_{\rm I}(z) = - \frac{C_1 \, \rho_{\rm m0}}{D_+(z)} ,
\label{F-def}
\eeq
where $D_+(z)$ is the linear growth factor normalized to unity today
and the normalization constant is 
$C_1 = 5 \times 10^{-14} h^{-2} M_{\odot}^{-1}$Mpc$^3$.
This redshift dependence follows from the linear theory model described in
\citet{Hirata2004}, but as in \citet{Bridle2007} and \citet{Heymans2013}, we use
the nonlinear density contrast in Eq.(\ref{tgamma_I-def}), because its form is more
general than a linear theory over $\delta_L$.

A first improvement to the model (\ref{tgamma_I-def}) would be to include
a dependence on mass for the prefactor $\hat{F}_{\rm I}$ [as for the ordinary
number-density bias $b(M,z)$]. Alternatively, one can include a dependence on
galaxy type (e.g., elliptical vs spiral galaxies), as done in \citet{Semboloni2008}
in N-body simulations (with a phenomenological model to align elliptical galaxies
with the host halo and spiral galaxies transverse to the halo spin) or as observed
from the comparison between simulations and galaxy surveys
\citep{Joachimi2013}. In particular, spiral galaxies are expected to be governed
by the angular momentum of the halo, which gives a quadratic dependence 
on the density field \citep{Catelan2001,Crittenden2001}.
This would give a small residual value for $\hat{F}_{\rm I}$ (if there remains a small
linear dependence on the halo ellipticity) and a higher-order quadratic term that
is not included in our study (and that we assume to be negligible, for instance
if the galaxy sample is mostly made of elliptical galaxies).
A second generalization would be to include a stochastic component to
$\hat{F}_{\rm I}$.
These improvements may be obtained from numerical simulations, but for our
general purpose we keep the simple parameterization (\ref{F-def}).
For specific galaxy samples with different normalizations $C_1$, our results should
be rescaled by the appropriate factors.

\subsection{Two-point shear correlation function}
\label{intrinsic-2pt}

\begin{figure}
\begin{center}
\epsfxsize=8.5 cm \epsfysize=6. cm {\epsfbox{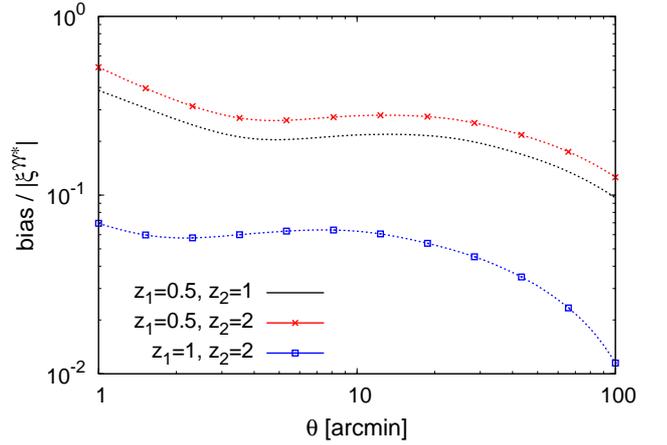}}
\end{center}
\caption{Relative intrinsic-alignment bias of the weak lensing shear two-point 
correlation, $\xi^{\rm I}/|\xi^{\gamma\gamma^*}|$, as a function of the angular
scale $\theta$, for three pairs of different source redshifts,
$(z_1,z_2)=(0.5,1), (0.5,2)$, and $(1,2)$. All total biases are negative.}
\label{fig_xi_intrinsic_gamma_z_z}
\end{figure}

\begin{figure}
\begin{center}
\epsfxsize=8.5 cm \epsfysize=6. cm {\epsfbox{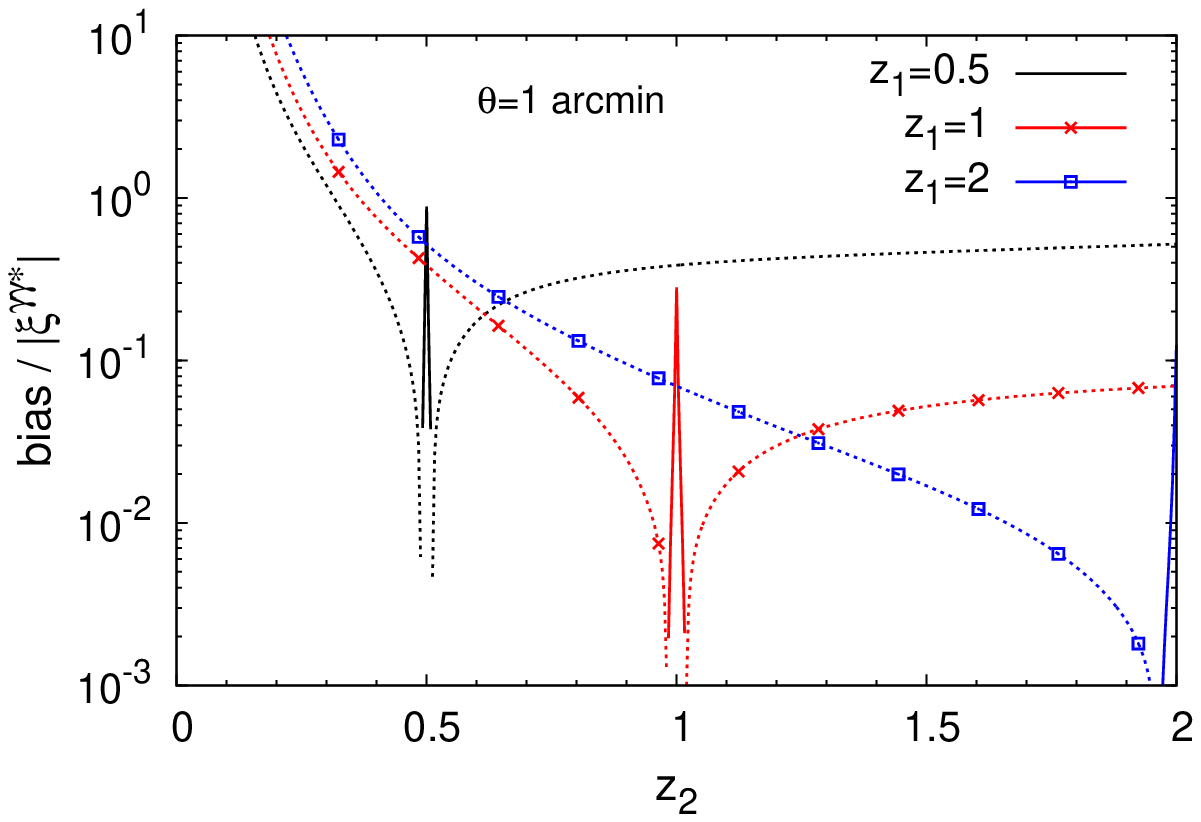}}
\epsfxsize=8.5 cm \epsfysize=6. cm {\epsfbox{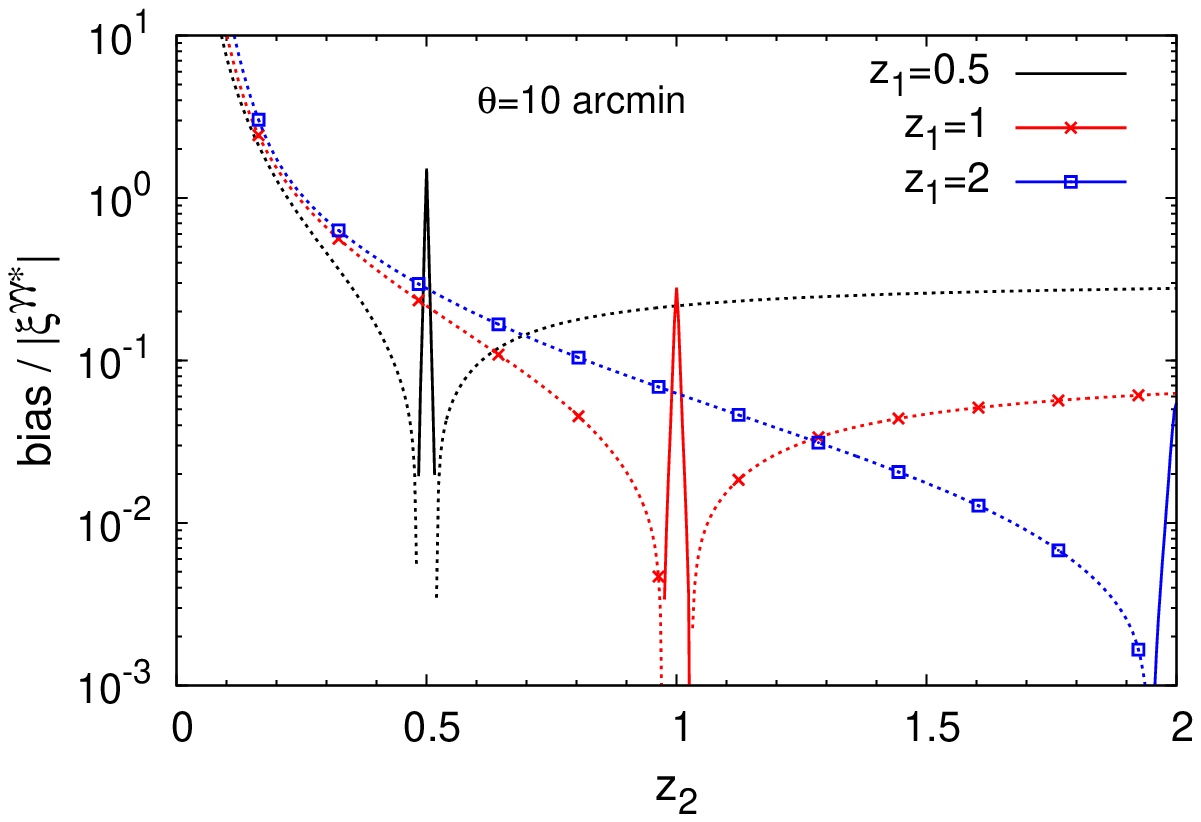}}
\epsfxsize=8.5 cm \epsfysize=6. cm {\epsfbox{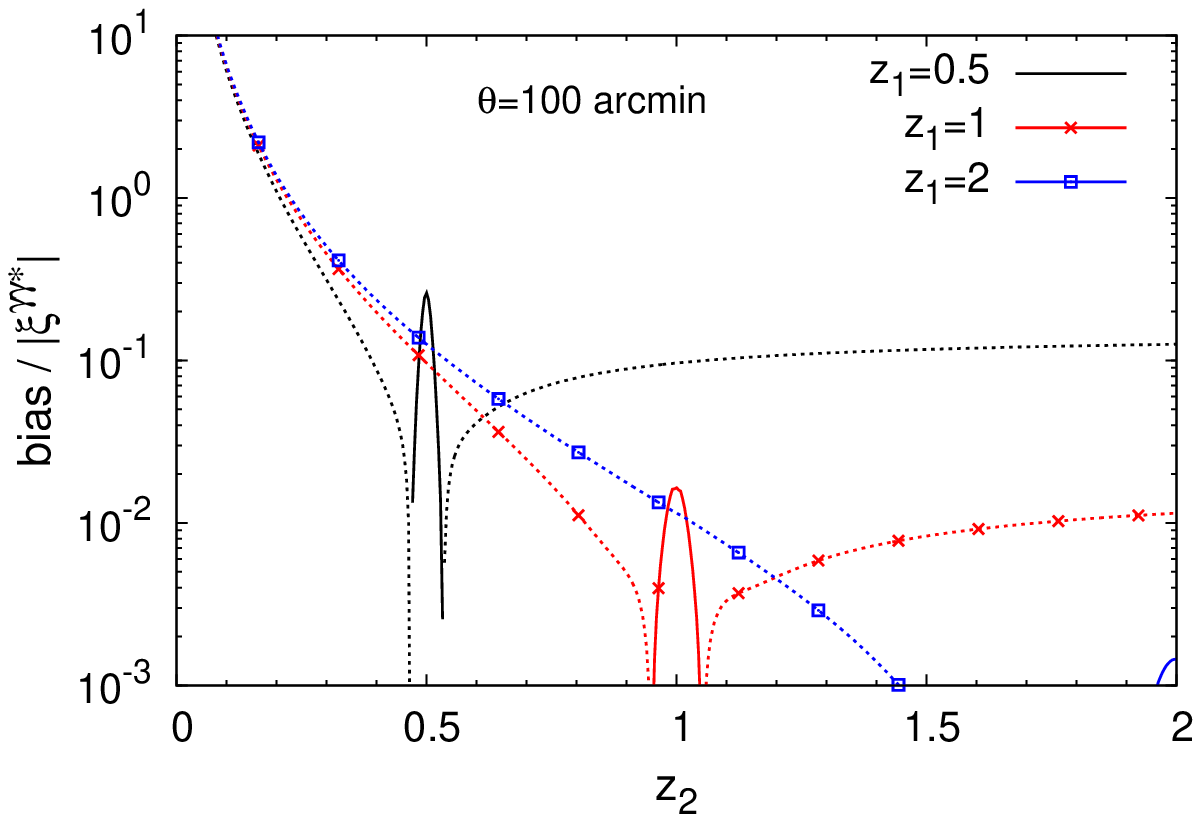}}
\end{center}
\caption{Relative intrinsic-alignment bias of the weak lensing shear two-point
correlation, $\xi^{\rm I}/|\xi^{\gamma\gamma^*}|$, as a function
of the second galaxy redshift $z_2$, for a fixed first galaxy redshift  $z_1=0.5,1$,
or $2$.
We consider the angular scales $\theta=1$, $10$, and $100$ arcmin, from the upper
to lower panel.}
\label{fig_xi_intrinsic_gamma_theta}
\end{figure}

Taking into account the intrinsic ellipticities, the average of the estimator 
$\hat\xi^{\gamma\gamma^*}$ in Eq.(\ref{hgamma-2}) now writes as
\beq
\lag \hat{\xi}^{\gamma\gamma^*}\rag = \frac{\lag (1+b_1\delta_1) (1+b_2\delta_2) 
(\gamma_1+\gamma_1^{\rm I}) (\gamma_2^*+\gamma_2^{\rm I*})\rag}
{1+b_1 b_2 \xi_{1,2}}
\eeq
As compared with Eq.(\ref{split-gamma}), this gives the additional contributions
 \beq
\xi^{\rm I} = \frac{\xi^{\rm GI} + \xi^{\rm II}}{1+b_1b_2\xi_{1,2}}   \;\;\; \mbox{and} \;\;\; 
\xi^{\delta\rm I} = \frac{\xi^{\delta\rm GI}}{1+b_1b_2\xi_{1,2}} ,
\label{xi-I}
\eeq
where the usual ``lensing-intrinsic'' and ``intrinsic-intrinsic'' contributions are
\beq
\xi^{\rm GI} = \lag \hat{\gamma}_1^{\rm I} \gamma_2^*\rag 
+  \lag \gamma_1\hat{\gamma}_2^{\rm I*}\rag ,
\label{xi-IG}
\eeq
and
\beq
\xi^{\rm II} = \lag \hat{\gamma}_1^{\rm I} \hat{\gamma}_2^{\rm I*}\rag .
\label{xi-II}
\eeq
The new ``source-lens clustering -- intrinsic'' contribution, which involves the coupling
of the source-lens clustering and intrinsic-alignment effects, is
\beq
\xi^{\delta \rm GI} = b_1 \lag \delta_1 \gamma_1 \hat{\gamma}_2^{\rm I*} \rag
+ b_2 \lag \hat{\gamma}_1^{\rm I} \delta_2 \gamma_2^* \rag \simeq 0 .
\label{xi-dI}
\eeq
It is negligible because of the vanishing of the lensing efficiency kernel $g(\chi',\chi)$
of Eq.(\ref{g-def}) at the source plane $\chi'=\chi$.

We do not treat the products $(1+b\delta)\gamma$ and $(1+b\delta)\gamma^{\rm I}$
in the same manner, because as explained below Eq.(\ref{tgamma_I-def}),
we consider $(1+b\delta)\gamma^{\rm I}$ as a single quantity
$\hat{\gamma}^{\rm I}$, while as in Sect.~\ref{galaxy-location-bias-shear} we
consider $(1+b\delta)$ and $\gamma$ as two separate quantities.
This is because $(1+b\delta)\gamma^{\rm I}$ is a one-point quantity,
whereas $(1+b\delta)\gamma$ is a two-point quantity that involves the source
and foreground fluctuations on its line of sight.

Then, from the intrinsic alignment model (\ref{tgamma_I-def}) we obtain
\beq
\lag \hat{\gamma}_1^{\rm I} \gamma_2^*\rag = \hat{F}_{\rm I}(z_1)
\int_0^{\chi_2} \dd\chi_{2'} \;  g_{2',2} \; \xi_{1,2'} ,
\label{xi-IG-1}
\eeq
and
\beq
\lag \hat{\gamma}_1^{\rm I} \hat{\gamma}_2^{\rm I*}\rag = \hat{F}_{\rm I}(z_1) 
\hat{F}_{\rm I}(z_2) \; \xi_{1,2} .
\label{xi-II-1}
\eeq

We show our results in Figs.~\ref{fig_xi_intrinsic_gamma_z_z} and
\ref{fig_xi_intrinsic_gamma_theta}.
In agreement with previous works, we find a bias of order $10\%$ for the shear
two-point correlation \citep{Heymans2013,Joachimi2013a}.
The relative effect increases for small source redshifts, $z_s \la 0.5$,
and can become greater than unity, because the shorter line of sight implies a
smaller gravitational lensing signal. Indeed, the comparison between 
Eq.(\ref{xi-gamma-1}) and Eq.(\ref{xi-IG-1}) gives
$\xi^{\gamma\gamma^*} \sim \chi g^2 x_0 \xi$ and 
$\xi^{\rm GI} \sim \hat{F}_{\rm I} g x_0 \xi$, where $x_0$ is the typical correlation
length, whence 
$\xi^{\rm GI}/\xi^{\gamma\gamma^*} \sim 1/(\chi g) \sim [c/(H_0 \chi)]^2$
increases for a small source redshifts.

When the two galaxy redshifts $z_1$ and $z_2$ are well separated,
$|z_2-z_1| \gg x_0 H_0/c$ (typically $|z_2-z_1| > 0.01$), the intrinsic-alignment
bias $\xi^{\rm I}$ is dominated by the lensing-intrinsic contribution 
(\ref{xi-IG}) and the intrinsic-intrinsic contribution (\ref{xi-II}) is negligible
(because the correlation between the two source galaxies is negligible).
Because the function $\hat{F}_{\rm I}(z)$ in Eq.(\ref{F-def}) is negative this yields a
negative bias.
When the two galaxy redshifts are very close, their intrinsic ellipticities are
significantly correlated and the intrinsic-intrinsic contribution (\ref{xi-II}) becomes
dominant. As seen from Eq.(\ref{xi-II-1}), this now gives a positive bias because
for these close pairs $\xi_{1,2}>0$. This gives rise to the positive spikes
in Fig.~\ref{fig_xi_intrinsic_gamma_theta} for $z_1 \simeq z_2$.

The comparison of Fig.~\ref{fig_xi_intrinsic_gamma_z_z} with
Fig.~\ref{fig_xi_gamma_z_z}, and of Fig.~\ref{fig_xi_intrinsic_gamma_theta} with
Fig.~\ref{fig_xi_gamma_theta}, shows that the intrinsic-alignment bias is much
greater than the source-lens clustering bias.
Thus, for practical purposes one does not need to worry about the source-lens
clustering bias of two-point statistics.
This remains true when one implements nulling techniques to eliminate the
intrinsic-alignment bias because this also removes the source-lens clustering bias
by the same effect.
Indeed, considering for instance $z_1<z_2$, this method corresponds to integrating
over a distribution of background source redshifts $n(z_2)$ so that
$\int \dd z_2 \, g_{2',2} n(z_2)=0$ at the plane $z_{2'}=z_1$ \citep{Joachimi2008}.
This damps all
contributions that arise at $z_{2'} \simeq z_1$, both the source-lens clustering
contribution $\lag \delta_1 \gamma_1 \gamma_2^* \rag  \supset \zeta_{1,1',2'}$
and the intrinsic-alignment contribution 
$\lag \hat{\gamma}_1^{\rm I} \gamma_2^*\rag \supset \xi_{1,2'}$.

\subsection{Three-point shear correlation function}
\label{intrinsic-3pt}

As for the case of the two-point estimator, the galaxy intrinsic alignments add new
contributions $\zeta_{\rm circ.}^{\rm I}$ and $\zeta_{\rm circ.}^{\delta\rm I}$ to the 
average of the three-point estimator $\hat\zeta^{\gamma\gamma\gamma}_{\rm circ.}$
in Eq.(\ref{hzeta-gamma-1-1}), which can be split as
\beq
\zeta_{\rm circ.}^{\rm I} = \frac{\zeta_{\rm circ.}^{\rm GGI} + \zeta_{\rm circ.}^{\rm GII} 
+ \zeta_{\rm circ.}^{\rm III}}{1 \!+\! b_1 b_2 \xi_{1,2} \!+\! b_2 b_3 \xi_{2,3} 
\!+\! b_1 b_3 \xi_{1,3} \!+\! b_1 b_2 b_3 \zeta_{1,2,3}} ,
\label{hzeta-I}
\eeq
and
\beq
\zeta_{\rm circ.}^{\delta\rm I} = \frac{\zeta_{\rm circ.}^{\delta\rm GGI} 
+ \zeta_{\rm circ.}^{\delta\rm GII} 
+ \zeta_{\rm circ.}^{\delta\delta\rm GGI}}{1 \!+\! b_1 b_2 \xi_{1,2} \!+\! b_2 b_3 \xi_{2,3} 
\!+\! b_1 b_3 \xi_{1,3} \!+\! b_1 b_2 b_3 \zeta_{1,2,3}} .
\label{hzeta-dI}
\eeq
For $z_1 \leq z_2 \leq z_3$, the non-negligible ``usual'' intrinsic contributions are
(because of the vanishing of the lensing kernel $g(\chi',\chi)$ at the source plane)
\[
\zeta_{\rm circ.}^{\rm GGI} = \lag \hat{\gamma}_1^{\rm I} \gamma_2 \gamma_3 
\rag_{\alpha} ,  \hspace{0.2cm}
\zeta_{\rm circ.}^{\rm GII} = \lag \hat{\gamma}_1^{\rm I} \hat{\gamma}_2^{\rm I} 
\gamma_3 \rag_{\alpha} , 
\]
\beq 
\zeta_{\rm circ.}^{\rm III} = \lag \hat{\gamma}_1^{\rm I} \hat{\gamma}_2^{\rm I}
\hat{\gamma}_3^{\rm I} \rag_{\alpha} ,
\label{zeta-III}
\eeq
where the subscript ``$\alpha$'' denotes the integration over the angles
$\alpha_{\vx_i}$ and the factor
$e^{-2\ii(\alpha_{\vx_1}+\alpha_{\vx_2}+\alpha_{\vx_3})}$, as in
Eq.(\ref{zeta-gamma-circ-def}).
When all redshifts are different, only the contribution
$\zeta_{\rm circ.}^{\rm GGI}$ is left:
\beq
z_1 \!<\! z_2 \!<\! z_3 \! :  \;\;\;  \zeta_{\rm circ.}^{\rm GGI} \simeq
\lag \hat{\gamma}_1^{\rm I} \gamma_2 \gamma_3 \rag_{\alpha} , \hspace{0.2cm}
\zeta_{\rm circ.}^{\rm GII} \simeq \zeta_{\rm circ.}^{\rm III} \simeq 0 .
\label{zeta-I-diff}
\eeq
The new ``source-lens clustering -- intrinsic'' contributions read as
(with $z_1\leq z_2\leq z_3$)
\beqa
\zeta_{\rm circ.}^{\delta\rm GGI} \!\! & = & \!\! b_1 \lag\delta_1\hat{\gamma}_2^{\rm I}
\rag_{\alpha} \lag\gamma_1\gamma_3\rag_{\alpha} + b_1 \lag\delta_1
\hat{\gamma}_3^{\rm I}\rag_{\alpha} \lag\gamma_1\gamma_2\rag_{\alpha} 
+ b_2 \lag\delta_2\hat{\gamma}_1^{\rm I}\rag_{\alpha} \nonumber \\
&& \!\! \times \lag\gamma_2\gamma_3\rag_{\alpha} 
+ b_2 \lag\delta_2\hat{\gamma}_3^{\rm I}\rag_{\alpha}
\lag\gamma_1\gamma_2\rag_{\alpha} + b_2 \lag\delta_2\gamma_3\rag_{\alpha}
\lag\hat{\gamma}_1^{\rm I}\gamma_2\rag_{\alpha} \nonumber \\
&& \!\! + b_3 \lag\delta_3\hat{\gamma}_1^{\rm I}\rag_{\alpha}
\lag\gamma_2\gamma_3\rag_{\alpha} 
+ b_3 \lag\delta_3\hat{\gamma}_2^{\rm I}\rag_{\alpha} 
\lag\gamma_1\gamma_3\rag_{\alpha} ,
\label{zeta-d-I-1}
\eeqa
\beq
\zeta_{\rm circ.}^{\delta\rm GII} = b_2 \lag\delta_2\hat{\gamma}_3^{\rm I}\rag_{\alpha}
\lag\hat{\gamma}_1^{\rm I}\gamma_2\rag_{\alpha} 
+ b _3 \lag\delta_3\hat{\gamma}_2^{\rm I}\rag_{\alpha}
\lag\hat{\gamma}_1^{\rm I}\gamma_3\rag_{\alpha} ,
\label{zeta-d-II-1}
\eeq
and
\beqa
\zeta_{\rm circ.}^{\delta\delta\rm GGI} \!\! & = & \!\! b_1 b_2 
\lag\delta_1\delta_2\hat{\gamma}_3^{\rm I}\rag_{\alpha} 
\lag\gamma_1\gamma_2\rag_{\alpha} 
+ b_1 b_3 \lag\delta_1\delta_3\hat{\gamma}_2^{\rm I}\rag_{\alpha} 
\lag\gamma_1\gamma_3\rag_{\alpha} \nonumber \\
&& \hspace{-0.6cm} + b_2 b _3 \lag\delta_2\delta_3\hat{\gamma}_1^{\rm I}\rag_{\alpha} 
\lag\gamma_2\gamma_3\rag_{\alpha} + b_2 b_3
\lag\delta_2\delta_3\rag_{\alpha} 
\lag\hat{\gamma}_1^{\rm I} \gamma_2\gamma_3\rag_{\alpha} ,
\label{zeta-d-d-I-1}
\eeqa
where the superscripts ``$\delta$'', ``G'', and ``I'', count the number of
terms $\delta$, $\gamma$, and $\hat{\gamma}^{\rm I}$.
When the three source redshifts are different, only the contribution 
$\zeta_{\rm circ.}^{\delta\rm GGI}$ is left:
\beqa
z_1 \!<\! z_2 \!<\! z_3& : &  \zeta_{\rm circ.}^{\delta\rm GGI} \simeq
b_2 \lag\delta_2\gamma_3\rag_{\alpha} 
\lag\hat{\gamma}_1^{\rm I}\gamma_2\rag_{\alpha} ,  \nonumber \\
&& \zeta_{\rm circ.}^{\delta\rm GII} \simeq \zeta_{\rm circ.}^{\delta\delta\rm GGI} 
\simeq 0 .
\label{zeta-d-I-diff}
\eeqa

\begin{figure}
\begin{center}
\epsfxsize=8.5 cm \epsfysize=6. cm {\epsfbox{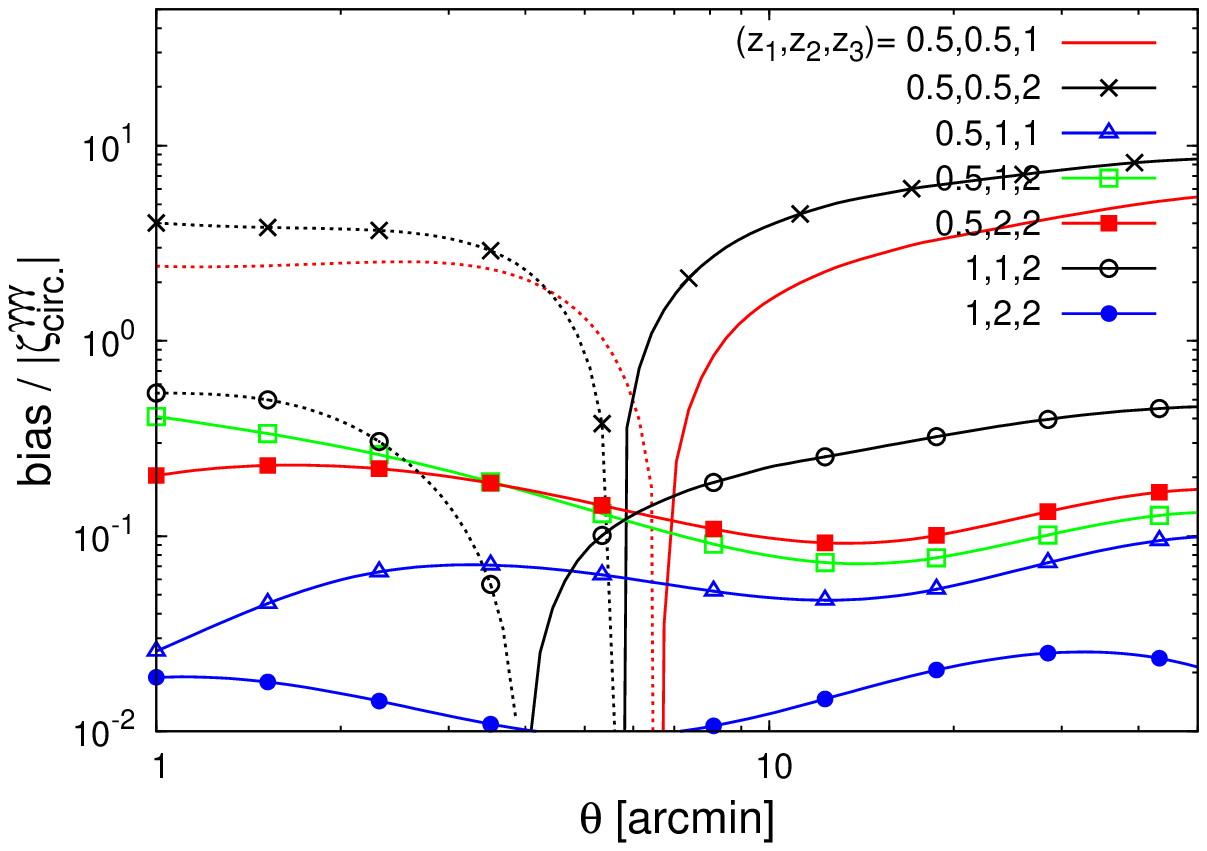}}
\end{center}
\caption{Relative intrinsic-alignment bias of the circular weak lensing shear
three-point correlation $\zeta^{\gamma\gamma\gamma}_{\rm circ.}$,
as a function of the angular scale $\theta$, for a few redshift triplets
$z_1 \leq z_2 \leq z_3$.}
\label{fig_zeta_intrinsic_gamma_z_z_z}
\end{figure}

\begin{figure*}
\begin{center}
\epsfxsize=8.5 cm \epsfysize=6. cm {\epsfbox{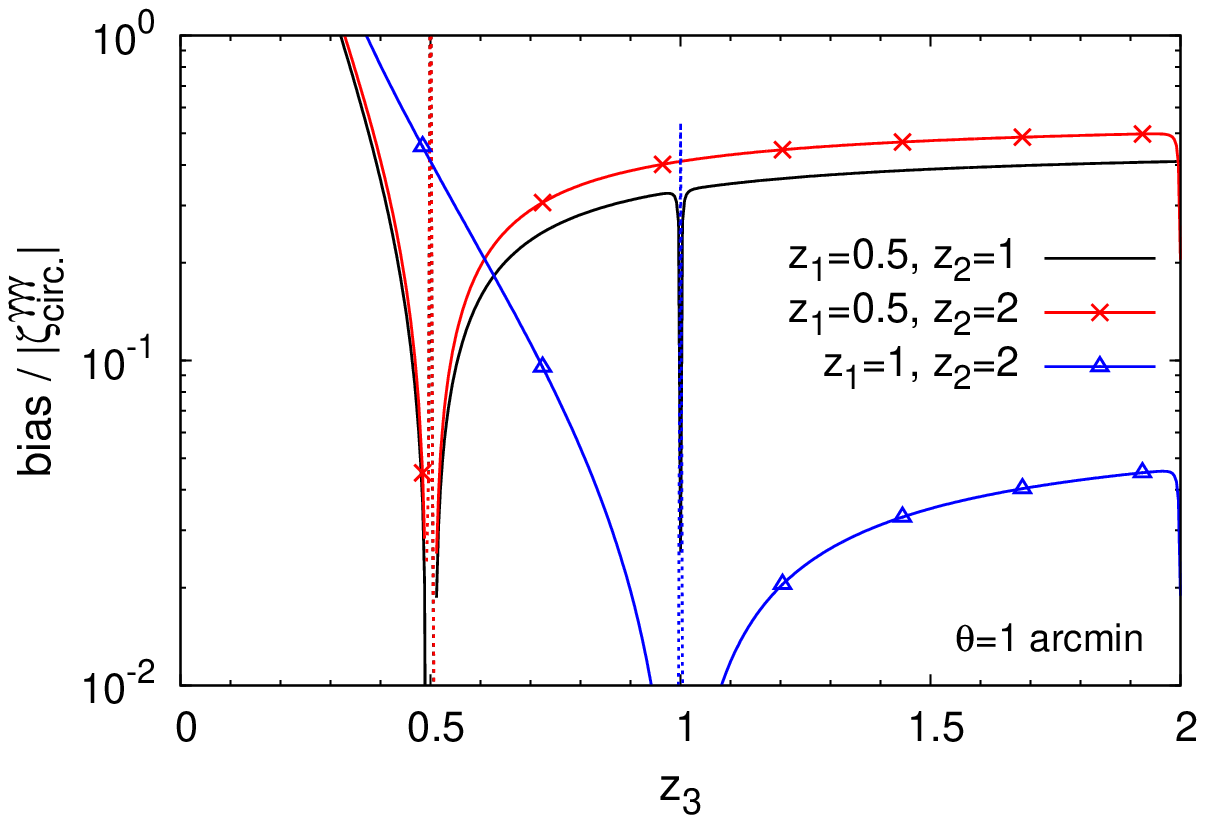}}
\epsfxsize=8.5 cm \epsfysize=6. cm {\epsfbox{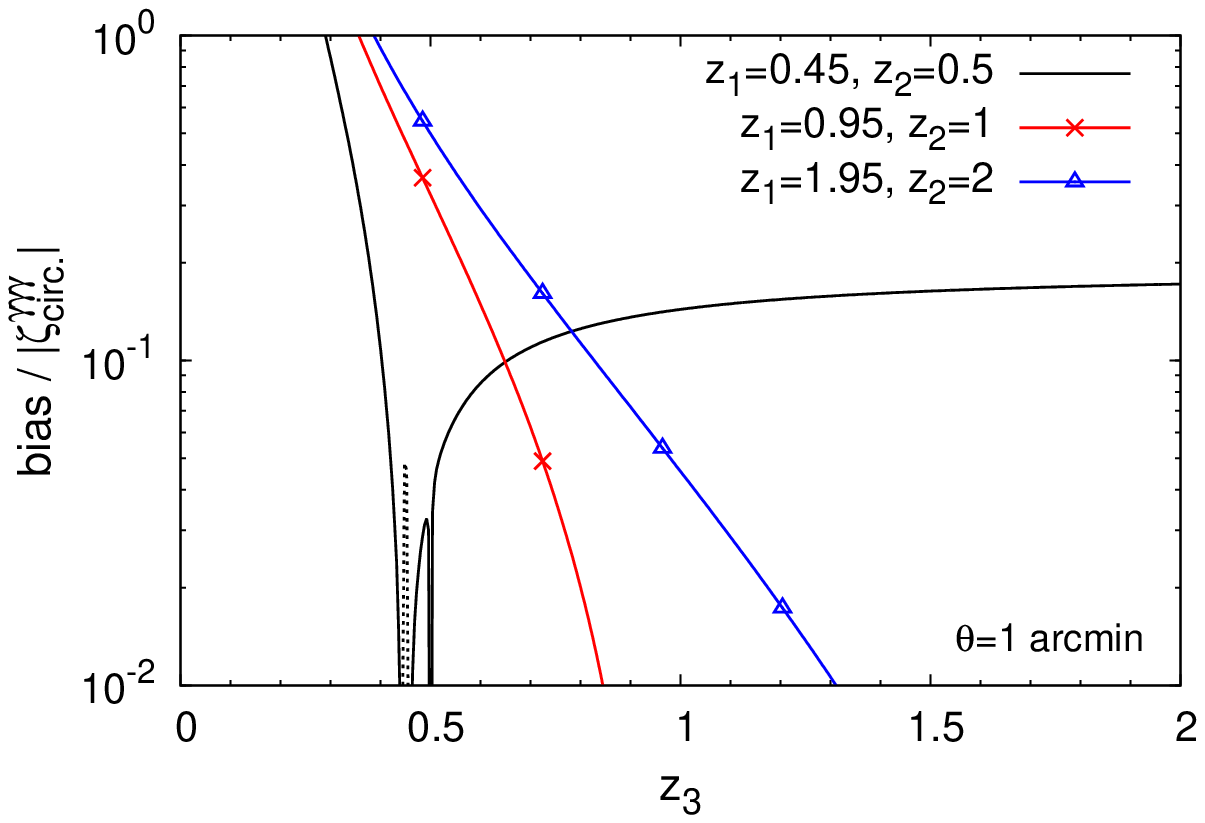}}\\
\epsfxsize=8.5 cm \epsfysize=6. cm {\epsfbox{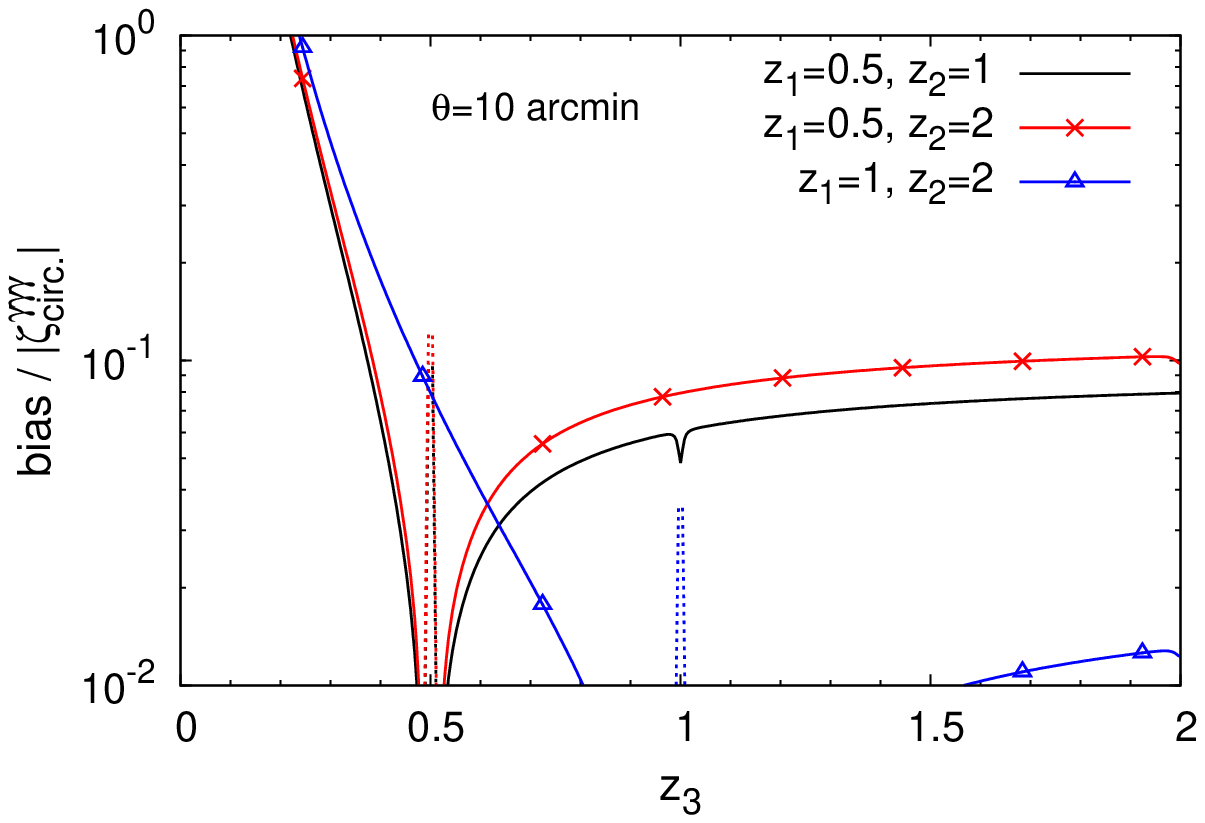}}
\epsfxsize=8.5 cm \epsfysize=6. cm {\epsfbox{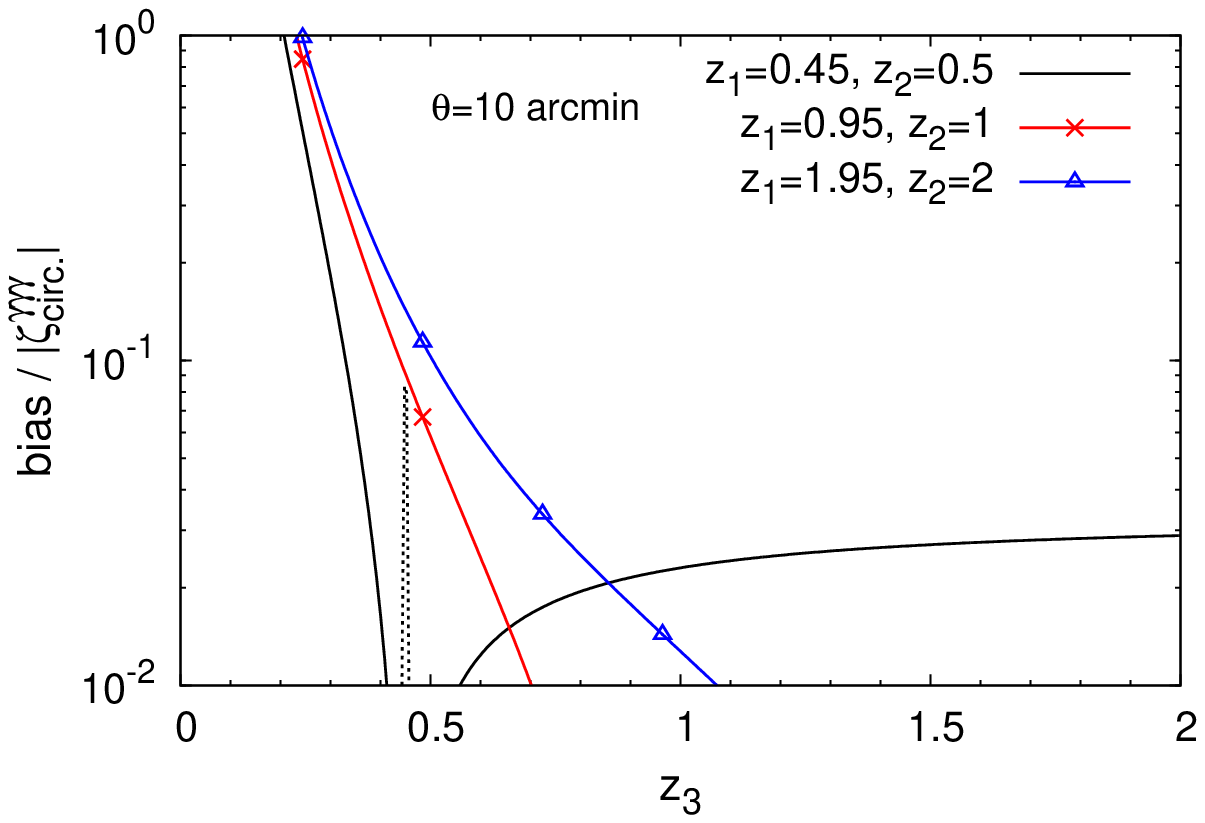}}
\end{center}
\caption{Relative intrinsic-alignment bias of the circular weak lensing shear
three-point correlation $\zeta^{\gamma\gamma\gamma}_{\rm circ.}$,
as a function of the third galaxy redshift $z_3$ for a fixed pair of redshifts
$\{z_1,z_2\}$. We consider the angular
scales $\theta=1$ (upper panels) and $10$ arcmin (lower panels).}
\label{fig_zeta_intrinsic_gamma_theta}
\end{figure*}

\subsubsection{Usual intrinsic-alignment bias}
\label{usual-intrinsic}

Using the intrinsic alignment model (\ref{tgamma_I-def}), the first lensing -- intrinsic 
contribution reads as [compare with Eq.(\ref{zeta-ggg-0})]
\beq
\zeta_{\rm circ.}^{\rm GGI}  = \lag \hat{\gamma}_1^{\rm I} \gamma_2 \gamma_3 
\rag_{\alpha} = \hat{F}_{\rm I}(z_1) \; g_{1,2} g_{1,3} \; 
\zeta_{\rm circ.}^{\rm 2D}(z_1) ,
\label{zeta-IGG-1}
\eeq
where $\zeta_{\rm circ.}^{\rm 2D}$ was defined in Eqs.(\ref{zetap-circ-B})
or (\ref{zetap-circ-zeta2D}).
Here, as in Sect.~\ref{three-point-gamma}, we used Limber's approximation.
The ``GII'' contribution, which is only relevant when
$z_1 \simeq z_2$ (we consider the ordering $z_1 \leq z_2 \leq z_3$),
writes as
\beq
\zeta_{\rm circ.}^{\rm GII} = \lag \hat{\gamma}_1^{\rm I} \hat{\gamma}_2^{\rm I} 
\gamma_3 \rag_{\alpha} = \hat{F}_{\rm I}(z_1) \hat{F}_{\rm I}(z_2) \, g_{12,3} \;
\zeta^{\rm 2.5D}_{\rm circ.} ,
\label{zeta-GII-1}
\eeq
where $g_{12,3} = g((\chi_1+\chi_2)/2,\chi_3)$ and we introduced
\beqa
\zeta^{\rm 2.5D}_{\rm circ.} & = & \int_{-\infty}^{\infty} \dd x_{3'\parallel} 
\int_0^{2\pi} \! \frac{\dd\alpha_{\vx_1}\dd\alpha_{\vx_2}\dd\alpha_{\vx_3}}{(2\pi)^3}
\nonumber \\
&& \hspace{-0.cm} \times \int \dd\vk_1\dd\vk_2\dd\vk_{3'} \;
e^{\ii [ \vk_1\cdot\vx_1+\vk_2\cdot\vx_2+\vk_{3'}\cdot\vx_{3'}]}
\nonumber \\
&& \hspace{-0.cm} \times \; \delta_D(\vk_1+\vk_2+\vk_{3'}) \; 
B(k_1,k_2,k_{3'}) \;  \nonumber \\
&& \hspace{-0.cm} \times \; e^{2\ii (\alpha_{\vk_1}+\alpha_{\vk_2}+\alpha_{\vk_{3'}}
- \alpha_{\vx_1} - \alpha_{\vx_2} - \alpha_{\vx_3})} .
\label{zeta-2.5D--1}
\eeqa
The superscript ``2.5D'' refers to the fact that this quantity is intermediate between 
the 2D correlation (\ref{zetap-circ-B}), which involves two integrations along the lines
of sight, and the 3D correlation, which has no integration along the lines of sight.
To simplify the computations, we again use the hierarchical ansatz
(\ref{bispectrum-def}) for the bispectrum. As described in
App.~\ref{intrinsic-shear-3pt-app}, this gives
\beq
\zeta^{\rm 2.5D}_{\rm circ.} = \zeta_{(12)}^{\rm 2.5D} + \zeta_{(13)}^{\rm 2.5D}
+ \zeta_{(23)}^{\rm 2.5D} ,
\eeq
where the three components are given by Eqs.(\ref{zetap-circ-2.5D-12-3})
and (\ref{zetap-circ-2.5D-13-3}) [and $\zeta_{(23)}^{\rm 2.5D} \simeq 
\zeta_{(13)}^{\rm 2.5D}$ when $z_1 \simeq z_2$].
The ``III'' contribution, which is only relevant when $z_1 \simeq z_2 \simeq z_3$,
writes as
\beq
\zeta_{\rm circ.}^{\rm III} = \lag \hat{\gamma}_1^{\rm I} \hat{\gamma}_2^{\rm I} 
\hat{\gamma}_3^{\rm I} \rag_{\alpha} =
\hat{F}_{\rm I}(z_1) \hat{F}_{\rm I}(z_2) \hat{F}_{\rm I}(z_3)
\, \zeta^{\rm 3D}_{\rm circ.} ,
\label{zeta-III-1}
\eeq
with
\beqa
\zeta^{\rm 3D}_{\rm circ.} & = & \int_0^{2\pi} \! 
\frac{\dd\alpha_{\vx_1}\dd\alpha_{\vx_2}\dd\alpha_{\vx_3}}{(2\pi)^3}
\int \dd\vk_1\dd\vk_2\dd\vk_3 \; B(k_1,k_2,k_3) \nonumber \\
&& \hspace{-0.cm} \times \; e^{\ii [ \vk_1\cdot\vx_1+\vk_2\cdot\vx_2+\vk_3\cdot\vx_3]}
\; \delta_D(\vk_1+\vk_2+\vk_3)  \nonumber \\
&& \hspace{-0.cm} \times \; e^{2\ii (\alpha_{\vk_1}+\alpha_{\vk_2}+\alpha_{\vk_3}
- \alpha_{\vx_1} - \alpha_{\vx_2} - \alpha_{\vx_3})} .
\label{zeta-3D--1}
\eeqa
Using again the hierarchical ansatz (\ref{bispectrum-def}) for the bispectrum,
we obtain
\beq
\zeta^{\rm 3D}_{\rm circ.} = \zeta_{(12)}^{\rm 3D} + \zeta_{(13)}^{\rm 3D}
+ \zeta_{(23)}^{\rm 3D} ,
\eeq
where the three components are given by permutations over
Eq.(\ref{zetap-circ-3D-12-2}), as described in App.~\ref{intrinsic-shear-3pt-app}.

We show our results in Figs.~\ref{fig_zeta_intrinsic_gamma_z_z_z} and
\ref{fig_zeta_intrinsic_gamma_theta}. The overall
behavior is similar to the one found in Figs.~\ref{fig_xi_intrinsic_gamma_z_z}
and \ref{fig_xi_intrinsic_gamma_theta}
for the two-point correlation function, with a relative bias that varies between
$1\%$ and more than unity, depending on the angular scale and the redshifts
of the sources. 
These orders of magnitude are consistent with previous works 
\citep{Semboloni2008} (but the different redshift distributions and galaxy models
make a precise comparison difficult).
Again, the relative bias increases for low source redshifts because
of the shorter lines of sight, which give rise to factors $\chi g$ in
the weak lensing signal that decrease as $\chi \rightarrow 0$.
Then, as noticed in \citet{Semboloni2008}, by removing or focusing on low-redshift 
galaxies, one can separate (or amplify the relative weight of) the cosmic-shear or
intrinsic-alignement signals.
This allows one to obtain on one hand a better constraint on cosmology from
weak lensing and on the other hand a better understanding of intrinsic alignments.
However, this requires reliable photometric or spectroscopic redshifts.

Because our three source redshifts never coincide, the ``III'' contribution is always
negligible as compared with the other two terms in Eq.(\ref{hzeta-I}).
For most redshifts in Fig.~\ref{fig_zeta_intrinsic_gamma_theta}, where 
the three source redshifts are significatly separated, the intrinsic alignment bias is
dominated by the contribution $\zeta_{\rm circ.}^{\rm GGI}$, associated with the
correlation of the intrinsic alignment of the foreground galaxy $z_1$ with the local
density field that gives rise to lensing distortions of the two more distant galaxies
$z_2$ and $z_3$. In our case, this yields a positive bias, whereas the cosmic shear
signal $\zeta^{\gamma\gamma\gamma}_{\rm circ.}$ is negative. As for the two-point
correlation case shown in Fig.~\ref{fig_xi_intrinsic_gamma_theta}, this opposite
sign is due to the negative sign in the model (\ref{F-def}).
When the lowest two source redshifts become close (i.e., $z_3 \simeq z_1$ in
Fig.~\ref{fig_zeta_intrinsic_gamma_theta}), the contribution
$\zeta_{\rm circ.}^{\rm GII}$ becomes dominant and gives a negative bias.
This gives rise to the (negative) spikes in Fig.~\ref{fig_zeta_intrinsic_gamma_theta}.

The comparison of Fig.~\ref{fig_zeta_intrinsic_gamma_theta} with
Fig.~\ref{fig_zeta_gamma_theta} shows that the intrinsic-alignment bias can
be either greater or smaller than the source-lens clustering bias, depending
on the source redshifts and angular scale. For low redshifts, $z_i \la 0.5$, the
intrinsic-alignment bias is typically greatest, while for high redshifts, $z_i \ga 1$,
the source-lens clustering bias is greatest. 
This is different from the case of the
two-point shear correlation, where the intrinsic-alignment bias was typically much
greater than the source-lens clustering bias, which can be neglected for practical
purposes. We can note that the scaling with the amplitude of the matter power
spectrum (or correlation function) is different:
\beqa
\mbox{2-pt function:} \!\!\! && \mbox{signal} \sim \lag\gamma\gamma\rag , \;\;
\mbox{source-lens clust.} \sim \lag\delta\gamma\gamma\rag , \nonumber \\
&& \mbox{intrinsic align.} \sim \lag\delta\gamma\rag ,
\label{bias-2pt}
\eeqa
\beqa
\mbox{3-pt function:} \!\!\! && \mbox{signal} \sim \lag\gamma\gamma\gamma\rag , \;\;
\mbox{s.-l. clust.} \sim \lag\delta\gamma\rag \lag\gamma\gamma\rag , 
\nonumber \\
&& \mbox{intrinsic align.} \sim \lag\delta\gamma\gamma\rag .
\label{bias-3pt}
\eeqa
Thus, whereas for the two-point estimator, the signal and the intrinsic-alignment bias
are of the same order $\sim \xi$ while the source-lens clustering bias is of higher
order $\sim \zeta \sim \xi^2$; for the three-point estimator, the signal, the
intrinsic-alignment bias, and the source-lens clustering bias, are all of the same order
$\sim \zeta \sim \xi^2$. 
Then, for three-point statistics both the intrinsic-alignment bias and the source-lens
clustering bias should be taken into account, if we aim at an accuracy better than
$10\%$.

\begin{figure}
\begin{center}
\epsfxsize=8.5 cm \epsfysize=6. cm {\epsfbox{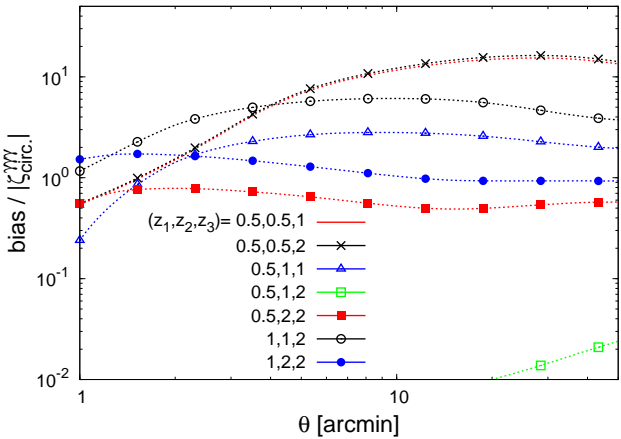}}
\end{center}
\caption{Relative source-lens clustering -- intrinsic-alignment bias of the circular
weak lensing shear three-point correlation
$\zeta^{\gamma\gamma\gamma}_{\rm circ.}$,
as a function of the angular scale $\theta$, for a few redshift triplets
$z_1 \leq z_2 \leq z_3$.}
\label{fig_zeta_slc_I_gamma_z_z_z}
\end{figure}

\begin{figure*}
\begin{center}
\epsfxsize=8.5 cm \epsfysize=6. cm {\epsfbox{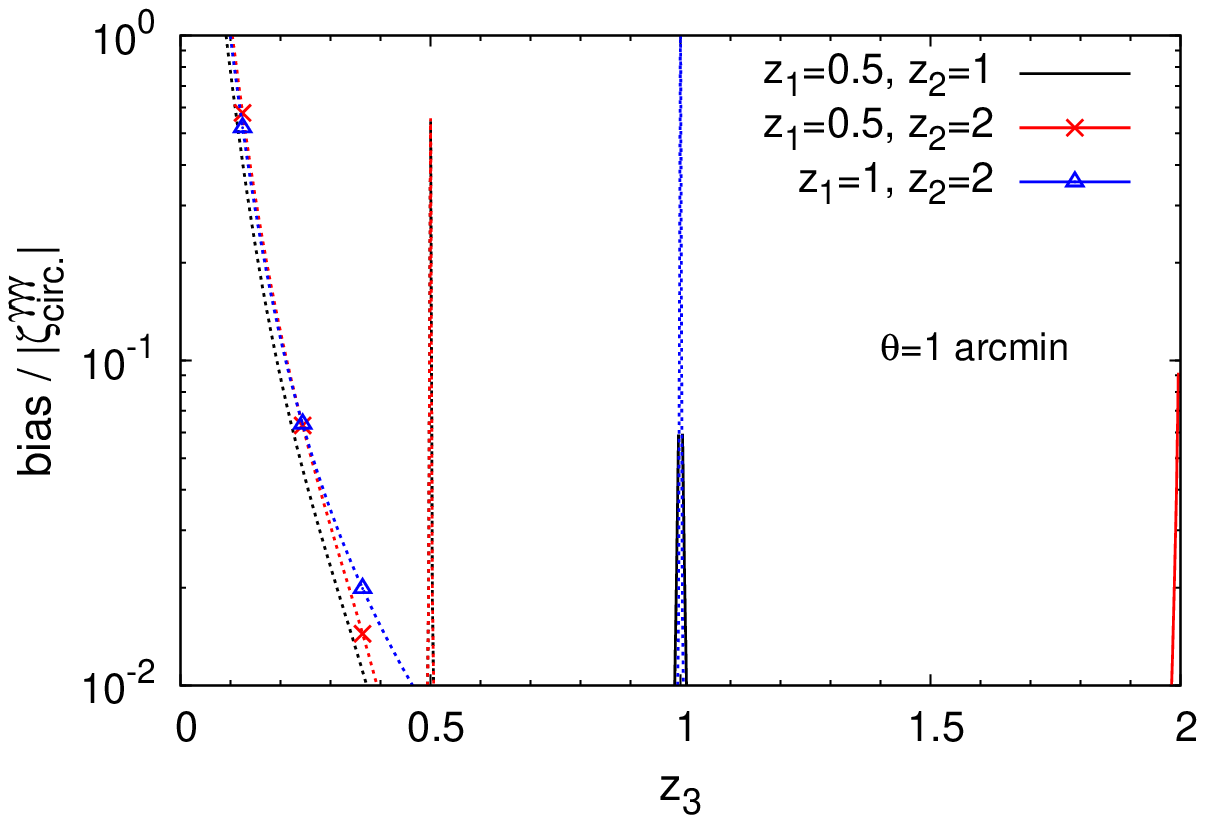}}
\epsfxsize=8.5 cm \epsfysize=6. cm {\epsfbox{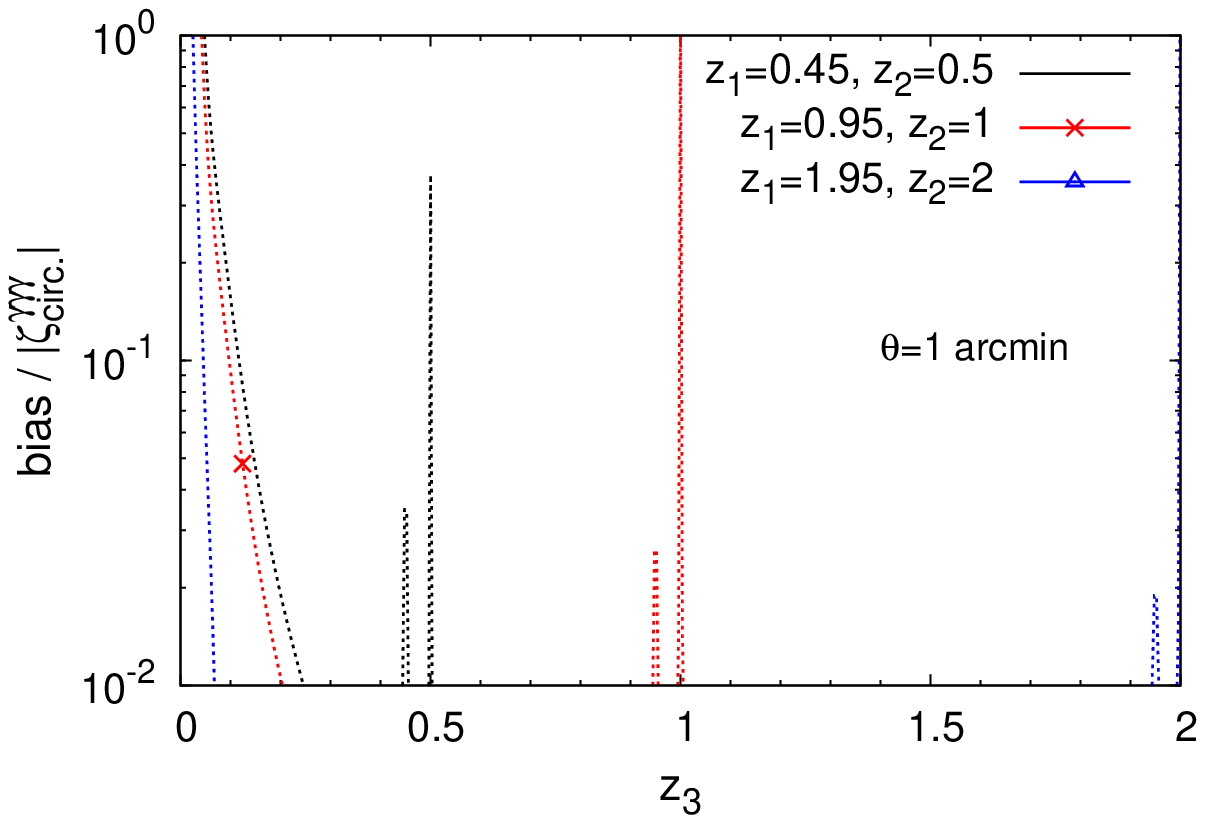}}\\
\epsfxsize=8.5 cm \epsfysize=6. cm {\epsfbox{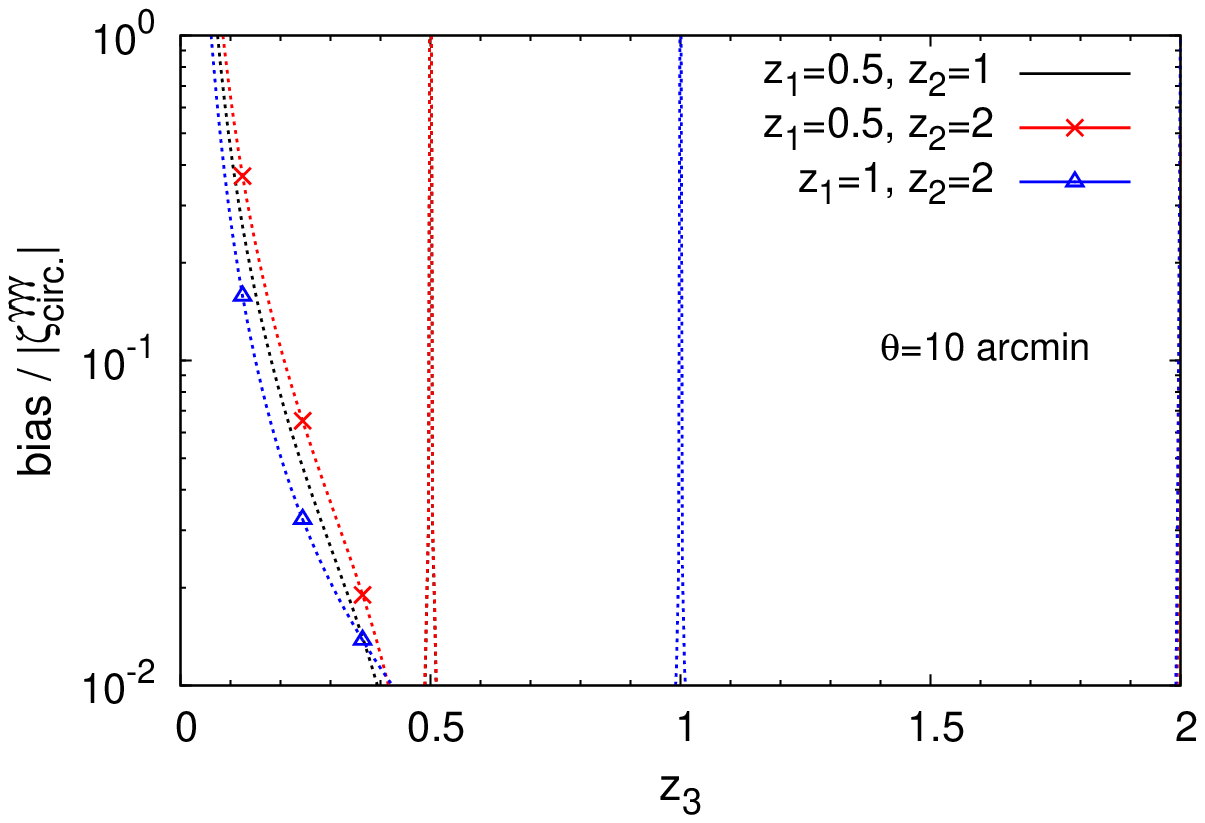}}
\epsfxsize=8.5 cm \epsfysize=6. cm {\epsfbox{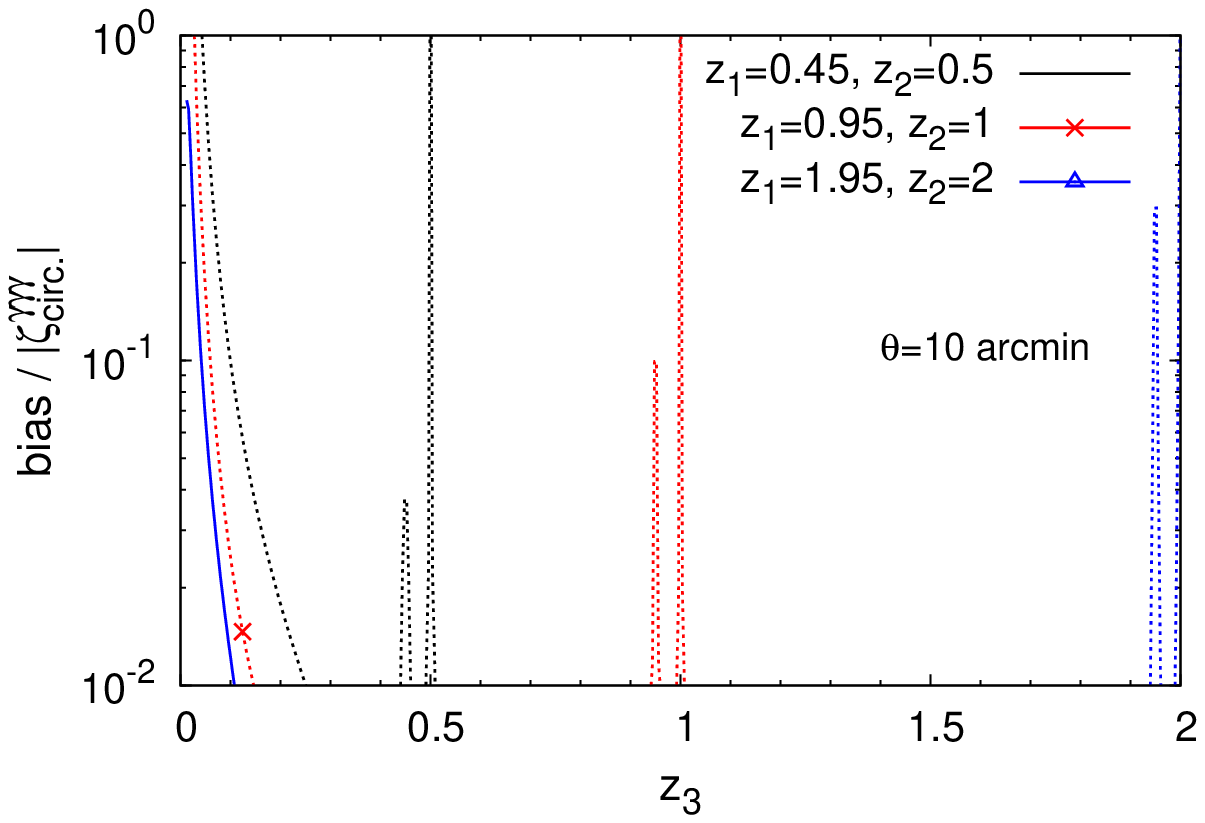}}
\end{center}
\caption{Relative source-lens clustering -- intrinsic-alignment bias of the circular weak 
lensing shear three-point correlation $\zeta^{\gamma\gamma\gamma}_{\rm circ.}$,
as a function of the third galaxy redshift $z_3$ for a fixed pair of redshifts
$\{z_1,z_2\}$. We consider the angular
scales $\theta=1$ (upper panels) and $10$ arcmin (lower panels).}
\label{fig_zeta_slc_I_theta}
\end{figure*}

Another difference with the case of two-point statistics is that nulling techniques
devised for the removal of intrinsic-alignment bias no longer automatically 
remove the source-lens clustering bias.
Indeed, considering source redshifts in non-overlapping bins with $z_1<z_2<z_3$,
the intrinsic-alignment bias is dominated by the contribution 
$\zeta_{\rm circ.}^{\rm GGI} = 
\lag\hat{\gamma}_1^{\rm I}\gamma_2\gamma_3 \rag_{\alpha}$
of Eq.(\ref{zeta-III}), and as for the two-point case it is sufficient to use
a weight $T(\chi_3)$ such that $\int \dd\chi_3 T(\chi_3) g(\chi_{3'},\chi_3)$ vanishes
at $\chi_{3'}=\chi_1$ to remove this bias \citep{Shi2010}.
This also removes the contributions 
$\lag\delta_1\gamma_2\rag_{\alpha} \lag\gamma_1\gamma_3\rag_{\alpha} 
+ \lag\delta_1\gamma_3\rag_{\alpha} \lag\gamma_1\gamma_2\rag_{\alpha}$
of the source-lens clustering bias in Eq.(\ref{hzeta-gamma-3}), but this
does not erase the third contribution 
$\lag\delta_2\gamma_3\rag_{\alpha} \lag\gamma_1\gamma_2\rag_{\alpha}$
that involves galaxy-density fluctuations at the second redshift $z_2$.
Thus, to remove both the intrinsic-alignment bias and the source-lens clustering bias
we need the weight $T(\chi_3)$ to satisfy a second constraint,
$\int \dd\chi_3 T(\chi_3) g(\chi_{3'},\chi_3)=0$ at $\chi_{3'}=\chi_2$,
to erase the correlations in both planes $z_1$ and $z_2$.
We leave a study of this problem to future work.

\subsection{Coupled source-lens clustering -- intrinsic-alignment bias}
\label{coupled}

The coupled source-lens clustering -- intrinsic alignment contributions 
(\ref{zeta-d-I-1})-(\ref{zeta-d-d-I-1}) involve new terms of the
form $\lag\delta_1\hat{\gamma}_2^{\rm I}\rag_{\alpha}$, 
$\lag\hat{\gamma}_1^{\rm I}\gamma_2\rag_{\alpha}$, and
$\lag\delta_1\delta_2\hat{\gamma}_3^{\rm I}\rag_{\alpha}$.
The first two terms can be read from Eqs.(\ref{xi-d1-g2}) and (\ref{xi-g1-g2}),
where an integration factor along the line of sight, $\int\dd\chi \, g$, is replaced by
a factor $\hat{F}_{\rm I}$. Thus, Eq.(\ref{xi-d1-g2}) becomes
\beqa
\lag\delta_1\hat{\gamma}_2^{\rm I}\rag_{\alpha} & = & \hat{F}_{\rm I}(z_2) \! 
\int_0^{2d} \frac{\dd r \; \xi(x_{1,2\parallel},r)}{\pi d \sin(\varphi)} - \hat{F}_{\rm I}(z_2)
\! \int_0^d \frac{\dd r}{d} \nonumber \\
&& \times \int_{d-r}^{d+r} \frac{\dd r' \; 2 \, \xi(x_{1,2\parallel},r')}{\pi d \sin(\varphi')}  ,
\label{xi-d1-gI2}
\eeqa
while Eq.(\ref{xi-g1-g2}) becomes
\beqa
\lag\hat{\gamma}_1^{\rm I}\gamma_2\rag_{\alpha} \!\! & = & \!\! - \hat{F}_{\rm I}(z_1) 
g_{1,2} \; \biggl\lbrace \int_0^{2d} \frac{\dd r \, r \, \xi^{\rm 2D}(r)}{\pi d^2} 
\left[ 2 \varphi - \frac{d}{r \sin\varphi} \right] \nonumber \\
&& \hspace{-0.7cm} - \int_0^d \frac{\dd r \, 2r}{d^2} \left( \! \xi^{\rm 2D}(r) 
\left[ \frac{d\!-\!r}{d}\right]^2 \!+\! \int_{d-r}^{d+r} \frac{\dd r' \, r' \, \xi^{\rm 2D}(r')}{\pi d^2} 
\right. \nonumber \\
&& \hspace{-0.7cm} \left. \times \left[ 2 \varphi' - \frac{d^2}{r \, r' \sin\varphi'} \right] 
\right) \biggl \rbrace .
\label{xi-gI1-g2}
\eeqa
From the property (\ref{Map-U}), the third term writes as
\beqa
\lag\delta_1\delta_2\hat{\gamma}_3^{\rm I}\rag_{\alpha} & = & \hat{F}_{\rm I}(z_3) 
\int_0^{2\pi} \frac{\dd\alpha_{\vx_1}\dd\alpha_{\vx_2}\dd\alpha_{\vx_3}}{(2\pi)^3} 
\biggl \lbrace \zeta(\vx_1,\vx_2,\vx_3)  \nonumber \\
&&  - \int_0^d \frac{\dd r \, 2r}{d^2} \zeta(\vx_1,\vx_2;x_{3\parallel},r) \biggl \rbrace .
\eeqa

We show the amplitude of this coupled contribution $\zeta_{\rm circ.}^{\delta \rm I}$
in Figs.~\ref{fig_zeta_slc_I_gamma_z_z_z} and \ref{fig_zeta_slc_I_theta}.
The comparison of Fig.~\ref{fig_zeta_slc_I_gamma_z_z_z} with
Fig.~\ref{fig_zeta_intrinsic_gamma_z_z_z} shows that when two source redshifts
are identical source-lens clustering effects amplify the intrinsic-alignment bias.
This yields a bias of order unity for a large range of source redshifts and angular
scales. However, when the three source redshifts are different (the single case
$\{z_1,z_2,z_3\}=\{0.5,1,2\}$ in these two figures), the coupled
source-lens clustering -- intrinsic alignment effects reduce to the contribution
(\ref{zeta-d-I-diff}) and only give rise to a bias of about $1\%$.
This is not surprising, because for this redshift triplet we can see in
Figs.~\ref{fig_zeta_gamma_z_z_z} and \ref{fig_zeta_intrinsic_gamma_z_z_z}
that the pure source-lens clustering and intrinsic alignment effects are of order
$10\%$, and we could expect their combination to be approximately $1\%$.

Figure~\ref{fig_zeta_slc_I_theta} confirms that, except for the coincident source
redshifts that give rise to high peaks for this bias, the coupled source-lens
clustering -- intrinsic alignment bias is usually very small. As the other sources
of bias, its relative amplitude with respect to the signal increases for low source
redshifts, and it becomes of order unity at $z_i \la 0.1$, but the usual
intrinsic alignment bias has already grown further in such cases.
Thus, for most practical purposes, the coupled contribution
$\zeta_{\rm circ.}^{\delta \rm I}$ is negligible or subdominant.

Eqs.(\ref{zeta-I-diff}) and (\ref{zeta-d-I-diff}) show that when the three source
redshifts are different, both the pure intrinsic-alignment and the coupled source-lens
clustering -- intrinsic alignment contributions involve the lowest-redshift
intrinsic alignment $\hat{\gamma}_1$. However, the first bias involves the
three-point correlation 
$\lag \hat{\gamma}_1^{\rm I} \gamma_2 \gamma_3 \rag_{\alpha}$ whereas the
second one involves the two-point correlation 
$\lag\hat{\gamma}_1^{\rm I}\gamma_2\rag_{\alpha}$.
This means that to remove the intrinsic-alignment bias through a nulling
technique \citep{Shi2010} one can integrate
with a suitable weight $T_2(\chi_2)$ or $T_3(\chi_3)$ over the redshift distribution
of either the intermediate or the farthest source plane, whereas the coupled bias can
only be removed by the integration with the weight $T_2(\chi_2)$ over the
intermediate source plane [or with a new weight $T'_3(\chi_3)$ that removes the
contributions from the plane $z_2$].
Therefore, to remove both contributions with a single weight, one must 
integrate over the intermediate source plane.

\section{Conclusion}
\label{Conclusion}

In this paper, we have estimated the source-lens clustering bias of two-point
and three-point weak lensing estimators. This arises from the fact that 
galaxies are not located at random in space: their distribution is correlated with
the density fluctuations that give rise to the weak lensing signal we wish to
measure.
Our approach does not rely on perturbation theory, which only applies to very
large angular scales, or on the measure of a ``convergence map'' from
galaxy surveys. Following the strategy used in practice in weak lensing surveys,
we consider estimators of the real-space two-point and three-point weak lensing
correlation functions, obtained by summing over pairs or triplets of source
galaxies over the survey area. This leads to a (typically positive) source-lens
clustering bias that is different from the (typically negative) source-lens clustering
bias associated with one-point estimators of the $\kappa$-map.
We study both the case of convergence
correlations (which are not observed in current surveys but enable simpler
computations) and shear correlations (which correspond to the current
observational probes but lead to somewhat heavier computations).

For two-point weak lensing correlation functions,
the source-lens clustering bias is typically several orders of magnitude below
the weak lensing signal, because 

(a) it only arises from density fluctuations close to at least one of the source galaxies, 
whereas the lensing signal is generated by density fluctuations along the lines
of sight up to the lowest galaxy redshift.

(b) the lensing efficiency kernel $g(\chi',\chi)$ vanishes at the source plane,
which yields additional suppression factors $x_0/(c/H_0)$, where
$x_0$ is the typical correlation length.
Moreover, the bias is smaller for the shear than for the convergence because
of some cancellations due to the spin-2 factor $e^{2\ii\alpha}$.

(c) it is quadratic over the density correlation ($\sim \xi^2+\xi^3$) whereas the
signal is linear ($\sim \xi$).

The only case where the bias is significant, and can even be larger than the signal,
is when we cross-correlate the gravitational lensing distortions of
a very low-redshift galaxy, $z_2 \la 0.05$, with a higher redshift galaxy ($z_1 \ga 0.5$),
as in tomographic studies where we bin the galaxy population in several redshift
bins. Then, the bias is set by the three-point correlation between the
low-redshift galaxy, at $z_2$, and density fluctuations on the two lines of sight
at nearby redshifts, $z_{2'} \simeq z_{1'} \simeq z_2$.
This increase in the bias, with respect to other galaxy configurations, is due to the
facts that one lensing kernel, $g_{1',1}$, is no longer small (because we are far from
the distant galaxy redshift), three-point correlations are larger at low redshift,
and the lensing signal itself is smaller, as the common depth of the two lines of sight,
up to $\min(z_1,z_2)$, is smaller.

In principles, one may subtract the main contamination
$\zeta^{\delta\gamma\gamma^*}$
from the estimator $\hat\xi^{\gamma\gamma^*}$ to measure the weak lensing
signal from Eq.(\ref{split-gamma}).
However, in cases where this bias is relevant, it involves the three-point
galaxy-matter-matter correlation function, which is not predicted to much better
than $10\%$ (we must take the inaccuracies of the matter three-point
correlation and of the galaxy bias into account).
Therefore, in practical analysis of cosmic shear surveys, it is probably more
convenient not to include the cross-correlation between low redshifts ($z_2 \la 0.05$)
and higher redshift galaxies ($z_1 \ga 0.5$).
This should not remove much of the weak lensing signal, which usually comes
from galaxies at $z \sim 1$, where the galaxy number density of the survey
is greatest. A simple implementation would be to discard any pair $\{i,j\}$
in the estimator (\ref{hxi-1}) whenever $\zeta^{\delta\gamma\gamma^*}$ is larger
than ten percent of $\xi^{\gamma\gamma^*}$.

For three-point weak lensing correlation functions, the source-lens clustering
bias is typically of order $10\%$ of the signal, both for the convergence and
the shear.
In contrast to the case of two-point estimators, as soon as the three source
galaxy redshifts are not identical the bias is not suppressed by a factor
$x_0/(c/H_0)$ associated with the vanishing of the lensing kernel $g(\chi',\chi)$
at the source plane. Moreover, the bias is not dominated by three-point
galaxy-matter correlation functions but by two-point correlations and on large
scales the signal and the bias show the same scaling $\propto \xi^2$ over
the density correlation.
These features explain the greater significance of the source-lens clustering
bias for three-point estimators than for two-point estimators.

Next, we have considered the intrinsic-alignment bias of these weak lensing
estimators. Using a simple intrinsic-alignment model, which assumes a linear
relationship between the galaxy and halo ellipticities, that has been used
in previous works for two-point statistics, we have extended this approach to
three-point statistics.
We find that the intrinsic-alignment bias is about $10\%$ of the signal for
estimators of either the two-point or three-point shear correlations,
but there is a significant dependence on scale and redshift.
In particular, the intrinsic-alignment bias becomes greater than unity for
low source redshifts $z_s \la 0.3$. As noticed in \citet{Semboloni2008},
low-redshift galaxies could be used to understand intrinsic alignments while
high-redshift galaxies would be used for measures of the cosmic shear,
but this requires reliable redshifts.

We have also investigated the coupling of source-lens clustering and
intrinsic-alignment effects. They are negligible for two-point estimators.
For three-point estimators, they are negligible or smaller than the usual
intrinsic-alignment bias, provided the three source redshifts are
different (otherwise they can be larger than the usual intrinsic-alignment bias).

Thus, for two-point statistics one can neglect the source-lens clustering bias
and focus on the intrinsic-alignment bias. Moreover, in this case
nulling techniques simultaneously remove the source-lens clustering and
intrinsic-alignment biases. 
For three-point statistics, such biases of $10\%$ are sufficiently small to be
neglected in current surveys.
However, it may be necessary to take them into account in future surveys
such as Euclid \citep{Refregier2010}.

One approach is to explicitly include these effects in the analysis by
building a model for these two biases (which may contain a few free
nuisance parameters that are estimated from the same observations).
For the source-lens clustering bias, 
a convenient feature is that it should be possible to obtain reasonably
good estimates of its amplitude because it is dominated by two-point
density-density and galaxy-density correlations (rather than three- or
higher-order correlations). However, it shows a linear dependence on the
galaxy bias [in the generic case where the three galaxy redshifts are
different, as in Eq.(\ref{hzeta-gamma-3})], within a linear galaxy bias
model. This may limit the accuracy of predictions for this source-lens
clustering bias to about $10\%$ (because of the measure of the galaxy
bias and the limited validity of the linear bias approximation), but in practice
this bias parameter may be treated as a free parameter to be constrained
by the same observations.
The intrinsic-alignment bias involves the three-point density correlation
(for a linear model), which is more difficult to model with a high accuracy in the
nonlinear regime. Moreover, the intrinsic alignment is much less well
understood than the galaxy number density bias. In particular, its dependence
on galaxy type and mass it not constrained to a very high accuracy.
There remains work to be done in this direction.

A second approach is to use a nulling technique, where one integrates over
the background source redshift distribution with a suitable weight $T(z_3)$ to remove
the dependence on density fluctuations at the redshift $z_1$ of the lowest-redshift 
galaxy.
However, in contrast with the case of two-point statistics, this no longer
simultaneously removes all contributions to the source-lens clustering bias,
because there remains a contribution associated with fluctuations at the redshift
of the intermediate galaxy $z_2$.
Therefore, one may need to impose a second constraint on $T(z_3)$ to remove
the dependence on both planes $z_1$ and $z_2$.
It remains to be seen whether this is really needed to lower the bias to a sufficiently
low level, and whether the loss of information entailed by such a procedure is
not to high. We leave this to future works.

\acknowledgement

We thank T. Nishimichi for sending us the skewness and kurtosis measured in
numerical simulations \citep{Nishimichi2011} that we used in
Fig.~\ref{fig_Sn} to check the model (\ref{Sn-def}).
We thank M. Kilbinger for discussions.
This work is supported in part by the French Agence Nationale de la Recherche under Grant ANR-12-BS05-0002.

\appendix

\section*{Appendix}

Throughout the paper and in the computations below, indices $i=1,2,3,$ refer
to the source galaxies $i$ at 3D positions $\vx_i$. These positions can also be
denoted as $(x_{i\parallel},\vx_{i\perp})$, where we use the flat sky approximation,
and $x_{\parallel}$ and $\vx_{\perp}$ are the longitudinal and transverse coordinates,
or as $(\chi_i,\vx_{i\perp})$, where $\chi_i$ is the radial coordinate.
Coordinates with a prime, such as $\vx_{i'}$, refer to the points along the line of sight
to the galaxy $i$, over which we integrate.
In a similar fashion, wavenumbers $\vk_i$ and $\vk_{i'}$ are associated with
Fourier transforms at points $\vx_i$ or $\vx_{i'}$.

Thus, cosmic shear signals as in Eqs.(\ref{xi-1}), (\ref{zeta-kappa-1}), and
(\ref{zetap-circ-1}), only involve correlations $\xi_{1',2'}$ ,$\zeta_{1',2',3'}$ or
bispectra $B(k_{1'},k_{2'},k_{3'})$, between points along the lines of sight.
In contrast, source-lens clustering and intrinsic-alignment contributions,
which involve correlations between a foreground galaxy $i$ and the nearby
density fluctuations $\vx_{i'}$ or $\vx_{j'}$ on the line of sight toward the same galaxy
or another background galaxy $j$, involve mixed correlations $\zeta_{1,1',2'}$,
$\xi_{3',1}$, or bispectra $B(k_1,k_2,k_{3'})$, in Eqs.(\ref{zeta-1}), (\ref{zeta-d1d2k3}),
or (\ref{zeta-circ-2.5D-1}).

In the computations below, the Dirac factor associated with Fourier space statistical
averages, as in $\lag \tdelta(\vk_1)\tdelta(\vk_2)\rag = \delta_D(\vk_1+\vk_2) P(k_1)$,
is usually written in its exponential form as 
$\int \dd\vr \, e^{\ii\vr\cdot(\vk_1+\vk_2)} /(2\pi)^3$,
as in Eq.(\ref{zeta1-1}) obtained from Eq.(\ref{zeta-P-gamma-1}).
Coordinates $\vr_i$ or $\vr_{i\perp}$ also appear when we express power spectra
in terms of the real-space correlations, as in Eq.(\ref{Pkperp-2}).
Thus, $\vr$ is an auxiliary 3D configuration-space coordinate, while $\vr_{\perp}$ is
its transverse component, and indices or primes as in $r_i'$ are only used to
distinguish between several dummy variables in multiple integrals,
without reference to the distinction between source locations
and line-of-sight points or to a particular galaxy.

\section{Two-point cosmic shear correlation }
\label{Density-shear}

\subsection{Density-shear two-point correlation}
\label{density-shear-2pt}

The product of two-point correlations in Eq.(\ref{xi-deltagamma-1}) reads as
\beq
\lag \delta_1 \gamma^*_2\rag \lag \delta_2 \gamma_1\rag = 
\int \dd\chi_{1'}  \dd\chi_{2'} \, g_{1',1} g_{2',2} \; 
\xi^{\delta\gamma^*\delta\gamma}_{1,2';2,1'} 
\eeq
where we introduced
\beqa
\xi^{\delta\gamma^*\delta\gamma}_{1,2';2,1'} & = & \int\dd\vk_1 \dd\vk_2 \; P(k_1) P(k_2)
\; e^{\ii \vk_1\cdot(\vx_1-\vx_{2'})+\ii \vk_2\cdot(\vx_2-\vx_{1'})} \nonumber \\
&& \times \; e^{2\ii (\alpha_{\vk_2}-\alpha_{\vk_1})} .
\label{xi-dgdg-1}
\eeqa
This involves the spin-2 correlation defined in Eq.(\ref{2_xi-def}).
Writing the power spectrum in terms of the two-point correlation function, integrating
over the longitudinal direction and the polar angles of $\vr_{\!\perp}$ and
$\vk_{\perp}$, Eq.(\ref{2_xi-def}) also reads as
\beqa
\xi^{(2)}(\vx) & = & \int \! \dd\vr_{\!\perp}  \xi(x_{\parallel},r_{\!\perp}) 
\int\! \frac{\dd\vk_{\perp}}{(2\pi)^2} \, e^{\ii\vk_{\perp}\cdot(\vx_{\perp}-\vr_{\!\perp})} 
\,  e^{2\ii (\alpha_{\vk}-\alpha_{\vx})} \nonumber \\
& = & - \int_0^{\infty} \!\!\! \dd r_{\!\perp} r_{\!\perp} \xi(x_{\parallel}, r_{\!\perp}) 
\int_0^{\infty} \!\!\! \dd k_{\perp} k_{\perp} J_0(k_{\perp}r_{\!\perp}) \nonumber \\
&& \times \; J_2(k_{\perp}x_{\perp}) .
\eeqa
[Here we note $\xi(x_{\parallel},x_{\perp}) = 
\xi\left(\sqrt{x_{\parallel}^2+x_{\perp}^2}\right)$.]
The integral over two Bessel functions satisfies
\beqa
\lefteqn{ \int_0^{\infty} \dd k \; k \; J_{n-1}(k x_1) J_{n+1}(k x_2) }  \label{Jn-1Jn+1} \\
&& = \frac{2n \, x_1^{n-1}}{x_2^{n+1}} \; \Theta(x_1<x_2) - \frac{1}{x_1} 
\delta_D(x_1-x_2)  \hspace{0.4cm} \mbox{if} \;\; n \geq 0 , \nonumber \\
&& = \frac{2 |n| \, x_2^{|n|-1}}{x_1^{|n|+1}} \; \Theta(x_2<x_1) - \frac{1}{x_1} 
\delta_D(x_1-x_2)  \hspace{0.4cm} \mbox{if} \;\; n \leq -1 , \nonumber
\eeqa
where $\Theta$ is the Heaviside function.
This yields Eq.(\ref{2_xi-1}) and Eq.(\ref{xi-dgdg-1}) writes as
\beq
\xi^{\delta\gamma^*\delta\gamma}_{1,2';2,1'} = \xi^{(2)}(\vx_{2',1}) \, \xi^{(2)}(\vx_{1',2}) \,
e^{2\ii (\alpha_{\vx_{1',2}}-\alpha_{\vx_{2',1}})} ,
\label{xi-dgdg-2}
\eeq
where we note $\vx_{1,2}=\vx_2-\vx_1$. Since we have
$\alpha_{\vx_{1',2}}= \alpha_{\vx_{1,2}}$ and
$\alpha_{\vx_{2',1}}= \alpha_{\vx_{2,1}}=\alpha_{\vx_{1,2}}+\pi$, Eq.(\ref{xi-dgdg-2})
gives Eq.(\ref{xi-deltagamma-2}).

\subsection{Density-shear-shear three-point correlation}
\label{density-shear-shear-3pt}

Substituting the expression (\ref{bispectrum-def}) of the bispectrum into
Eq.(\ref{zeta-gamma-2}), we are led to compute the quantity
\beqa
 \zeta_{1,1',2'}^{(1',2')} \!+\! \zeta_{1,1',2'}^{(1,1')} \!+\! \zeta_{1,1',2'}^{(1,2')} 
 & \! = \! & \int \!\! \dd\vk_1 \dd\vk_{1'}\dd\vk_{2'} \, 
 \delta_D(\vk_1 \!+\! \vk_{1'} \!+\! \vk_{2'} \!)  \nonumber \\
&& \hspace{-3.2cm} \times \, e^{\ii (\vk_1\cdot\vx_1+\vk_{1'}\cdot\vx_{1'}+\vk_{2'}\cdot\vx_{2'})} 
\, e^{2\ii (\alpha_{\vk_{1'}}-\alpha_{\vk_{2'}})} \nonumber \\
&& \hspace{-3.2cm} \times \, [ P(k_{1'}) P(k_{2'}) + P(k_1) P(k_{1'}) + P(k_1) P(k_{2'}) ] ,
\label{zeta-P-gamma-1}
\eeqa
where the superscripts in the three terms in the left hand side refer to the
arguments of the power spectra in the three terms in the bracket in the right hand
side. Here and in the following, we use the flat sky approximation and the
fact that the three points, $1$, $1'$, and $2'$, are at the same redshift (i.e., within radial
distances of order $\sim 8 h^{-1}$Mpc).
By symmetry, the first term vanishes,
\beqa
\zeta_{1,1',2'}^{(1',2')} & = & \int \!\! \dd\vk_{1'}\dd\vk_{2'} \,
e^{\ii \vk_{1'}\cdot(\vx_{1'}-\vx_1)+\ii\vk_{2'}\cdot(\vx_{2'}-\vx_1)} 
\, e^{2\ii (\alpha_{\vk_{1'}}-\alpha_{\vk_{2'}})} \nonumber \\
&& \hspace{0cm} \times \, P(k_{1'}) P(k_{2'})  \nonumber \\
& = & 0 .
\label{zeta3-1}
\eeqa
Indeed, $\vx_1$ and $\vx_{1'}$ are along the same line of sight, hence their projected
separation in the transverse plane is zero, $\vx_{1'\!\perp}-\vx_{1\perp}=0$,
and the angular integration over the polar angle $\alpha_{\vk_{1'}}$ vanishes.

Using the exponential representation of the Dirac factor, the second term reads
as
\beqa
\zeta_{1,1',2'}^{(1,1')} & = & \int \!\! \frac{\dd\vr}{(2\pi)^3} 
\int \!\! \dd\vk_1 \dd\vk_{1'}\dd\vk_{2'}
\, P(k_1) P(k_{1'}) \, e^{2\ii (\alpha_{\vk_{1'}}-\alpha_{\vk_{2'}})} \nonumber \\
&& \times \,
e^{\ii\vk_1\cdot(\vr+\vx_1)+\ii\vk_{1'}\cdot(\vr+\vx_{1'})+\ii\vk_{2'}\cdot(\vr+\vx_{2'})} .
\label{zeta1-1}
\eeqa
Using Eqs.(\ref{P-xi-def}) and (\ref{2_xi-def}) in Eq.(\ref{zeta1-1}) gives
\beqa
\zeta_{1,1',2'}^{(1,1')} & \! = \! & \int \!\! \frac{\dd\vr_{\!\perp}}{(2\pi)^2} \, 
\xi(x_{2',1\parallel},|\vr_{\!\perp}\!\!+\!\vx_{1\perp}|) \, 
\xi^{(2)}(x_{2',1'\parallel},|\vr_{\!\perp}\!\!+\!\vx_{1'\!\perp}|) \nonumber \\
&& \times \int \!\! \dd\vk_{2'\!\perp} \, e^{\ii \vk_{2'\!\perp}\cdot(\vr_{\!\perp}+\vx_{2'\!\perp})} \, 
e^{2\ii (\alpha_{\vr_{\!\perp}+\vx_{1'\!\perp}}-\alpha_{\vk_{2'\!\perp}})} .
\label{zeta1-2}
\eeqa
Using the fact that $\vx_{1'\!\perp}=\vx_{1\perp}$, we make the change of variables
$\vr'_{\perp}=\vr_{\!\perp}+\vx_{1\perp}$. Then, using the Jacobi-Anger expansion,
\beq
e^{\ii\vk_{\perp}\cdot\vx_{\perp}} = \sum_{n=-\infty}^{\infty} \ii^n \; 
J_n(k_{\perp} x_{\perp}) \; e^{\ii n (\alpha_{\vx_{\perp}}-\alpha_{\vk_{\perp}})} ,
\label{Jacobi}
\eeq
and Eq.(\ref{Jn-1Jn+1}), we can perform the integration over $\vk_{2'\!\perp}$,
which gives
\beqa
\zeta_{1,1',2'}^{(1,1')} & = & \xi_{2',1} \; \xi^{(2)}_{2',1'} - 2 \int_{x_{2,1\perp}}^{\infty} 
\frac{\dd r_{\!\perp}}{r_{\!\perp}} \; \xi(x_{2',1\parallel},r_{\!\perp}) \nonumber \\
&& \times \; \xi^{(2)}(x_{2',1'\parallel},r_{\!\perp}) .
\label{zeta1-3}
\eeqa
As compared with the factor $\xi_{2',1} \xi_{2',1'}$ that arises for the convergence,
as in Eq.(\ref{zeta-1}), the source-lens clustering bias of the cosmic shear is suppressed
by the spin-2 factor $e^{2\ii \alpha}$. It replaces one correlation $\xi$ by a 
correlation $\xi^{(2)}$, which is smaller because of the subtraction in
Eq.(\ref{2_xi-1}), and it yields a second subtraction in Eq.(\ref{zeta1-3}).

In a similar fashion, the third term of Eq.(\ref{zeta-P-gamma-1}) also reads as
\beqa
\zeta_{1,1',2'}^{(1,2')} & = & \int \!\! \frac{\dd\vr_{\!\perp}}{(2\pi)^2} \int \!\! \dd\vk_{1'\!\perp} 
\int \!\! \dd\vk_1  \, e^{\ii k_{1\parallel}(x_{1\parallel}-x_{1'\parallel})
+\ii\vk_{1\perp}\cdot(\vr_{\!\perp}+\vx_{1\perp})} \nonumber \\
&& \times \, e^{\ii \vk_{1'\!\perp}\cdot(\vr_{\!\perp}+\vx_{1'\!\perp})} \, 
e^{2\ii (\alpha_{\vk_{1'\!\perp}}-\alpha_{\vr_{\!\perp}+\vx_{2'\!\perp}})} \; P(k_1) \nonumber \\
&& \times \, \xi^{(2)}(x_{1',2'\parallel},|\vr_{\!\perp} \!+\! \vx_{2'\!\perp}|)  .
\label{zeta2-1}
\eeqa
Then, making the change of variable $\vr'_{\perp}= \vr_{\!\perp}+\vx_{2'\!\perp}$
and using the expansion (\ref{Jacobi}) we can integrate over angles.
Next, using the property (\ref{Jn-1Jn+1}) and the summation rule
$\sum_{n=-\infty}^{\infty} J_n(x)^2 = 1$, we obtain
\beqa
\zeta_{1,1',2'}^{(1,2')} & \! = \! & -2\pi \!\! \int_0^{\infty} \!\!\!\! \dd r_{\!\perp} r_{\!\perp} 
\int_{-\infty}^{\infty} \!\!\! \dd k_{1\parallel} \int_0^{\infty} \!\!\! \dd k_{1\perp} k_{1\perp}  \, 
e^{\ii k_{1\parallel}x_{1'\!,1\parallel}} P(k_1) \nonumber \\
&&  \hspace{-1.2cm} \times \, \xi^{(2)}(x_{1'\!,2'\parallel},r_{\!\perp}) \;
\biggl \lbrace  - \frac{1}{r_{\!\perp}} \delta_D(r_{\!\perp}-x_{1,2\perp}) + 
\Theta(x_{1,2\perp} \!\!<\! r_{\!\perp})  \nonumber \\
&&  \hspace{-1.2cm} \times \! \sum_{n=1}^{\infty} 2 n 
\frac{x_{1,2\perp}^{n-1}}{r_{\!\perp}^{n+1}} J_{n-1}(k_{1\perp}r_{\!\perp}) 
J_{n-1}(k_{1\perp}x_{1,2\perp}) + \Theta(r_{\!\perp} \!\!<\! x_{1,2\perp}) \nonumber \\
&&  \hspace{-1.2cm} \times \! \sum_{n=1}^{\infty} 2 n 
\frac{r_{\!\perp}^{n-1}}{x_{1,2\perp}^{n+1}} J_{n+1}(k_{1\perp}r_{\!\perp}) 
J_{n+1}(k_{1\perp}x_{1,2\perp}) \biggl \rbrace .
\label{zeta2-2}
\eeqa
Then, expressing $P(k_1)$ in terms of the two-point correlation function,
as in
\beqa
\int_{-\infty}^{\infty} \dd k_{\parallel} \; e^{\ii k_{\parallel} x_{\parallel}} P(k) & = & 
\int_{-\infty}^{\infty} \dd k_{\parallel} \int \frac{\dd\vr}{(2\pi)^3} \; 
e^{\ii k_{\parallel} x_{\parallel} -\ii \vk\cdot\vr} \xi(r)  \nonumber \\
&& \hspace{-1.2cm}= \int_0^{\infty} \frac{\dd r_{\!\perp} \; r_{\!\perp}}{2\pi} \; 
\xi(x_{\parallel},r_{\!\perp}) \, J_0(k_{\perp} r_{\!\perp}) ,
\label{Pkperp-1}
\eeqa
and using the property
\beqa
\lefteqn{\int_0^{\infty} \dd k \, k J_0(a k) J_n(b k) J_n(c k) = 
\Theta(|b-c|<a<b+c) \; \frac{\cos(n\varphi)}{\pi b c \sin(\varphi)}} 
\nonumber \\
&& \mbox{with} \;\;\;   \varphi = \mbox{Arccos}\left[ \frac{b^2+c^2-a^2}{2bc} \right] ,
\label{J0JnJn}
\eeqa
where $a>0,b>0,c>0$, $n$ is integer, and $\Theta$ is a unit top-hat with obvious
notations (i.e., unity when the conditions are satisfied and zero otherwise), we can 
integrate over $\vk_1$ and we obtain
\beqa
\zeta_{1,1',2'}^{(1,2')} & = & \xi_{1',1} \, \xi^{(2)}_{1',2'} - \int_0^{\infty} 
\frac{\dd r_{\!\perp}}{r_{\!\perp}} 
\xi(x_{1',1\parallel},r_{\!\perp}) \int_{|r_{\!\perp}-x_{1,2\perp}|}^{r_{\!\perp}+x_{1,2\perp}}
\frac{\dd r'_{\perp}}{r'_{\perp}} \nonumber \\
&& \times \; \xi^{(2)}(x_{1',2'\parallel},r'_{\perp}) 
\frac{2}{\pi r_{\!\perp}^2 \sqrt{r_{\!\perp}^{'2}-(r_{\!\perp}-x_{1,2\perp})^2}}
\nonumber \\
&& \times \; 
\frac{(x_{1,2\perp}^2-r_{\!\perp}^2-r_{\!\perp}^{'2})^2-2 r_{\!\perp}^2 r_{\!\perp}^{'2}}
{\sqrt{(r_{\!\perp}+x_{1,2\perp})^2-r_{\!\perp}^{'2}}}  .
\label{zeta2-3}
\eeqa
Again, as compared with the factor $\xi_{1',1} \, \xi_{1',2'}$ that arises for the
convergence, the spin-2 factor $e^{2\ii\alpha}$ suppresses the source-lens clustering
bias by replacing a factor $\xi$ by $\xi^{(2)}$ and introducing another subtraction.

\section{Three-point cosmic shear correlation }
\label{shear-3pt-app}

Using the exponential representation of the Dirac distribution, Eq.(\ref{zetap-circ-B})
also writes as
\beqa
\zeta_{\rm circ.}^{\rm 2D} & \! = \! & \int_{-\infty}^{\infty} \dd x_{2'\parallel} \dd x_{3'\parallel} 
\int_0^{2\pi} \! \frac{\dd\alpha_{\vx_1}\dd\alpha_{\vx_2}\dd\alpha_{\vx_3}}{(2\pi)^3}
\int \frac{\dd\vr}{(2\pi)^3} \nonumber \\
&& \hspace{-0.8cm} \times \int \! \dd\vk_{1'}\dd\vk_{2'}\dd\vk_{3'} \, 
e^{\ii [ \vk_{1'}\cdot(\vx_{1'}+\vr)+\vk_{2'}\cdot(\vx_{2'}+\vr)+\vk_{3'}\cdot(\vx_{3'}+\vr)]}
\nonumber \\
&& \hspace{-0.8cm} \times \; B(k_{1'},k_{2'},k_{3'}) \; 
e^{2\ii (\alpha_{\vk_{1'}}+\alpha_{\vk_{2'}}+\alpha_{\vk_{3'}}
- \alpha_{\vx_1} - \alpha_{\vx_2} - \alpha_{\vx_3})} .
\label{zetap-circ-1}
\eeqa
Integrating one after the other over the longitudinal components,
$\{x_{2'\parallel},x_{3'\parallel}\}$, $\{k_{2'\parallel},k_{3'\parallel}\}$,
$\{r_{\parallel},k_{1'\parallel}\}$, the angles 
$\{\alpha_{\vx_1},\alpha_{\vx_2},\alpha_{\vx_3}\}$,
$\{\alpha_{\vk_{1'}},\alpha_{\vk_{2'}},\alpha_{\vk_{3'}}\}$,
and $\alpha_{\vr}$, we obtain
\beqa
\zeta_{\rm circ.}^{\rm 2D} & \! = \! & - (2\pi)^4 \int_0^{\infty} \dd r_{\!\perp} \dd k_{1'\!\perp} 
\dd k_{2'\!\perp} \dd k_{3'\!\perp} \; r_{\!\perp} k_{1'\!\perp} k_{2'\!\perp} k_{3'\!\perp} 
\nonumber \\
&& \times \; B(k_{1'\!\perp},k_{2'\!\perp},k_{3'\!\perp}) \; J_2(k_{1'\!\perp} d) 
J_2(k_{2'\!\perp} d) J_2(k_{3'\!\perp} d) \nonumber \\
&& \times \; J_0(k_{1'\!\perp} r_{\!\perp}) J_0(k_{2'\!\perp} r_{\!\perp})
J_0(k_{3'\!\perp} r_{\!\perp}) ,
\label{zetap-circ-2}
\eeqa
where $d=\chi_{1'} \theta$ is the radius of the circumcircle at radial distance
$\chi_{1'}$. Using the ansatz (\ref{bispectrum-def}), this reads as
\beqa
\zeta_{\rm circ.}^{\rm 2D} & \! = \! & - S_3 \int_0^{\infty} \dd r_{\!\perp} 
\dd r_{1\perp} \dd r_{2\perp} \dd k_{1'\!\perp} \dd k_{2'\!\perp} 
\dd k_{3'\!\perp} \; r_{\!\perp} r_{1\perp} r_{2\perp} \nonumber \\
&& \times \; k_{1'\!\perp} k_{2'\!\perp} k_{3'\!\perp} \; \xi^{\rm 2D}(r_{1\perp}) \;
\xi^{\rm 2D}(r_{2\perp}) \; J_0(k_{1'\!\perp} r_{1\perp}) \nonumber \\
&& \times \; J_0(k_{2'\!\perp} r_{2\perp}) \; J_2(k_{1'\!\perp} d) J_2(k_{2'\!\perp} d) 
J_2(k_{3'\!\perp} d) \nonumber \\
&& \times \; J_0(k_{1'\!\perp} r_{\!\perp}) J_0(k_{2'\!\perp} r_{\!\perp})
J_0(k_{3'\!\perp} r_{\!\perp}) ,
\label{zetap-circ-3}
\eeqa
where we used Eq.(\ref{Pkperp-1}) to write
\beqa
P(k_{\perp}) & = & \int_{-\infty}^{\infty} \frac{\dd x_{\parallel}}{2\pi} 
\int_{-\infty}^{\infty} \dd k_{\parallel} \; e^{\ii k_{\parallel} x_{\parallel}} P(k)
\nonumber \\
& = & \int_0^{\infty} \frac{\dd r_{\!\perp} \; r_{\!\perp}}{(2\pi)^2} \; 
\xi^{\rm 2D}(r_{\!\perp}) \, J_0(k_{\perp} r_{\!\perp}) .
\label{Pkperp-2}
\eeqa
Using the property (\ref{J0JnJn}) and
\beqa
\lefteqn{\int_0^{\infty} \dd k \, J_1(a k) J_0(b k) J_0(c k) = 
\Theta( |b \!-\! c| \!<\! a \!<\! b+c) \; \frac{\varphi}{\pi a} 
+ \frac{\Theta(a \!>\! b \!+\! c)}{a} } \nonumber \\
&& \mbox{with} \;\;\;   \varphi = \mbox{Arccos}\left[ \frac{b^2+c^2-a^2}{2bc} \right] ,
\label{J0J0J1}
\eeqa
with the relation $J_0(z)+J_2(z)=2 J_1(z)/z$, we obtain
\beqa
\int_0^{\infty} \dd k \, k J_2(a k) J_0(b k) J_0(c k) & = & 
\frac{2\Theta(a \!>\! b \!+\! c)}{a^2}  \nonumber \\
&& \hspace{-1cm} + \frac{\Theta( |b \!-\! c| \!<\! a \!<\! b+c)}{\pi a^2} 
\left( 2\varphi - \frac{a^2}{bc\sin\varphi}\right)  \nonumber \\
&& \hspace{-4cm} \mbox{with} \;\;\;   \varphi = \mbox{Arccos}\left[ \frac{b^2+c^2-a^2}{2bc} \right] .
\label{J2J0J0}
\eeqa
Then, using Eqs.(\ref{Jn-1Jn+1}) and (\ref{J2J0J0}), we can integrate 
Eq.(\ref{zetap-circ-3}) over wavenumbers, which yields
\beqa
\zeta_{\rm circ.}^{\rm 2D} & \!\! = \! & S_3 \left( \int_0^{2d} \frac{\dd r \, r \, \xi^{\rm 2D}(r)}
{\pi d^2} \left[ 2 \varphi - \frac{d}{r \sin(\varphi)} \right] \right)^{\!\!2} \!\!
- S_3 \! \int_0^d \! \frac{\dd r \, 2r}{d^2} \nonumber \\ 
&& \times \left( \int_{d-r}^{d+r} \frac{\dd r' \, r' \, \xi^{\rm 2D}(r')}{\pi d^2}
\left[ 2 \varphi' - \frac{d^2}{r \, r' \sin(\varphi')} \right] \right. \nonumber \\
&& \left. + \int_0^{d-r} \frac{\dd r' \; 2 r'}{d^2} \; \xi^{\rm 2D}(r') \right)^{\!\!2} ,
\label{zetap-circ-4}
\eeqa
where the angles $\varphi$ and $\varphi'$ are given by Eq.(\ref{phi-phip}).

\section{Three-point lensing-intrinsic shear correlations }
\label{intrinsic-shear-3pt-app}

To compute the lensing-intrinsic three-point correlations (\ref{zeta-2.5D--1})
and (\ref{zeta-3D--1}) we proceed as in App.~\ref{shear-3pt-app}.
Using the exponential representation of the Dirac distribution, Eq.(\ref{zeta-2.5D--1})
also writes as
\beqa
\zeta_{\rm circ.}^{\rm 2.5D} & \! = \! & \int_{-\infty}^{\infty} \dd x_{3'\parallel} 
\int_0^{2\pi} \! \frac{\dd\alpha_{\vx_1}\dd\alpha_{\vx_2}\dd\alpha_{\vx_3}}{(2\pi)^3}
\int \frac{\dd\vr}{(2\pi)^3} \nonumber \\
&& \hspace{-0.cm} \times \int \! \dd\vk_1\dd\vk_2\dd\vk_{3'} \, 
e^{\ii [ \vk_1\cdot(\vx_1+\vr)+\vk_2\cdot(\vx_2+\vr)+\vk_{3'}\cdot(\vx_{3'}+\vr)]}
\nonumber \\
&& \hspace{-0.cm} \times \; B(k_1,k_2,k_{3'}) \; 
e^{2\ii (\alpha_{\vk_1}+\alpha_{\vk_2}+\alpha_{\vk_{3'}}
- \alpha_{\vx_1} - \alpha_{\vx_2} - \alpha_{\vx_3})} .
\label{zeta-circ-2.5D-1}
\eeqa
Integrating one after the other over the longitudinal components,
$\{x_{3'\parallel},k_{3'\parallel}\}$, the angles 
$\{\alpha_{\vx_1},\alpha_{\vx_2},\alpha_{\vx_3}\}$,
$\{\alpha_{\vk_1},\alpha_{\vk_2},\alpha_{\vk_{3'}}\}$,
and $\alpha_{\vr}$, we obtain
\beqa
\zeta_{\rm circ.}^{\rm 2.5D} & \! = \! & - (2\pi)^2 \!\! \int_{-\infty}^{\infty} \!\! \dd r_{\parallel} 
\dd k_{1\parallel} \dd k_{2\parallel} \int_0^{\infty} \!\! \dd r_{\!\perp} \dd k_{1\perp} 
\dd k_{2\perp} \dd k_{3'\!\perp} \; r_{\!\perp} k_{1\perp} k_{2\perp} \nonumber \\
&& \hspace{-0.5cm} \times \; k_{3'\!\perp} \; 
e^{\ii[ k_{1\parallel}(x_{1\parallel}+r_{\parallel})
+k_{2\parallel}(x_{2\parallel}+r_{\parallel}) ]} \, B(k_1,k_2,k_{3'\!\perp})
J_2(k_{1\perp} d) \nonumber \\
&& \hspace{-0.5cm} \times \; J_2(k_{2\perp} d) J_2(k_{3'\!\perp} d)
\; J_0(k_{1\perp} r_{\!\perp}) J_0(k_{2\perp} r_{\!\perp}) J_0(k_{3'\!\perp} r_{\!\perp}) ,
\label{zetap-circ-2.5D-2}
\eeqa
where $d=(\chi_1+\chi_2)\theta/2$ is the radius of the circumcircle at radial distance
$(\chi_1+\chi_2)/2$ (this three-point correlation is only relevant
when the two redshifts $z_1$ and $z_2$ are very close).
Next, using the hierarchical ansatz (\ref{bispectrum-def}), we can split
$\zeta_{\rm circ.}^{\rm 2.5D}$ into three contributions. The first term, associated
with the product $P(k_1)P(k_2)$ in the bispectrum ansatz, writes as
\beqa
\zeta_{(12)}^{\rm 2.5D} & \! = \! & (2\pi)^2 \frac{S_3}{3} 
\!\! \int_{-\infty}^{\infty} \!\! \dd r_{\parallel} \dd k_{1\parallel} \dd k_{2\parallel} 
\int_0^{\infty} \!\! \dd r_{\!\perp} \dd k_{1\perp} 
\dd k_{2\perp} \; r_{\!\perp} k_{1\perp} k_{2\perp} \nonumber \\
&& \hspace{-0.5cm} \times \; \left[ \frac{\delta_D(r_{\perp}\!-\!d)}{d} - 
\frac{2\Theta(r_{\perp}\!<\!d)}{d^2} \right] 
e^{\ii[ k_{1\parallel}(x_{1\parallel}+r_{\parallel})
+k_{2\parallel}(x_{2\parallel}+r_{\parallel}) ]} \nonumber \\
&& \hspace{-0.5cm} \times \; J_2(k_{1\perp} d) J_2(k_{2\perp} d) 
\; J_0(k_{1\perp} r_{\!\perp}) J_0(k_{2\perp} r_{\!\perp}) P(k_1) P(k_2) ,
\label{zetap-circ-2.5D-12-1}
\eeqa
where we used Eq.(\ref{Jn-1Jn+1}) to integrate over $k_{3'\!\perp}$.
Next, writing the power spectra in terms of the two-point correlation functions,
we can integrate over $\{k_{1\parallel},k_{2\parallel}\}$. This yields
\beqa
\zeta_{(12)}^{\rm 2.5D} & \! = \! & \frac{S_3}{3} 
\!\! \int_{-\infty}^{\infty} \!\! \dd r_{\parallel}
\int_0^{\infty} \!\! \dd r_{\!\perp} \dd k_{1\perp} 
\dd k_{2\perp} \dd r_{1\perp} \dd r_{2\perp} 
\; r_{\!\perp} k_{1\perp} k_{2\perp} r_{1\perp} r_{2\perp} \nonumber \\
&& \hspace{-0.5cm} \times \; \left[ \frac{\delta_D(r_{\perp}\!-\!d)}{d} - 
\frac{2\Theta(r_{\perp}\!<\!d)}{d^2} \right] \,
\xi(x_{1\parallel} \!+\! r_{\parallel},r_{1\perp}) \,
\xi(x_{2\parallel} \!+\! r_{\parallel},r_{2\perp}) \nonumber \\
&& \hspace{-0.5cm} \times \; J_2(k_{1\perp} d) J_2(k_{2\perp} d) 
\; J_0(k_{1\perp} r_{\!\perp}) J_0(k_{2\perp} r_{\!\perp}) 
\; J_0(k_{1\perp} r_{1\perp})  \nonumber \\
&& \hspace{-0.5cm} \times \; J_0(k_{2\perp} r_{2\perp}) .
\label{zetap-circ-2.5D-12-2}
\eeqa
Using Eq.(\ref{J2J0J0}) we can integrate over $\{k_{1\perp},k_{2\perp}\}$, which
yields
\beqa
\zeta_{(12)}^{\rm 2.5D} & \! = \! & \frac{S_3}{3} \int_{-\infty}^{\infty} 
\dd r_{\parallel} \biggl\lbrace \int_0^{2d} \frac{\dd r_1 \, r_1}{\pi d^2}
\xi(x_{1\parallel} \!+\! r_{\parallel},r_1)
\left[ 2 \varphi_1 - \frac{d}{r_1 \sin\varphi_1} \right]  \nonumber \\
&& \hspace{-0.7cm} \times \; \int_0^{2d} \frac{\dd r_2 \, r_2}{\pi d^2} 
\xi(x_{2\parallel} \!+\! r_{\parallel},r_2)
\left[ 2 \varphi_2 - \frac{d}{r_2 \sin\varphi_2} \right] 
- \int_0^d \! \frac{\dd r_{\perp} \, 2r_{\perp}}{d^2} \nonumber \\
&& \hspace{-0.7cm} \times \; \left( \int_{d-r_{\perp}}^{d+r_{\perp}} 
\frac{\dd r_1' \, r_1'}{\pi d^2} \xi(x_{1\parallel} \!+\! r_{\parallel},r_1')
\left[ 2 \varphi_1' - \frac{d^2}{r_{\perp} \, r_1' \sin\varphi_1'} \right] \right. 
\nonumber \\
&& \hspace{-0.7cm} \left. + \int_0^{d-r_{\perp}} \frac{\dd r_1' \; 2 r_1'}{d^2} 
\, \xi(x_{1\parallel} \!+\! r_{\parallel},r_1') \right) 
\left( \int_{d-r_{\perp}}^{d+r_{\perp}} \frac{\dd r_2' \, r_2'}{\pi d^2}
\xi(x_{2\parallel} \!+\! r_{\parallel},r_2') \right. \nonumber \\
&& \hspace{-0.7cm} \left. \times 
\left[ 2 \varphi_2' - \frac{d^2}{r_{\perp} \, r_2' \sin\varphi_2'} \right]
\!+\! \int_0^{d-r_{\perp}} \! \frac{\dd r_2' \; 2 r_2'}{d^2} \, 
\xi(x_{2\parallel} \!+\! r_{\parallel},r_2') \! \right) \! \biggl \rbrace ,
\label{zetap-circ-2.5D-12-3}
\eeqa
where the angles $\varphi_i$ and $\varphi_i'$ are given by
\beq
\varphi_i = \mbox{Arccos} \left( \frac{r_i}{2d} \right) , \;\;\;
\varphi_i' = \mbox{Arccos} \left( \frac{r_{\perp}^2+r_i'^2-d^2}{2 r_{\perp} r_i'}
\right) ,
\label{phi_i-phi_ip}
\eeq
as in Eq.(\ref{phi-phip}).
The second term in Eq.(\ref{zetap-circ-2.5D-2}), associated with the product
$P(k_1)P(k_3)$, writes as
\beqa
\zeta_{(13)}^{\rm 2.5D} & \! = \! & (2\pi)^2 \frac{S_3}{3} 
\!\! \int_{-\infty}^{\infty} \!\! \dd r_{\parallel} \dd k_{1\parallel} \dd k_{2\parallel} 
\int_0^{\infty} \!\! \dd r_{\!\perp} \dd k_{1\perp} 
\dd k_{3'\!\perp} \; r_{\!\perp} k_{1\perp} k_{3'\!\perp} \nonumber \\
&& \hspace{-0.5cm} \times \; \left[ \frac{\delta_D(r_{\perp}\!-\!d)}{d} - 
\frac{2\Theta(r_{\perp}\!<\!d)}{d^2} \right] 
e^{\ii[ k_{1\parallel}(x_{1\parallel}+r_{\parallel})
+k_{2\parallel}(x_{2\parallel}+r_{\parallel}) ]} \nonumber \\
&& \hspace{-0.5cm} \times \; J_2(k_{1\perp} d) J_2(k_{3'\!\perp} d) 
\; J_0(k_{1\perp} r_{\!\perp}) J_0(k_{3'\!\perp} r_{\!\perp}) P(k_1) P(k_{3'\!\perp}) ,
\label{zetap-circ-2.5D-13-1}
\eeqa
where we used Eq.(\ref{Jn-1Jn+1}) to integrate over $k_{2\perp}$.
Next, writing the power spectra in terms of the two-point correlation functions,
with Eq.(\ref{Pkperp-2}) for $P(k_{3'\!\perp})$, we can integrate over
$\{k_{2\parallel},r_{\parallel},k_{1\parallel}\}$. This yields
\beqa
\zeta_{(13)}^{\rm 2.5D} & \! = \! & \frac{S_3}{3} 
\int_0^{\infty} \!\! \dd r_{\!\perp} \dd k_{1\perp} 
\dd k_{3'\!\perp} \dd r_{1\perp} \dd r_{3\perp} 
\; r_{\!\perp} k_{1\perp} k_{3'\!\perp} r_{1\perp} r_{3\perp} \nonumber \\
&& \hspace{-0.5cm} \times \; \left[ \frac{\delta_D(r_{\perp}\!-\!d)}{d} - 
\frac{2\Theta(r_{\perp}\!<\!d)}{d^2} \right] \,
\xi(x_{1\parallel} \!-\! x_{2\parallel},r_{1\perp}) \,
\xi^{\rm 2D}(r_{3\perp}) \nonumber \\
&& \hspace{-0.5cm} \times \; J_2(k_{1\perp} d) J_2(k_{3'\!\perp} d) 
\; J_0(k_{1\perp} r_{\!\perp}) J_0(k_{3'\!\perp} r_{\!\perp}) 
\; J_0(k_{1\perp} r_{1\perp})  \nonumber \\
&& \hspace{-0.5cm} \times \; J_0(k_{3'\!\perp} r_{3\perp}) .
\label{zetap-circ-2.5D-13-2}
\eeqa
Using Eq.(\ref{J2J0J0}) we can integrate over $\{k_{1\perp},k_{3'\!\perp}\}$, which
yields
\beqa
\zeta_{(13)}^{\rm 2.5D} & \! = \! & \frac{S_3}{3}
\int_0^{2d} \frac{\dd r_1 \, r_1}{\pi d^2}
\xi(x_{1\parallel} \!-\! x_{2\parallel},r_1)
\left[ 2 \varphi_1 - \frac{d}{r_1 \sin\varphi_1} \right]  \nonumber \\
&& \hspace{-0.7cm} \times \; \int_0^{2d} \frac{\dd r_3 \, r_3}{\pi d^2} 
\xi^{\rm 2D}(r_3) \left[ 2 \varphi_3 - \frac{d}{r_3 \sin\varphi_3} \right] 
- \frac{S_3}{3} \int_0^d \! \frac{\dd r_{\perp} \, 2r_{\perp}}{d^2} \nonumber \\
&& \hspace{-0.7cm} \times \; \left( \int_{d-r_{\perp}}^{d+r_{\perp}} 
\frac{\dd r_1' \, r_1'}{\pi d^2} \xi(x_{1\parallel} \!-\! x_{2\parallel},r_1')
\left[ 2 \varphi_1' - \frac{d^2}{r_{\perp} \, r_1' \sin\varphi_1'} \right] \right. 
\nonumber \\
&& \hspace{-0.7cm} \left. + \int_0^{d-r_{\perp}} \frac{\dd r_1' \; 2 r_1'}{d^2} 
\, \xi(x_{1\parallel} \!-\! x_{2\parallel},r_1') \right) 
\left( \int_{d-r_{\perp}}^{d+r_{\perp}} \frac{\dd r_3' \, r_3'}{\pi d^2}
\xi^{\rm 2D}(r_3') \right. \nonumber \\
&& \hspace{-0.7cm} \left. \times 
\left[ 2 \varphi_3' - \frac{d^2}{r_{\perp} \, r_3' \sin\varphi_3'} \right]
\!+\! \int_0^{d-r_{\perp}} \! \frac{\dd r_3' \; 2 r_3'}{d^2} \, 
\xi^{\rm 2D}(r_3') \! \right) ,
\label{zetap-circ-2.5D-13-3}
\eeqa
where the angles $\varphi_i$ and $\varphi_i'$ are given by Eq.(\ref{phi_i-phi_ip}).
The third contribution $\zeta_{(23)}^{\rm 2.5D}$ is obtained in the same manner,
and within our approximation $z_1 \simeq z_2$ we have
$\zeta_{(23)}^{\rm 2.5D} \simeq \zeta_{(13)}^{\rm 2.5D}$.

In a similar fashion, integrating over the angles
$\{\alpha_{\vx_1},\alpha_{\vx_2},\alpha_{\vx_3}\}$,
$\{\alpha_{\vk_1},\alpha_{\vk_2},\alpha_{\vk_3}\}$,
and $\alpha_{\vr}$, Eq.(\ref{zeta-3D--1}) writes as
\beqa
\zeta_{\rm circ.}^{\rm 3D} & \! = \! & - 2\pi \!\! \int_{-\infty}^{\infty} \!\! \dd r_{\parallel} 
\dd k_{1\parallel} \dd k_{2\parallel} \dd k_{3\parallel} \int_0^{\infty} \!\!\! \dd r_{\!\perp} 
\dd k_{1\perp} \dd k_{2\perp} \dd k_{3\perp} \; r_{\!\perp} k_{1\perp} k_{2\perp} \nonumber \\
&& \hspace{-0.5cm} \times \; k_{3\perp} \; 
e^{\ii[ k_{1\parallel}(x_{1\parallel}+r_{\parallel})
+k_{2\parallel}(x_{2\parallel}+r_{\parallel}) +k_{3\parallel}(x_{3\parallel}+r_{\parallel}) ]} 
\, B(k_1,k_2,k_3) \nonumber \\
&& \hspace{-0.5cm} \times \; J_2(k_{1\perp} d) J_2(k_{2\perp} d) J_2(k_{3\perp} d)
J_0(k_{1\perp} r_{\!\perp}) J_0(k_{2\perp} r_{\!\perp}) J_0(k_{3\perp} r_{\!\perp}) , 
\nonumber \\
&&
\label{zetap-circ-3D-2}
\eeqa
where $d=(\chi_1+\chi_2+\chi_3)\theta/3$ is the radius of the circumcircle at radial
distance $(\chi_1+\chi_2+\chi_3)/3$ (this three-point correlation is only relevant
when the three redshifts are very close).
Next, using again the hierarchical ansatz (\ref{bispectrum-def}), we can split
$\zeta_{\rm circ.}^{\rm 3D}$ into three contributions. The first term, associated
with the product $P(k_1)P(k_2)$ in the bispectrum ansatz, writes as
\beqa
\zeta_{(12)}^{\rm 3D} & \! = \! & (2\pi)^2 \frac{S_3}{3} 
\!\! \int_{-\infty}^{\infty} \!\! \dd k_{1\parallel} \dd k_{2\parallel} 
\int_0^{\infty} \!\! \dd r_{\!\perp} \dd k_{1\perp} 
\dd k_{2\perp} \; r_{\!\perp} k_{1\perp} k_{2\perp} \nonumber \\
&& \hspace{-0.5cm} \times \; \left[ \frac{\delta_D(r_{\perp}\!-\!d)}{d} - 
\frac{2\Theta(r_{\perp}\!<\!d)}{d^2} \right] 
e^{\ii[ k_{1\parallel}(x_{1\parallel}-x_{3\parallel})
+k_{2\parallel}(x_{2\parallel}-x_{3\parallel}) ]} \nonumber \\
&& \hspace{-0.5cm} \times \; J_2(k_{1\perp} d) J_2(k_{2\perp} d) 
\; J_0(k_{1\perp} r_{\!\perp}) J_0(k_{2\perp} r_{\!\perp}) \; P(k_1) P(k_2) ,
\label{zetap-circ-3D-12-1}
\eeqa
where we used Eq.(\ref{Jn-1Jn+1}) to integrate over $k_{3\perp}$
and we also integrated over $\{k_{3\parallel},r_{\parallel}\}$.
Next, writing the power spectra in terms of the two-point correlation functions,
we can integrate over wavenumbers by using Eq.(\ref{J2J0J0}). This yields
\beqa
\zeta_{(12)}^{\rm 3D} & \! = \! & \frac{S_3}{3}
\int_0^{2d} \frac{\dd r_1 \, r_1}{\pi d^2} \; \xi(x_{1,3\parallel},r_1)
\left[ 2 \varphi_1 - \frac{d}{r_1 \sin\varphi_1} \right]  \nonumber \\
&& \hspace{-0.7cm} \times \int_0^{2d} \! \frac{\dd r_2 \, r_2}{\pi d^2} 
\, \xi(x_{2,3\parallel},r_2) \left[ 2 \varphi_2 - \frac{d}{r_2 \sin\varphi_2}
 \right] 
- \frac{S_3}{3} \! \int_0^d \! \frac{\dd r_{\perp} \, 2r_{\perp}}{d^2} \nonumber \\
&& \hspace{-0.7cm} \times \; \left( \int_{d-r_{\perp}}^{d+r_{\perp}} 
\frac{\dd r_1' \, r_1'}{\pi d^2} \, \xi(x_{1,3\parallel},r_1')
\left[ 2 \varphi_1' - \frac{d^2}{r_{\perp} \, r_1' \sin\varphi_1'} \right] \right. 
\nonumber \\
&& \hspace{-0.7cm} \left. + \int_0^{d-r_{\perp}} \frac{\dd r_1' \; 2 r_1'}{d^2} 
\, \xi(x_{1,3\parallel},r_1') \right) 
\left( \int_{d-r_{\perp}}^{d+r_{\perp}} \frac{\dd r_2' \, r_2'}{\pi d^2} \,
\xi(x_{2,3\parallel},r_2') \right. \nonumber \\
&& \hspace{-0.7cm} \left. \times 
\left[ 2 \varphi_2' - \frac{d^2}{r_{\perp} \, r_2' \sin\varphi_2'} \right]
\!+\! \int_0^{d-r_{\perp}} \! \frac{\dd r_2' \; 2 r_2'}{d^2} \, 
\xi(x_{2,3\parallel},r_2') \! \right) ,
\label{zetap-circ-3D-12-2}
\eeqa
where $x_{i,j\parallel}=x_{i\parallel} - x_{j\parallel}$ and the angles $\varphi_i$ and
$\varphi_i'$ are given by Eq.(\ref{phi_i-phi_ip}).
The second and third contributions $\zeta_{(13)}^{\rm 3D}$ and
$\zeta_{(23)}^{\rm 3D}$ are also given by Eq.(\ref{zetap-circ-3D-12-2}) through
permutations over the indices $\{1,2,3\}$.

\section{Comparison of models for the lensing three-point functions}
\label{comparison-3pt}

\begin{figure}
\begin{center}
\epsfxsize=8.5 cm \epsfysize=6. cm {\epsfbox{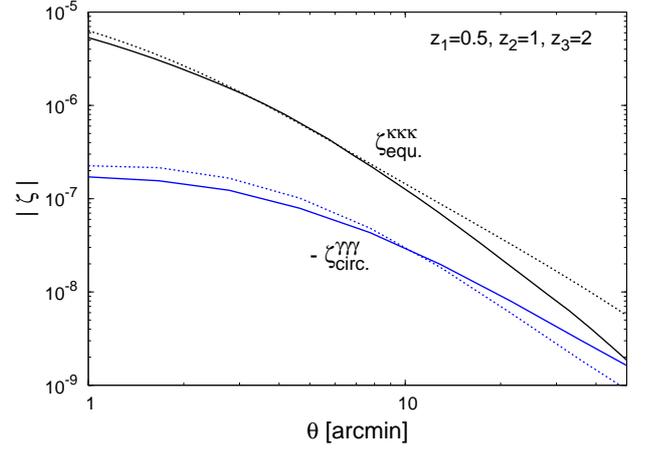}}
\end{center}
\caption{Convergence and shear three point correlations 
$\zeta^{\kappa\kappa\kappa}_{\rm equ.}$ and
$\zeta^{\gamma\gamma\gamma}_{\rm circ.}$,
as a function of the angular scale $\theta$, for the redshift triplet
$z_1=0.5, z_2=1, z_3=2$.
We show the predictions from the model of \citet{Valageas2012a,Valageas2012b} 
(solid lines) and from the hierarchical ansatz (\ref{zeta-def}) (dotted lines).
Because $\zeta^{\gamma\gamma\gamma}_{\rm circ.}$ is negative we plot
$-\zeta^{\gamma\gamma\gamma}_{\rm circ.}$.}
\label{fig_zeta_ansatz_z_z_z}
\end{figure}

We compare in Fig.~\ref{fig_zeta_ansatz_z_z_z} the predictions for three-point
weak lensing correlations from the model of \citet{Valageas2012a,Valageas2012b},
which combines one-loop perturbation theory with a halo model
\citep{Valageas2011e} and has been checked against ray-tracing numerical
simulations, with the predictions from the hierarchical ansatz (\ref{zeta-def}).
We consider the source redshift triplet $\{0.5,1,2\}$ but other redshifts give
similar results.

The shear three-point correlation is typically smaller than the convergence
one because of the spin-2 factor $e^{2\ii\alpha}$, which leads to some
cancellations as seen from the counterterms in Eq.(\ref{zetap-circ-zeta2D})
for the circular average (\ref{zeta-gamma-circ-def}).
This effect is larger on smaller scales where the slope of
$\zeta^{\kappa\kappa\kappa}_{\rm equ.}$ is lower.

Figure~\ref{fig_zeta_ansatz_z_z_z} shows that the hierarchical ansatz
(\ref{zeta-def}) provides the correct order of magnitude for weak lensing
three-point functions on scales $\theta \la 50'$.
More precisely, both approximations agree to better than a factor $1.5$
for $\theta<10'$ and a factor $3$ for $\theta<40'$, 
for $\zeta^{\kappa\kappa\kappa}_{\rm equ.}$; and
to better than a factor $1.5$
for $\theta<30'$ for $\zeta^{\gamma\gamma\gamma}_{\rm circ.}$.
Because most of the cosmological information from weak lensing three-point
correlations measured in galaxy surveys comes from $\theta \la 10'$, as
the amplitude of the signal decreases on larger scales, the hierarchical
ansatz (\ref{zeta-def}) would be sufficient to estimate the relative importance of the
source-lens clustering bias (which itself is dominated by contributions that only
depend on the two-point density correlation) or of other sources of noise.
This provides significantly faster numerical computations.

\bibliographystyle{aa} 
\bibliography{ref3}

\end{document}